\documentclass[11pt,a4paper]{article}
\usepackage{amsmath,amsfonts,amssymb,amsthm,amstext,amscd,array}
\usepackage{tikz}
\usepackage{mathrsfs}
\usepackage{bbm}
\usepackage{bm}
\usepackage[arrow,matrix,curve]{xy}

\def\nn{\nonumber}

\newcommand{\sgn}{{\rm sgn}}

\newcommand{\Z}{\zed}

\newcommand{\mbf}[1]{{\boldsymbol {#1} }}
\def\ii{{\,{\rm i}\,}}
\def\dd{{\rm d}}

\def\Id{{\rm id}}

\def\sfd{{\sf d}}
\def\sfs{{\sf s}}
\def\sft{{\sf t}}

\def\CE{{\tt CE}}

\newcommand{\unit}{\mathbbm{1}}   			% identity map/matrix

\def\mx{{\mbf x}}

\def\mbp{{\mbf p}}

\newcommand{\frg}{\mathfrak{g}}				% frak-letters
\newcommand{\fru}{\mathfrak{u}}				% frak-letters

\newcommand{\CV}{\mathcal{V}}
\newcommand{\CCV}{\mathscr{V}}
\newcommand{\CCC}{\mathscr{C}}
\newcommand{\CCG}{\mathscr{G}}
\newcommand{\CCB}{\mathscr{B}}
\newcommand{\CCW}{\mathscr{W}}
\newcommand{\CCP}{\mathscr{P}}
\newcommand{\CCF}{\mathscr{F}}

\newcommand{\CX}{\mathcal{X}}
\newcommand{\CY}{\mathcal{Y}}
\newcommand{\CZ}{\mathcal{Z}}
\newcommand{\CK}{\mathcal{K}}
\newcommand{\Ccal}{\mathcal{C}}

\newcommand{\eq}{\begin{equation}}
\newcommand{\eqend}{\end{equation}}
\newcommand{\eqa}{\begin{eqnarray}}
\newcommand{\nonueqa}{\begin{eqnarray*}}
\newcommand{\eqaend}{\end{eqnarray}}
\newcommand{\nonueqaend}{\end{eqnarray*}}

\newcommand{\bma}[1]{\begin{array}{#1}}
\newcommand{\ema}{\end{array}}
\newcommand{\bc}{\begin{center}}
\newcommand{\ec}{\end{center}}

\renewcommand{\thefootnote}{\fnsymbol{footnote}}
\newcommand{\newsection}{\setcounter{equation}{0}\section}

\newcommand{\complex}{{\mathbb C}} %% complex numbers
\newcommand{\zed}{{\mathbb Z}} %% integers
\newcommand{\real}{{\mathbb R}} %% real numbers

\newcommand{\torus}{{\mathbb T}} %% torus
\newcommand{\quat}{{\mathbb H}} %% torus
\def\alg{{\mathcal A}}

\def\Lcal{{\mathcal L}}

\newif\ifold             \oldtrue

\def\nn{\nonumber}

\def\e{{\,\rm e}\,}

\def\be{\begin{equation}}
\def\ee{\end{equation}}
\def\bea{\begin{eqnarray}}
\def\eea{\end{eqnarray}}
\def\bd{\begin{displaymath}}
\def\ed{\end{displaymath}}

\def\s{\sigma}

\newcommand{\ph}{\phantom{ }}

\newcommand{\beq}{\begin{eqnarray}}
\newcommand{\eeq}{\end{eqnarray}}

\newcommand{\sg}{\sigma}

\makeatletter
\newdimen\normalarrayskip              % skip between lines
\newdimen\minarrayskip                 % minimal skip between lines
\normalarrayskip\baselineskip
\minarrayskip\jot
\newif\ifold             \oldtrue            
\def\arraymode{\ifold\relax\else\displaystyle\fi} % mode of array entries
\def\@arrayskip{\ifold\baselineskip\z@\lineskip\z@
     \else
     \baselineskip\minarrayskip\lineskip2\minarrayskip\fi}
\def\@arrayclassz{\ifcase \@lastchclass \@acolampacol \or
\@ampacol \or \or \or \@addamp \or
   \@acolampacol \or \@firstampfalse \@acol \fi
\edef\@preamble{\@preamble
  \ifcase \@chnum
     \hfil$\relax\arraymode\@sharp$\hfil
     \or $\relax\arraymode\@sharp$\hfil
     \or \hfil$\relax\arraymode\@sharp$\fi}}
\def\@array[#1]#2{\setbox\@arstrutbox=\hbox{\vrule
     height\arraystretch \ht\strutbox
     depth\arraystretch \dp\strutbox
     width\z@}\@mkpream{#2}\edef\@preamble{\halign \noexpand\@halignto
\bgroup \tabskip\z@ \@arstrut \@preamble \tabskip\z@ \cr}%
\let\@startpbox\@@startpbox \let\@endpbox\@@endpbox
  \if #1t\vtop \else \if#1b\vbox \else \vcenter \fi\fi
  \bgroup \let\par\relax
  \let\@sharp##\let\protect\relax
  \@arrayskip\@preamble}
\makeatother

\newcommand{\wg}{\wedge}

\newcommand{\p}{\partial}
\newcommand{\mc}{\mathcal}

\newcommand{\trm}{\textrm}

\begin{document}
\begin{titlepage}
\begin{flushright}

\baselineskip=12pt

HWM--12--08\\
EMPG--12--12
%\hfill{ }\\
%\today
\end{flushright}

\begin{center}

\vspace{2cm}

\baselineskip=24pt

{\Large\bf Membrane Sigma-Models \\ and Quantization of
Non-Geometric Flux Backgrounds}

\baselineskip=14pt

\vspace{1cm}

{\bf Dionysios Mylonas}${}^{1,2}$ \footnote{Email: \ {\tt dm281@hw.ac.uk}}, {\bf
  Peter Schupp}${}^{1,2,3}$ \footnote{Email: \ {\tt
    p.schupp@jacobs-university.de}} and {\bf Richard
  J. Szabo}${}^{1,2}$ \footnote{Email: \ {\tt R.J.Szabo@hw.ac.uk}}
\\[4mm]
\noindent ${}^1${\it Department of Mathematics, Heriot--Watt University\\ Colin Maclaurin Building,
  Riccarton, Edinburgh EH14 4AS, U.K.}
\\[4mm]
\noindent ${}^2${\it Maxwell Institute for
Mathematical Sciences\\ Edinburgh, U.K.}
\\[4mm]
\noindent ${}^3${\it Jacobs University
  Bremen\\ 28759 Bremen, Germany}
\\[30mm]

\end{center}

\begin{abstract}

\baselineskip=12pt

We develop quantization techniques for describing the nonassociative geometry
probed by closed strings in flat non-geometric $R$-flux
backgrounds $M$. Starting from a suitable Courant
sigma-model on an open membrane with target space $M$, regarded as a topological sector of closed string dynamics
in $R$-space, we derive a twisted Poisson sigma-model on
the boundary of the membrane whose target space is the cotangent
bundle $T^*M$ and whose quasi-Poisson structure coincides
with those previously proposed. We argue that from the
membrane perspective the path integral over multivalued closed string
fields in $Q$-space is equivalent to integrating over open strings in $R$-space. The corresponding boundary correlation functions reproduce
Kontsevich's deformation quantization formula for the twisted
Poisson manifolds. For constant $R$-flux, we derive closed
formulas for the corresponding nonassociative star product and its
associator, and compare them with
previous proposals for a 3-product of fields on $R$-space. We
develop various versions of the Seiberg--Witten map which relate our
nonassociative star products to associative ones and add fluctuations
to the $R$-flux background. We show that the
Kontsevich formula coincides with the star product obtained by
quantizing the dual of a Lie 2-algebra via convolution in an
integrating Lie 2-group associated to the T-dual doubled geometry, and
hence clarify the relation to the twisted convolution products for topological nonassociative
torus bundles. We further demonstrate how our approach leads to a
consistent quantization of Nambu--Poisson 3-brackets.

\end{abstract}

\end{titlepage}
\setcounter{page}{2}

\newpage

{\baselineskip=12pt
\tableofcontents
}

\renewcommand{\thefootnote}{\arabic{footnote}}
\setcounter{footnote}{0}

\newsection{Introduction and summary}

\subsection*{Background and context}

String vacua with $p$-form field fluxes along the extra dimensions are
called flux compactifications and have been intensively studied in
recent years because of their ability to cure some of the problems
suffered by the more conventional Calabi--Yau
compactifications~\cite{Grana,DKrev,BKLSrev}. They have also provided
broader notions of string geometry, and compactifications on
non-geometric spaces have emerged as consistent string
vacua. Non-geometric backgrounds can arise from taking T-duality
transformations of conventional geometric backgrounds, and their
non-geometric nature is reflected in the fact that the transition
functions between local patches typically involve string duality
transformations.

A prototypical example of a non-geometric space is obtained by T-dualising a
three-torus $\torus^3$ with non-vanishing three-form
$H$-flux~\cite{Hull:2004in,Shelton:2005cf,Dabholkar:2005ve}. T-dualising along one cycle gives rise to a twisted torus
with geometric $f$-flux related to a metric connection on the tangent
bundle of the
nilmanifold. However, T-dualising along two cycles gives rise to a
space with non-geometric $Q$-flux whereby one of the cycles is only periodic up to T-duality,
which mixes momentum and winding modes; the resulting geometry is thus
only well-defined locally and is called a T-fold. An approach to
describing these backgrounds mathematically in the context of open
string theory was put forward
in~\cite{Mathai:2004qq,Ellwood:2006my,Grange:2006es,Bouwknegt:2008kd} using the
language of noncommutative geometry: The T-fold can be regarded as a
fibration of noncommutative two-tori over the base circle. A more
radical possibility comes about when one T-dualises along all three
circles, which takes a geometric torus $\torus^3$ with uniform
$H$-flux to a space with non-geometric $R$-flux; it provides a
realisation of the expectations that there should exist a
generalization of T-duality that applies to torus bundles with
non-isometric torus action. The precise geometric meaning of
$R$-spaces has remained somewhat elusive. In the context of open string
theory it is described in~\cite{Bouwknegt:2004ap,Ellwood:2006my} using
nonassociative twisted convolution algebras, which are defined
in~\cite{Bouwknegt:2007sk} as objects internal to a tensor category
with a weak monoidal structure: The $R$-space can be regarded
generally as a bundle of noncommutative and nonassociative tori.

The open string picture of the geometry of T-folds is a result of the
$B$-field experienced by the fibre directions, which depends on the
coordinate of the base circle. It can be understood from the way in
which noncommutative geometry arises in open string theory as a result
of a constant two-form $B$-field on D-branes~\cite{Seiberg:1999vs}. Canonical quantization of the open
string sigma-model results in commutation relations for the string endpoints given by
\beq 
\big[X^i (\tau, \sigma )\,,\, X^j (\tau, \sigma '\, ) \big] \Big{|}_{\sigma =\sigma ' =0, 2\pi } = \ii\theta^{ij} \label{12} \eeq
where $\theta= -2\pi \, \alpha' \, (1+\mc{F}^2)^{-1}\,
\mc{F}$, $\mc{F}=B-F$ with $F$ the gauge field strength two-form on
the D-brane. The D-brane worldvolume becomes a noncommutative space
with a low-energy effective field theory that is given by a
noncommutative gauge theory in a double-scaling limit $\alpha'\to0,B\to\infty$ which
decouples the open strings from closed strings, i.e. from gravity~\cite{Seiberg:1999vs}. The
resulting twisted gauge symmetries suggest that these noncommutative
field theories are models for general relativity which contain
emergent and noncommutative gravity (see e.g.~\cite{Szabo:2006wx} for
a review).

However, it is more natural to seek
noncommutative gravity structures emerging in the sector of \emph{closed
  strings}. A similar computation for closed strings reveals that the equal-time
equal-position commutator is a function of the worldsheet coordinates,
and therefore not a well-defined target space object. However, the
authors of~\cite{Blumenhagen:2010hj,Blumenhagen:2011ph} define a
3-bracket as the Jacobiator (which they call the ``{cyclic double commutator}'')
\beq [X^i , X^j , X^k ] := \lim_{\sg_i \to\sg} \, \big[ [ X^i (\tau,
\sg_1 ), X^j (\tau, \sg_2) ], X^k (\tau, \sg_3) \big ] +\trm{cyclic}
\ . \label{14} \eeq
Calculating this for the SU(2) WZW model and for a linearized conformal field theory in
flat space with $H$-flux they obtain the non-trivial result
\beq [X^i,X^j,X^k]=\ii \alpha\, \theta^{ijk} \ , \label{15}\eeq
where $\theta^{ijk}$ is proportional to the background flux,
and $\alpha=0$ for the $H$-flux background while $\alpha=1$ after an odd number of 
T-duality transformations.

In~\cite{Lust:2010iy} the same type of nonassociativity emerges in a somewhat different context, via T-duality on a three-torus $M=\torus^3$
with non-vanishing $H$-flux. The basic mechanism is that while on
geometric spaces in the presence of 3-flux (either NS--NS $H$-flux or
metric $f$-flux) the closed string sigma-model fields commute,
$[X^i(\tau,\sigma), X^j(\tau,\sigma)]=0$, on non-geometric spaces
(T-folds or $R$-spaces) they no longer commute, $[X^i(\tau,\sigma),
X^j(\tau,\sigma)]\neq 0$. Via T-duality, this maps back to the
original geometric space as a noncommutativity relation
\beq
\big[X^i(\tau,\sigma)\,,\,X_j^*(\tau,\sigma)\big]\neq 0
\eeq
between string coordinates $X^i\in M$ and the dual coordinates
$X^*_i\in M^*$. This suggests that T-duality and doubled geometry is
the natural framework to investigate closed string
noncommutative geometry. This point was noticed already some time ago for tori
with constant $B$-fields
in~\cite{Landi:1998ii,Lizzi:1999nm}, while noncommutative
correspondence spaces associated to T-folds are described
in~\cite{Brodzki:2007hg,Bouwknegt:2008kd} in the context of open strings. In~\cite{Lust:2010iy} this was
demonstrated by applying T-duality along one direction of a twisted three-torus with
non-vanishing geometric flux, which reveals that there is a non-trivial
commutation relation between the coordinates in the doubled geometry
that is determined by the winding number in the base direction. This led to the
conjecture that after three T-dualities on the original $\torus^3$
with 3-flux the coordinates satisfy a nonassociative phase space
coordinate algebra whose Jacobiator reproduces the 3-bracket
(\ref{15}), confirming that the
source of the nonassociativity structure is the non-trivial
$H$-flux~\cite{Lust:2010iy}, as in~\cite{Bouwknegt:2004ap};
in~\cite{Lust:2012fp} it was pointed out that these phase space
relations define a twisted Poisson structure. In this
setting, double field theory is the natural framework for
investigating the effects of T-duality in the context of double
geometry (see e.g.~\cite{Hohm:2011gs} for a review); the doubled
geometry in this case is the double torus $M\times M^*$ including
the momentum coordinates together with the dual winding
coordinates. It provides a means for formulating effective string
actions which are invariant under T-duality
transformations. Noncommutative and nonassociative phase space structures were also found
on geometric twisted tori in~\cite{Chatzistavrakidis:2012yp} from a
different point of view as solutions of matrix theory compactification
conditions.

In~\cite{Lust:2010iy} it was further argued that a
framework for understanding closed string nonassociative geometry in
the context of double field theory is provided by an analogy with open
string noncommutativity. Using T-duality, a noncommutative field
theory on D-branes may be mapped to an ordinary field theory on
D-branes intersecting at angles. A closed string on the doubled
$\torus^3$ geometry has boundary conditions satisfied by momentum
states on the double twisted torus as
\beq
X(\tau ,
\s+2\pi)=\e^{2\pi\ii\theta} \, X(\tau,\s) \qquad \mbox{and} \qquad X^* (\tau+2\pi
,\s)=\e^{2\pi\ii\theta} \, X^*(\tau ,\s) \ ,
\eeq
where $\theta=- n\, H$ is
determined by the dual momentum $n\in\zed$ in the direction along
which T-duality acts, while the directions are exchanged for the dual
$H$-flux background ($X\leftrightarrow X^*$). This situation resembles
the open string case: T-duality switches between the directions of
commutative and noncommutative boundary conditions in a manner dictated by the dual momentum. Recall that noncommutativity arises in open string theory as a result of a $B$-field background and a field strength $F$ in the gauge theory on a D-brane. By means of the Seiberg--Witten map this noncommutative gauge theory may be mapped to an ordinary gauge theory~\cite{Seiberg:1999vs}. This is similar to the role that T-duality assumes here, and an interesting question is whether there is a map in closed string theory that exchanges a nonassociative field theory with an associative field theory.

\subsection*{Summary of results}

In this paper we shall investigate the origins of noncommutative and
nonassociative geometry
for closed strings in $R$-space, and relate it with open string
noncommutativity. We will develop two equivalent nonassociative
quantizations of constant $R$-flux backgrounds and connect them to the
previous studies reviewed above. Our starting point is the observation
that the open string sigma-model with a closed $B$-field
background in the limit that decouples it from gravity is given by the
Poisson sigma-model, whose boundary correlation functions naturally
lead to Kontsevich's star product for the deformation quantization of fields
along the associated Poisson structure~\cite{Cattaneo:1999fm}. We will argue that the
suitable analog for closed strings in flux backgrounds is a
higher version of the Poisson sigma-model called the Courant sigma-model; this is a
sigma-model on an open three-dimensional \emph{membrane}, with the boundaries regarded as the closed strings, whose target space
is $M$ and whose field content is valued in a Courant algebroid. In the present case the pertinent Courant
algebroid over $M$ is the standard Courant algebroid $C= TM\oplus
T^*M$. This setting is natural from the point of view of double field theory
alluded to above: If $Y$ denotes the double twisted torus which is T-dual
to $M\times M^*$, then the structure
algebra of $C$ given by the twisted Courant--Dorfman
bracket coincides with the usual Lie bracket on vector fields in the
tangent Lie algebroid $TY$~\cite{Hull:2009sg}. The relevance of Courant algebroids in non-geometric flux
compactifications and to gauge symmetries in double field theory has also been noted
in~\cite{Halmagyi:2009te,Hull:2009zb,Blumenhagen:2012pc}. We will show that with
the structure functions of $C$ appropriate to the pure constant
$R$-flux background, the membrane sigma-model reduces to a twisted
Poisson sigma-model on the boundary whose target space is the
\emph{cotangent bundle} $T^*M$ of the original target space $M$. The
twist is given by a non-flat $U(1)$-gerbe on \emph{momentum space},
and the resulting linear twisted Poisson structure coincides exactly
with that proposed in~\cite{Lust:2010iy,Lust:2012fp}. Our membrane
sigma-model thus gives a straightforward dynamical explanation of these
nonassociative phase space relations and also a geometric
interpretation for the effective target space geometry seen by closed
strings in the $R$-flux compactification; it appears to conform with
the general expectation that the background fields in non-geometric
spaces have non-trivial dependence on the extra dual coordinates in
the doubled geometry representation~\cite{Dabholkar:2005ve}.

With the twisted closed string boundary conditions considered
in~\cite{Lust:2010iy,Condeescu:2012sp}, we will then argue that the
closed string path integral is equivalent to that of an \emph{open}
string twisted Poisson sigma-model on a disk. This sort of open/closed
string duality suggests that considering strings in non-geometric
spaces as membranes in geometric spaces appears to rectify the problem of the
non-decoupling of gravity, observed
in~\cite{Herbst:2001ai} for open strings in $H$-space and
in~\cite{Ellwood:2006my} for open strings in
$R$-space. As mentioned above, the
resulting boundary correlation functions naturally define a quantization of the
$R$-flux background, and in fact one can reproduce the entire setting of
Kontsevich's global deformation quantization for twisted Poisson
structures. We develop this formalism for arbitrary (not necessarily
constant) $R$-flux and describe the resulting nonassociative star
product, the corresponding associator, as well as their various
derivation properties through the formality maps; for constant
$R$-flux we derive explicit closed formulas which resemble the Moyal--Weyl formula. We shall also see
that our formalism appears to have the right features to define a
proper nonassociative quantization of Nambu--Poisson 3-brackets, at
least for constant trivectors, which are relevant for the quantum
geometry of M-branes (see e.g.~\cite{DeBellis:2010pf,Saemann:2011yi,Saemann:2012ex}). We will further demonstrate how the 3-product of
fields proposed in~\cite{Blumenhagen:2011ph} (see
also~\cite{Takhtajan:1993vr}) arises in special subsectors of our
general formalism.

Using this approach we are also able to clarify the meaning of
Seiberg--Witten maps in this setting. The deformation quantization of
twisted Poisson structures leads to noncommutative gerbes in the sense
of~\cite{Aschieri:2002fq}. The nonassociative star product can be ``untwisted'' to a family of associative star products that are all related by Seiberg--Witten maps. This formulation has a large gauge symmetry given by star commutators with gauge parameters that live on phase space.
In the particular setting that we study in this paper, we can also
work directly with the nonassociative star product by restricting the
class of admissible gauge fields: Seiberg--Witten maps can then be
used to describe fluctuations at the boundary of the membrane (or at
the endpoints of the open strings) and are closely related to
quantized general coordinate transformations. We find two particularly interesting
examples. Firstly, a dynamical Seiberg--Witten map from the
associative canonical star product on phase space to the
nonassociative $R$-twisted star product; this map can be computed
explicitly to all orders in closed form and may be the first example
of its kind. Secondly, Nambu--Poisson maps that can be used to add fluctuations to the (constant) $R$-flux background.

Our second approach to the deformation quantization of the $R$-flux
background introduces the
appropriate mathematical language to deal with the target space
nonassociativity which should prove helpful in the development of
nonassociative deformations of gravity, and ultimately of double field
theory. Our twisted Poisson structure is linear and we show that it
has the structure of a Lie 2-algebra, which is a categorified version
of an ordinary Lie algebra in which the Jacobi identity is weakened to
a natural transformation. The pertinent Lie 2-algebra that we use is
in fact related to the noncommutative $Q$-space background where
closed string noncommutativity originally appears, and it can be
regarded as a reduction of the structure algebra of the Courant
algebroid $C$ over a point. This Lie 2-algebra can be integrated to a
Lie 2-group $\CCG$ that is a categorification of the Heisenberg group which
defines the double twisted torus $Y$. By using the nonassociative convolution product
induced by horizontal multiplication in $\CCG$, we induce a
nonassociative star product on the algebra of functions on phase space
by embedding it as an algebra object in the category $\CCG$, in the
spirit of~\cite{Bouwknegt:2007sk}; this mapping can be regarded as a
higher version of the Weyl--Wigner quantization map which is familiar from
conventional approaches to noncommutative field theory~\cite{Szabo:2001kg}. We demonstrate that this star
product is identical to the nonassociative Kontsevich star product;
this alternative derivation lends further credibility to our membrane
perspective of the closed string dynamics in $R$-space. In this
manner, we can precisely relate the closed string noncommutative and nonassociative
backgrounds with the noncommutative and nonassociative torus bundles that
were proposed by~\cite{Bouwknegt:2004ap} to capture the effective
geometry of strings in $R$-flux compactifications.

\subsection*{Outline}

The outline of the remainder of this paper is as follows. In
Section~\ref{TTHA} we develop the Courant sigma-model which we propose
as a description of a sector of the closed string dynamics in
$R$-space; we reduce the membrane path integral to a string path
integral corresponding to a twisted Poisson sigma-model, and argue
that it can be quantized as an open string theory with worldsheet a
disk. In Section~\ref{Defquant} the corresponding boundary correlation functions are considered
 which develop Kontsevich's global deformation
quantization of fields along the twisted Poisson structure. In
Section~\ref{BCHquant} we construct our alternative deformation quantization of
the $R$-flux background via convolution in a suitable Lie 2-group and
demonstrate that it coincides with the approach based on Kontsevich's
formula. Two appendices at the end of the paper are delegated to some
of the more technical aspects of our analysis. In
Appendix~\ref{HigherLA} we review in some detail all notions
regarding the higher algebraic and geometric structures that are
employed in the main text. In Appendix~\ref{Graphs} we present some technical details of
the explicit computation of Kontsevich's formula.

\newsection{AKSZ sigma-models in $\mbf R$-space\label{TTHA}}

In this section we propose sigma-models for closed strings in
$R$-flux backgrounds. The Poisson sigma-model with target space $M$
describes
the topological sector of string theory in two-form $B$-field
backgrounds. To incorporate non-trivial three-form fluxes, one instead
needs a coupling to \emph{membranes}, which motivates the need for
using higher mathematical structures for the twistings that arise in these
instances. The effective dynamics in three-form flux backgrounds is
thus provided by suitable
Courant sigma-models with target space $M$ which describe topological sectors of membrane
theories. We will first review how the sigma-model appropriate to $H$-space can be reduced on
the boundary of an open membrane to a twisted Poisson sigma-model with
target space
$M$~\cite{Park:2000au,Hofman:2002rv,Hofman:2002jz,Bouwknegt:2011vn}. Then
we will show that for constant $R$-flux the appropriate Courant sigma-model
reduces to a string theory with target space the \emph{cotangent
  bundle} of $M$ with twisted Poisson structure which coincides with
that found in~\cite{Lust:2010iy,Lust:2012fp}. This geometric interpretation of the
$R$-flux background is related to the doubled geometry
description of non-geometric flux
compactifications~\cite{Hull:2004in,Dabholkar:2005ve}, and also to the
description of the twisted Poisson structure on $T^*M$ as an ordinary
Poisson structure on the loop space of $M$~\cite{Saemann:2012ex}. 

\subsection{Poisson and Courant sigma-models\label{PCsigma}}

AKSZ
sigma-models whose target spaces comprise a symplectic Lie $n$-algebroid $E$
over a manifold $M$ may be constructed using higher Chern--Simons
action functionals~\cite{Kotov:2007nr,Fiorenza:2011jr} (see Appendix~\ref{NWQuant}
for the relevant details concerning algebroids). A
simple case is the cotangent Lie algebroid $E=T^*M$ over a Poisson
manifold $M$ with Poisson bivector $\Theta=\frac12\, \Theta^{ij}(x)\,
\partial_i\wedge \partial_j$ where $x=(x^i) \in M$ are local coordinates with
$\partial_i:=\frac\partial{\partial x^i}$; this is a symplectic Lie
1-algebroid with the canonical symplectic structure on the cotangent
bundle $T^*M$. Let $\Sigma_2$ be a two-dimensional string worldsheet. The AKSZ
construction defines a topological field theory on $C^\infty(T\Sigma_2,
T^*M)$ (regarded as a space of Lie algebroid morphisms). A Poisson Lie algebroid-valued differential form on $\Sigma_2$
is given by the smooth embedding $X=(X^i):\Sigma_2\to M$ of the string
worldsheet in target space,
and an auxilliary one-form field on the worldsheet $\xi=(\xi_i) \in\Omega^1(\Sigma_2,X^*T^*M)$. The corresponding AKSZ
action is
\beq
S_{\rm AKSZ}^{(1)} = \int_{\Sigma_2}\, \Big(\xi_i\wedge \dd X^i+\frac12\,
\Theta^{ij}(X)\, \xi_i\wedge \xi_j \Big) \ ,
\eeq
which coincides with the action of the Poisson
sigma-model~\cite{Ikeda:1993fh,Schaller:1994es,Cattaneo:1999fm,Cattaneo:2001ys}. The Poisson
sigma-model is the most general two-dimensional topological field
theory that can be obtained from the AKSZ construction.

Note that although on-shell the bivector field $\Theta$ is required to have
vanishing Schouten--Nijenhuis bracket with itself (in particular so
that it defines a differential $\dd_\Theta$ on the algebra of multivector fields, see Appendix~\ref{SNbrackets}), the perturbative
expansion of~\cite{Cattaneo:1999fm} still makes sense when $\Theta$ is
a twisted Poisson bivector and reproduces the Kontsevich
formality maps for nonassociative star
products~\cite{Cattaneo:2001vu}; the topological nature of the Poisson
sigma-model allows for it to be perturbatively expanded around a non-vacuum solution.

A Courant structure is the first higher analog of a
Poisson structure. The
corresponding AKSZ sigma-model has target space comprising a symplectic
Lie 2-algebroid with a ``degree 2 symplectic form'', which is the same thing as a Courant algebroid $E$
over a manifold $M$~\cite{Royt1}. In~\cite{Royt1} it is shown that Courant
algebroids $E\to M$ are in a canonical bijective correspondence with
AKSZ sigma-models on a three-dimensional membrane worldvolume ${\Sigma_3}$. A
Courant algebroid-valued differential form on ${\Sigma_3}$ is given
by the smooth embedding of the membrane
worldvolume $X=(X^i):{\Sigma_3}\to M$ in target space, a one-form $\alpha=(\alpha^I)\in\Omega^1(\Sigma_3,X^*E)$, and an auxilliary two-form field on the worldvolume
$\phi=(\phi_i)\in\Omega^2({\Sigma_3},X^*T^*M)$. The structure functions of the Lie
2-algebroid are specified by choosing a local basis of sections
$\{\psi_I\}$ of $E\to M$ such that the fibre 
metric $h_{IJ}:=\langle \psi_I, \psi_J\rangle$ is constant. We
define the anchor matrix $P_I{}^i$ by $\rho(\psi_I)=P_I{}^i(x)\,
\partial_i$, and the three-form $T_{IJK}(x):=[\psi_I,\psi_J,\psi_K]_E$. Then the
canonical three-dimensional topological field theory associated to the
Courant algebroid $E\to M$ is described by the AKSZ
action
\beq
S_{\rm AKSZ}^{(2)} = \int_{\Sigma_3}\, \Big(\phi_i\wedge \dd X^i+\frac12\,
h_{IJ}\, \alpha^I\wedge \dd\alpha^J-P_I{}^i(X)\, \phi_i\wedge \alpha^I+
\frac16\, T_{IJK}(X)\, \alpha^I\wedge\alpha^J \wedge\alpha^K \Big) \ ,
\label{SAKSZ2}\eeq
which is the action of the Courant sigma-model~\cite{Ikeda:2002wh,Hofman:2002rv,Royt2}.

\subsection{$H$-space sigma-models\label{Htwist}}

The Courant algebroid of exclusive interest in geometric flux
compactifications of string theory is the {\em
  standard Courant algebroid} $C=TM\oplus T^*M$ twisted by a closed
NS--NS three-form flux $H=\frac16\, H_{ijk}(x)\, \dd x^i\wedge\dd x^j\wedge \dd x^k$. The structure maps of $C$ comprise the skew-symmetrization of the \emph{$H$-twisted
Courant--Dorfman bracket} given by~\cite{Severa:2001qm}
\bea
 \big[(Y_1, \alpha_1)\,,\,(Y_2,\alpha_2)\big ]_H &:=& \big([Y_1,Y_2]_{TM}
 \,,\,
 \Lcal_{Y_1}\alpha_2-\Lcal_{Y_2}\alpha_1 \\ && \qquad \qquad
 \qquad -\, \mbox{$\frac12$}\, \dd
 \big(\alpha_2(Y_1)-\alpha_1(Y_2)\big)+ H(Y_1,Y_2,-)
 \big) \nonumber
\eea
for vector fields $Y_1,Y_2\in C^\infty(M,TM)$ and one-form fields
$\alpha_1,\alpha_2\in \Omega^1(M)$, the metric is the natural dual pairing between $TM$ and $T^*M$,
\begin{equation}
 \big\langle (Y_1,\alpha_1)\,,\,
 (Y_2,\alpha_2)\big\rangle=\alpha_2(Y_1)+ \alpha_1(Y_2)~,
\end{equation}
and the anchor map is the trivial projection $\rho:C\rightarrow
TM$ onto the first factor; the map $\sfd:C^\infty(M)\to C^\infty(M,C)$ is given by $\sfd
f=\frac12\, \dd f$. This is an {\em exact} Courant algebroid, i.e. it fits into the short exact sequence
\begin{equation}
 0 \ \longrightarrow \ T^*M \ \xrightarrow{ \ \rho^* \ } \ C\
 \stackrel{\rho}{\longrightarrow} \ TM\ \longrightarrow \ 0~ ,
\end{equation}
where $\rho^*:T^*M\to C^*$ is the transpose of the anchor map $\rho$
followed by the identification $C^*\cong C$ induced by the pairing on
the Courant algebroid.
Every exact Courant algebroid on $M$ is
isomorphic to one of the form $C=TM\oplus T^*M$ with the structure
maps given as above; the isomorphism classes are parametrized by
elements $[H]\in H^3(M,\real)$ of the
degree~$3$ real cohomology of the target space.

To determine the structure maps of the exact Courant
algebroid in a convenient basis, we suppose henceforth that the tangent bundle
$TM\cong M\times\real^d$ is trivial, where $d=\dim(M)$; this
assumption will avoid the appearence of geometric $f$-fluxes and
other fluxes, as eventually we will want to apply triple T-duality to
take us directly into the pure $R$-flux background. Then in local
coordinates $x= (x^i)$ for $M$, a natural frame for
$TM\oplus T^*M$ is given by
\beq
\varrho_i=\partial_i \qquad \mbox{and} \qquad \chi^i=\dd
x^i
\eeq
for $i=1,\dots,d$. Writing $\varrho_i$ for $(\varrho_i,0)$ and $\chi^i$ for
$(0,\chi^i)$ for simplicity, the metric is given by
\beq
\langle \varrho_i,\chi^j\rangle=\delta_i{}^j \ .
\label{YPmetric}\eeq
The corresponding twisted Courant--Dorfman algebra is isomorphic to the
algebra with the sole non-trivial brackets
\beq
[\varrho_i,\varrho_j]_H=H_{ijk}\, \chi^k \ .
\label{PijH}\eeq
The non-vanishing ternary brackets are given by (see Appendix~\ref{NWQuant})
\beq
[\varrho_i,\varrho_j,\varrho_k]_H = H_{ijk} \ .
\label{PijkH}\eeq
As reviewed in~\cite{Lust:2012fp}, the brackets (\ref{PijH}) and
(\ref{PijkH}) for constant $H$-flux mimic the phase space quasi-Poisson algebra of a charged particle in
the background field of a magnetic monopole~\cite{Jackiw:1984rd}.

We now write
\beq
(\alpha^I) = (\alpha^1,\dots,\alpha^{2d}):= (\alpha^1,\dots,\alpha^d,\xi_1,\dots,\xi_d)
\label{alphaIsplit}\eeq
where
$(\alpha^i)\in\Omega^1({\Sigma_3},X^*TM)$ and $(\xi_i)\in\Omega^1({\Sigma_3},
X^*T^*M)$; throughout, upper case indices $I,J,\dots\in\{1,\dots,2d\}$ run over directions
of the doubled geometry, while lower case indices
$i,j,\dots\in\{1,\dots,d\}$ run over directions of the original
configuration space. Then the action (\ref{SAKSZ2}) becomes
\beq
S_{\rm WZ}^{(2)} = \int_{\Sigma_3}\, \Big(\phi_i\wedge \dd X^i+
\alpha^i \wedge \dd\xi_i -\phi_i\wedge \alpha^i+
\frac16\, H_{ijk}(X)\, \alpha^i\wedge\alpha^j\wedge\alpha^k \Big) \ .
\eeq
When $\Sigma_2:=\partial{\Sigma_3}\neq\emptyset$, this is the action of the
canonical open topological membrane theory~\cite{Park:2000au}; in this case
we can take the consistent Dirichlet boundary
conditions $\alpha^i=\phi_i=0$ on $\Sigma_2$ (we could also take
$X^i=\xi_i=0$ and hybrids thereof; see~\cite{Hofman:2002rv} for a
discussion of the resulting modifications). One can also modify the action by
adding a boundary term of the form
\beq
S_{\rm WZ}^\partial = \oint_{\Sigma_2} \, \Big( \xi_i\wedge \dd
X^i+\frac12\, \Theta^{ij}(X)\, \xi_i\wedge \xi_j+ \Gamma^i{}_j(X)\,
\xi_i\wedge \alpha^j+\frac12\, \Xi_{ij}(X)\, \alpha^i\wedge
\alpha^j\Big) \ .
\label{SWZpartial}\eeq
In~\cite{Hofman:2002rv,Hofman:2002jz} only the $\Theta$-deformation is
kept, corresponding to a canonical transformation on the Courant
algebroid which gives
the boundary/bulk open topological membrane action
\beq
\widetilde{S}_{\rm WZ}^{\,(2)} = \int_{\Sigma_3}\, \Big(\phi_i\wedge \big(
\dd X^i- \alpha^i \big) +
\alpha^i \wedge \dd\xi_i +
\frac16\, H_{ijk}(X)\, \alpha^i\wedge\alpha^j\wedge\alpha^k \Big) +
\oint_{\Sigma_2} \, \frac12 \ \Theta^{ij}(X)\, \xi_i\wedge \xi_j
\ . \nonumber \\
\eeq
In this
case the consistent boundary conditions require that $\Theta=\frac12\, \Theta^{ij}(x)\,
\partial_i\wedge\partial_j$ is an
$H$-twisted Poisson bivector on $M$, i.e. its Schouten--Nijenhuis
bracket with itself is
given by
\beq
[\Theta,\Theta]_{\rm S}=\mbox{$\bigwedge^3$}\Theta^\sharp(H) \ ,
\eeq
and the Jacobi identity for the corresponding bracket is violated (see
Appendix~\ref{SNbrackets}); here $\bigwedge^3\Theta^\sharp(H)$
denotes the natural way to turn the three-form $H$ into a three-vector
by using $\Theta$ to ``raise the indices''. After integrating out the
two-form fields $\phi_i$ we arrive at the AKSZ action
\beq
\widetilde{S}_{\rm AKSZ}^{\,(1)} = \oint_{\Sigma_2} \, \Big(\xi_i\wedge \dd X^i+\frac12\,
\Theta^{ij}(X)\, \xi_i\wedge \xi_j \Big) +\int_{\Sigma_3} \ \frac16\, 
H_{ijk}(X)\, \dd X^i\wedge\dd X^j\wedge\dd X^k \ ,
\label{WZ-Poisson}\eeq
which is the action of the $H$-twisted
Poisson sigma-model with target space
$M$~\cite{Park:2000au,Klimcik:2001vg}. Note that including the last
term of (\ref{SWZpartial}) would result in an additional global $B$-field
coupling $\frac12\, \Xi_{ij}(x) \, \dd X^i\wedge\dd X^j$ on the string worldsheet.

\subsection{$R$-space sigma-models\label{Rtwist}}

The relevance of the topological twisted Poisson sigma-model (\ref{WZ-Poisson}) in
the effective theory of strings in $R$-flux backgrounds was noted
in~\cite{Ellwood:2006my,Halmagyi:2008dr}. Here we shall start
with the general Courant
sigma-model (\ref{SAKSZ2}) and the argument of~\cite{Halmagyi:2009te}
that the appropriate theory in $R$-space is
described by a non-topological membrane sigma-model, not a string theory; the
membrane action in this case is not generally equivalent to the action
of a string theory on
the boundary of a membrane. This would also corroborate the
observation of~\cite{Ellwood:2006my} that the $R$-space geometry does
not seem to exist as a low-energy effective description of string
theory, in the sense that open strings in $R$-space cannot be
consistently decoupled from gravity; the absence of a topological
limit and the non-decoupling of gravity for open strings in $H$-space
was also observed in~\cite{Herbst:2001ai}. In a sense to be
elucidated below, the
membrane theory geometrizes the non-geometric $R$-flux background, in a way
reminescent of the manner in which M-theory geometrizes string
dualities. 
In~\cite{Andriot:2011uh,Andriot:2012wx} potential target space effective actions for non-geometric $Q$-flux
and $R$-flux backgrounds were constructed in the context of double
field theory, which provides a geometrical role to the non-geometric
fluxes related to gauge transformations (diffeomorphisms); it would be
interesting to derive these effective descriptions from membrane
sigma-models of the sort described here.
Although the $R$-space is not even locally geometric as a Riemannian manifold~\cite{Shelton:2005cf}, in this
paper we work only at tree-level in the low-energy effective field theory on target
space where we can treat the $R$-space
locally as the original $d$-dimensional manifold $M$. 

The Courant algebroid pertinent to the $R$-flux background is again
the {standard Courant algebroid} $C=TM\oplus T^*M$, but now twisted by a
trivector flux $R=\frac16\,
R^{ijk}(x) \,\partial_i\wedge\partial_j\wedge\partial_k$ satisfying a
suitable integrability condition. The bracket on $C$ 
is the skew-symmetrization of Roytenberg's $R$-twisting of the {Courant--Dorfman bracket} given by~\cite{Royt3,Halmagyi:2009te,Blumenhagen:2012pc}
\bea
 \big[(Y_1, \alpha_1)\,,\,(Y_2,\alpha_2)\big ]_R &:=& \big([Y_1,Y_2]_{TM}+R(\alpha_1,\alpha_2,-)
 \,,\, \\ && \qquad 
 \Lcal_{Y_1}\alpha_2-\Lcal_{Y_2}\alpha_1 -\, \mbox{$\frac12$}\, \dd
 \big(\alpha_2(Y_1)-\alpha_1(Y_2)\big)
 \big) \ , \nonumber
\eea
while the remaining structure maps are identical to those of Section~\ref{Htwist}.

Writing the generators of the natural frame for $TM\oplus T^*M$ as
$\varrho_i$ and $\chi^i$ as before, the corresponding Roytenberg algebra is isomorphic to the
algebra with the non-trivial brackets
\beq
[\chi^i,\chi^j]_R=R^{ijk}\, \varrho_k
\eeq
and the metric (\ref{YPmetric}). When $R$ is a constant flux this is the $d$-dimensional Heisenberg
algebra; this mimicks the commutation relations for closed
string fields which are obtained by applying three T-duality transformations
to the $H$-space $M=\torus^3$~\cite{Lust:2010iy,Lust:2012fp}, with the remaining non-trivial structure map
\beq
[\chi^i,\chi^j,\chi^k]_R = R^{ijk} \ .
\eeq
Below we will recover the commutation relations of~\cite{Lust:2010iy,Lust:2012fp}
dynamically from an associated twisted Poisson sigma-model.

With the same splitting (\ref{alphaIsplit}), the action
(\ref{SAKSZ2}) in the pure $R$-flux background becomes
\beq
{S}_R^{(2)} = \int_{\Sigma_3}\, \Big(\phi_i\wedge \big(
\dd X^i- \alpha^i \big) +
\alpha^i \wedge \dd\xi_i +
\frac16\, R^{ijk}(X)\, \xi_i\wedge\xi_j\wedge\xi_k \Big) +
\frac12\, \oint_{\Sigma_2} \, g^{ij}(X)\, \xi_i\wedge * \xi_j
\ , \nonumber \\
\label{SR2pure}\eeq
where $g^{-1}=\frac12\, g^{ij}(x)\,\partial_i\otimes\partial_j$ is the
inverse of a chosen
metric tensor on target space $M$, and $*$ is the Hodge duality operator
with respect to a chosen metric on the
worldsheet $\Sigma_2=\partial{\Sigma_3}$; we have again chosen Dirichlet boundary
 conditions $\alpha^i=\phi_i=0$ on $\Sigma_2$. As in~\cite{Halmagyi:2009te},
we have added a metric-dependent term on the boundary
$\Sigma_2$ of the membrane, which breaks the topological symmetry of
the Courant sigma-model, in order to ensure that the choice
$R^{ijk}\neq0$ is consistent with the equations of motion and also
with the
gauge symmetries of the field theory~\cite{Hofman:2002rv}. Note that only $g^{-1}$
 appears, not the metric $g$ itself; it will play the role of a
 metric on momentum space later on. 
 Integrating out the two-form fields $\phi_i$ leads to the action
\beq
{S}_R^{(2)} = \oint_{\Sigma_2}\, 
\xi_i\wedge\dd X^i + \int_{\Sigma_3} \ \frac16\, 
R^{ijk}(X)\, \xi_i\wedge\xi_j\wedge\xi_k +
\oint_{\Sigma_2} \ \frac12 \, g^{ij}(X)\, \xi_i\wedge *
\xi_j  \ . 
\label{SR2gen}\eeq

We will now specialize to the case where both the $R$-flux and the
target space 
metric are constant; this is the situation relevant to the
considerations of~\cite{Blumenhagen:2010hj,Lust:2010iy,Blumenhagen:2011ph,Lust:2012fp}. On the boundary
of the membrane,
the equations of motion for $X^i$ then force $\xi_i=\dd P_i$ to be an
exact form (modulo harmonic forms on $\Sigma_2$), where
$P_i\in C^\infty({\Sigma_3},X^*T^*M)$ is a section of the cotangent bundle
of $M$ restricted to ${\Sigma_3}$; this solution is also consistent with the
equations of motion in the bulk and henceforth we restrict the
configuration space for the path integral to this domain of fields. Then the
action (\ref{SR2gen}) reduces to a pure boundary action of the form
\beq
S_R^{(2)} = \oint_{\Sigma_2}\, \Big(\dd P_i\wedge \dd
X^i+\frac12\, R^{ijk}\, P_i\, \dd P_j\wedge \dd P_k\Big) +
\oint_{\Sigma_2} \ \frac12 \,g^{ij} \, \dd P_i\wedge*\dd P_j \ .
\label{SR2string}\eeq
This action can be recast in the form
\beq
S_R^{(2)}=\oint_{\Sigma_2} \, -\frac12\, \Theta^{-1}_{IJ}(X) \, \dd
X^I\wedge\dd X^J +
\oint_{\Sigma_2} \ \frac12\, g_{IJ}\, \dd X^I\wedge *\dd X^J \ ,
\label{SR2XI}\eeq
where the fields
\beq
X= (X^I) = (X^1,\dots,X^{2d}) := (X^1,\dots,X^d,P_1,\dots,P_d)
\eeq
embed the string worldsheet $\Sigma_2$ in the cotangent bundle of $M$,
i.e. the effective target space is now phase
space, and we
have introduced the block matrix on $T^*M$ given by
\beq
\Theta=\big(\Theta^{IJ}\big)=\begin{pmatrix} R^{ijk}\, p_k & \delta^i{}_j \\
  -\delta_i{}^j & 0 \end{pmatrix}
\label{Theta}\eeq
with local phase space coordinates
\beq
x= (x^I)=(x^1,\dots,x^{2d}):= (x^1,\dots,x^d,p_1,\dots,p_d) \ .
\eeq
The ``closed string metric''
\beq
\big(g_{IJ} \big) = \begin{pmatrix} 0 & 0 \\ 0 & g^{ij} \end{pmatrix}
\label{closedmetric}\eeq
acts on momentum
space but not on configuration space. The matrix $\Theta$ is always invertible and its inverse is given by
\beq
\Theta^{-1}= \big(\Theta^{-1}_{IJ}\big) = \begin{pmatrix} 0 &
  -\delta_i{}^j \\ \delta^i{}_j& R^{ijk}\, p_k \end{pmatrix} \ .
\eeq

We can linearize the action (\ref{SR2XI}) in the embedding fields
$X= (X^I) :\Sigma_2\to
T^*M$ by introducing auxilliary fields
$\eta_I\in\Omega^1(\Sigma_2 , X^*T^*(T^*M))$ to write
\beq
S_R^{(2)} = \oint_{\Sigma_2}\, \Big(\eta_I\wedge\dd X^I+\frac12\,
\Theta^{IJ}(X) \, \eta_I\wedge\eta_J\Big) +
\oint_{\Sigma_2} \ \frac12\, G^{IJ}\, \eta_I\wedge*\eta_J \ ,
\label{SR2PoissonXI}\eeq
where the ``open string metric''
\beq
\big(G^{IJ}\big) = \begin{pmatrix} g^{ij} & 0 \\ 0 & 0 \end{pmatrix}
\label{openmetric}\eeq
is related to (\ref{closedmetric}) by the usual closed-open string
relations~\cite{Seiberg:1999vs} that involve $\Theta$ and the ``$B$-field'' $\Theta^{-1}$
(note that $(g_{IJ})$ is not the inverse of $(G^{IJ})$).
This is the action of the non-topological generalized Poisson sigma-model
for the embedding of the string worldsheet $\Sigma_2$ into the
\emph{cotangent bundle} $T^*M$ of the manifold $M$ with bivector field
\beq
\Theta=\mbox{$\frac12$} \, \Theta^{IJ}(x) \, \partial_I\wedge\partial_J \ ,
\label{Bivec}\eeq
whose coefficient matrix $\Theta^{IJ}$ is given by (\ref{Theta}) and $\partial_I:=\frac\partial{\partial x^I}$; below we will write
phase space derivatives as $\partial_i:=\frac\partial{\partial x^i}$
and $\tilde\partial^i:=\frac\partial{\partial p_i}$. For completeness,
we express the action (\ref{SR2PoissonXI}) more explicitly in
phase space component form by decomposing the one-form fields
\beq
(\eta_I)=(\eta_1,\dots,\eta_{2d}):= (\eta_1,\dots,\eta_d,\pi^1,\dots
,\pi^d )
\eeq
and writing
\beq
S_R^{(2)} = \oint_{\Sigma_2}\, \Big(\eta_i\wedge\dd
X^i+\pi^i\wedge \dd P_i+\frac12\, R^{ijk}\, P_k\, \eta_i\wedge
\eta_j+\eta_i\wedge\pi^i\Big) + \oint_{\Sigma_2} \ \frac12\,
g^{ij}\, \eta_i\wedge*\eta_j \ .
\label{SR2expl}\eeq

The first order action (\ref{SR2expl}) is equivalent to the string sigma-model
(\ref{SR2string}). Note that only the momentum space components $P_i$ of the strings have
propagating degrees of freedom in $T^*M$; in this
sense the generalized Poisson sigma-model is still topological in the
original configuration space $M$. Moreover, the bivector field $\Theta$ defines a {twisted Poisson structure}
on the cotangent bundle, with twisting provided by a (trivial) non-flat $U(1)$-gerbe in \emph{momentum space}: Computing its Schouten--Nijenhuis
bracket with itself yields
\beq
[\Theta,\Theta]_{\rm S} = \mbox{$\bigwedge^3$}\Theta^\sharp(H) \ ,
\eeq
where
\beq
H=\mbox{$\frac16$}\, R^{ijk}\,\dd p_i\wedge\dd p_j\wedge \dd p_k
\label{Hmom}\eeq
is a closed three-form $H$-flux on the cotangent bundle $T^*M$; a $2$-connection on this
gerbe is given by the $B$-field
\beq
B= \mbox{$\frac16$}\, R^{ijk}\, p_k\, \dd p_i\wedge\dd p_j
\label{Bfieldmomsp}\eeq
with $H=\dd B$, which is gauge equivalent to the topological part of
the string sigma-model (\ref{SR2string}). In this way our membrane
sigma-model (\ref{SR2pure}) provides a geometric interpretation of the $R$-flux
background; this gerbe description will be exploited in Section~\ref{SWmaps}.

The antisymmetric brackets at linear order 
\beq
\{x^I,x^J\}_\Theta=\Theta^{IJ}(x)
\label{Alg2}\eeq
are given explicitly
by
\beq
\{x^i,x^j\}_\Theta= R^{ijk}\, p_k \ , \qquad \{x^i,p_j\}_\Theta=\delta^i{}_j \qquad
\mbox{and} \qquad \{p_i,p_j\}_\Theta=0 \ .
\label{Alg}\eeq
The corresponding Jacobiator is
\beq
\{x^I,x^J,x^K\}_\Theta := [\Theta,\Theta]_{\rm S}( x^I, x^J,
x^K) = \Pi^{IJK} \ ,
\eeq
where
\beq
\Pi^{IJK}=\mbox{$\frac13$}\, \big( \Theta^{KL}\, \p_L\Theta^{IJ} +
\Theta^{IL}\, \p_L\Theta^{JK} + \Theta^{JL}\, \p_L\Theta^{KI} \big) \
.
\label{Trivector}\eeq
The only non-vanishing components of this trivector field are
\beq
\{x^i,x^j,x^k\}_\Theta = R^{ijk} \ .
\label{Jacobiator}\eea
The expressions (\ref{Alg}) and (\ref{Jacobiator}) are precisely the
nonassociative phase space commutation relations for
quantized closed string coordinates which were derived
in~\cite{Lust:2010iy,Lust:2012fp}.

Although we are mostly interested in the case of constant $R^{ijk}$,
we can speculate on how to extend our discussion to non-constant
$R$-flux. By local orthogonal transformations the 3-vector $R$ can be
brought into canonical form wherein its only non-vanishing components
are $R^{ijk}(x)=|R(x)|^{1/3}\, \varepsilon^{ijk}$ for $i,j,k=1,2,3$,
where $\varepsilon^{ijk}$ is the totally antisymmetric tensor and $|R(x)|$ is the determinant of the 
matrix $R^{iJ}(x)$, $J=(jk)$. By a suitable coordinate transformation,
$R^{ijk}$ can thus be taken to be the constant tensor $\varepsilon^{ijk}$. Depending on how
the remaining structure functions of the Courant algebroid $C\to M$
transform, this may then yield a reduction of the membrane sigma-model
(\ref{SR2pure}) on $\Sigma_3$ to a string sigma-model on the boundary $\Sigma_2$ as before.

In any case,
the sigma-model (\ref{SR2expl}) and its associated brackets also make sense when
$R^{ijk}$ is a general function of $x\in M$, i.e. a generic trivector field
on configuration space. Quantizing these brackets thus provides a
means for quantizing generic Nambu--Poisson structures on $M$
with 3-bracket determined by the trivector $R$. It provides a
geometric way of incorporating nonassociativity into quantized
Nambu--Poisson manifolds, extending the (limited) techniques
of~\cite{DeBellis:2010pf,Saemann:2012ex} which relied on associative
algebras. Moreover, quantization of the membrane sigma-model provides
a dynamical realization of the nonassociative geometry which is an
alternative to the
reduced target space membrane models of~\cite{DeBellis:2010sy}. This
quantization is explored in detail below.

\subsection{Boundary conditions and correlation functions\label{BCs}}

It is natural to expect that the path integral for the $R$-twisted
Courant sigma-model provides a universal quantization formula for closed
strings in $R$-space, regarded as the boundaries of the membranes. For
the $H$-space open membrane sigma-model of Section~\ref{Htwist}, it
is argued in~\cite{Hofman:2002rv,Hofman:2002jz} that the path integral
defines a formal quantization for the corresponding twisted Poisson
structure, and an explicit prescription is given for quantizing appropriate
current algebra and $L_\infty$ brackets of the boundary strings from correlation functions of the open topological membrane
theory; for more general deformations of the exact Courant algebroid
$C=TM\oplus T^*M$, the path integral is argued to provide a universal
quantization formula for generic quasi-Lie bialgebras. The formulas for the worldsheet Poisson algebra nicely
resemble those which arise from transgressing the higher bracket
structures to loop space~\cite{Saemann:2012ex}. Unfortunately, the
complicated nature of the BV formalism which is necessary to quantize
the open topological membrane theory obstructs a complete
quantization. In particular, the Courant sigma-model with $R$-flux involves very complicated 2-algebroid gauge
symmetries; for the general Courant sigma-model the full gauge-fixed action
can be found in~\cite{Royt2}, and it involves both ghost fields and
ghosts-for-ghosts.

Here we would like to develop a quantization framework that is based on the
induced twisted Poisson sigma-model (\ref{SR2expl}), which involves
only Lie algebroid gauge symmetries, and
whose quantization on the disk is described
in~\cite{Cattaneo:1999fm,Cattaneo:2001ys}. For this, we will interpret
the membrane theory as an effective theory of \emph{open} strings with suitable boundary conditions imposed on the string embedding
fields. In~\cite{Hofman:2002rv} (see also~\cite{Bouwknegt:2011vn}) it
is proposed that the
boundary $\Sigma_2=\partial\Sigma_3$ can be taken to be an {open}
string 
worldsheet in the open topological membrane theory by regarding the membrane worldvolume
$\Sigma_3$ as a manifold with \emph{corners} (see e.g.~\cite{Joyce}), and allowing for different
boundary conditions on the various components of the boundary. In the
following we will take another approach that is directly related to
the way in which the twisted Poisson structure originates in
closed string theory on the $R$-flux
background~\cite{Lust:2010iy,Condeescu:2012sp} (see~\cite{Lust:2012fp}
for a review). We shall argue that
the corners of the membrane worldvolume can be mimicked via branch
cuts on a closed surface which give the multivalued
string maps responsible for the target space noncommutativity. In this
way the membrane serves to provide a sort of open/closed string
duality; the analogy between closed strings in non-geometric flux
backgrounds and open strings was also pointed out in~\cite{Lust:2010iy}.

The setting of~\cite{Lust:2010iy,Condeescu:2012sp} is that of closed
strings on the $Q$-space
duality frame obtained by applying two T-duality transformations to
the three-torus $M=\torus^3$ with constant NS--NS three-form flux $H=h\,\dd x^1\wedge \dd
x^2\wedge \dd x^3$. Locally, this space is a fibration of a two-torus
$\torus^2$ over a circle $S^1$; globally it is not well-defined as a
Riemannian manifold and is the simplest example of a \emph{T-fold}~\cite{Hull:2004in}. 
A representative class of twisted torus fibrations are provided by elliptic
T-folds where the monodromies act on the fibre coordinates as rotations. The closed string
worldsheet is the cylinder ${\cal C}=\real\times S^1$ with coordinates
$(\sigma^0,\sigma^1)$. The embedding field corresponding to
the base direction is denoted $X^3$, while for the fibre directions
we use complex fields denoted $Z,\overline{Z}=\frac1{\sqrt2}\,
(X^1\pm\ii X^2)$. As an extended closed string
wraps $\tilde p\,^3$ times around the base of the fibration, the fibre
directions need only close up to a monodromy corresponding to
an $SL(2,\zed)$ automorphism of the $\torus^2$-fibre. One thus arrives at the
\emph{twisted} boundary conditions
\beq
Z(\sigma^0,\sigma^1+2\pi)=\e^{2\pi\ii\theta}\, Z(\sigma^0,\sigma^1)
\qquad \mbox{and} \qquad X^3(\sigma^0,\sigma^1+2\pi)=X^3(\sigma^0,\sigma^1)
+2\pi\, \tilde p\,^3
\label{ZX3twisted}\eeq
where $\theta=-h\, \tilde p\,^3$; more precisely, one should impose
asymmetric boundary conditions for the left- and right-moving fields
in the fibre directions. To linear order in the flux, one can solve the
equations of motion of the closed string worldsheet sigma-model in the
usual way via oscillator mode expansions for the fibre coordinate fields subject to the
twisted boundary conditions (\ref{ZX3twisted}). Standard canonical
quantization then shows that the fibre directions acquire a
noncommutative deformation determined by the $H$-flux and the winding number
(or T-dual Kaluza--Klein momentum) $\tilde p\,^3$ in the $S^1$-direction, in
exactly the same way in which open string boundaries are deformed in the
presence of a $B$-field. Written in
terms of a real parametrization, we may express this closed string
noncommutativity generally in the $Q$-flux background via the Poisson brackets
\beq
\{ x^i,x^j\}_Q =
Q^{ij}{}_k\, \tilde p\,^k \qquad \mbox{and} \qquad \{ x^i,\tilde p\,^j\}_Q
=0 =\{ \tilde p\,^i,\tilde p\,^j\}_Q \ ,
\label{QfluxPoisson}\eeq
with constant flux $Q^{ij}{}_k=-2\pi\,h\, \varepsilon^{ij}{}_k$.
These brackets define a bonafide Poisson structure, since they are
just the
relations of a Heisenberg
algebra, as in the defining 2-brackets of the corresponding Courant algebroid. A
T-duality transformation to the $R$-flux background sends
$Q^{ij}{}_k\mapsto R^{ijk}$ and $\tilde p\,^k\mapsto p_k$, and maps the
Poisson brackets (\ref{QfluxPoisson}) to the twisted Poisson
structure (\ref{Alg}). This change of duality frame will be useful for
some of our later considerations.

This description of the $Q$-space is consistent with its
description from~\cite{Bouwknegt:2004ap} as a fibration of
stabilized noncommutative two-tori $\torus_{\theta}^2$, with Poisson bivector $\theta$
varying over the base $S^1$; low-energy effective open string
constructions are given in~\cite{Ellwood:2006my,Grange:2006es}, where
the noncommutative fibration is regarded as the algebra of open string
field theory in this background, and
also in~\cite{Bouwknegt:2008kd} within a $C^*$-algebra framework which
describes the T-fold as a topological approximation to a
$\torus^2$-equivariant gerbe with $2$-connection on $\torus^3$. The
interplay between open and closed string interpretations of the
noncommutative and nonassociative flux backgrounds was noted
in~\cite{Bouwknegt:2004ap,Ellwood:2006my}; a step towards
understanding the pertinent picture was recently carried out in the
context of matrix theory compactifications on twisted tori
in~\cite{Chatzistavrakidis:2012yp}, which constructs solutions with
noncommutative and nonassociative cotangent bundles.
To
explicitly relate the closed and open string pictures, we use the fact
that the boundary conditions (\ref{ZX3twisted}) define a twisted
sector of an orbifold conformal field theory on the quotient of
$M=\torus^3$ by the free action of a discrete abelian monodromy group; they
describe closed strings on the orbifold, which can be regarded as open
strings on the covering space $M$. When computing conformal field
theory correlation
functions, the monodromy can be implemented by inserting a suitable
twist field at a point $\sigma'\,^1\in S^1$ which creates a branch cut
along the
temporal direction $\real$ for the multivalued closed string
fields. We now extend the worldsheet $\Ccal=\real\times S^1$ to the
membrane worldvolume $\Sigma_3=\real\times(S^1\times \real)$ with coordinates
$(\sigma^0,\sigma^1,\sigma^2)$ such that the branch point at
$\sigma'\,^1\in S^1$ is blown up to a branch cut
$I=\{\sigma'\,^1\}\times\real\subset S^1\times\real$ extended along the $\sigma^2$-direction, i.e. the branch cut on the closed string worldsheet is blown
up to a ``branch surface'' on a closed membrane worldvolume. The
membrane fields are also taken to be multivalued and
non-differentiable across the branch cut $I$; hence Stokes' theorem on
$\Sigma_3$ receives contributions from the multivalued fields across
the cut whenever integration by parts is used to reduce worldvolume
integrals, as we did in Section~\ref{Rtwist}. This effectively reduces the
membrane to an ``open
string'' with worldsheet $\Sigma_2:=\partial{\Sigma_3} = \real\times I$
and coordinates $(\sigma^0,\sigma^2)$; classically, the mapping
$\Sigma_3\to\Sigma_2$ is a simple application of Stokes' theorem on
the equations of motion $\xi_i=\dd P_i$. In this way the branch cut $I$
plays the role of a ``corner'' separating $\Sigma_3$ into
regions~\cite{Joyce}. The more general orbifolds
of~\cite{Condeescu:2012sp} can be treated in an analogous way.

We can depict these two reductions from the membrane theory to the
closed and open string
theories via the following schematic diagram:

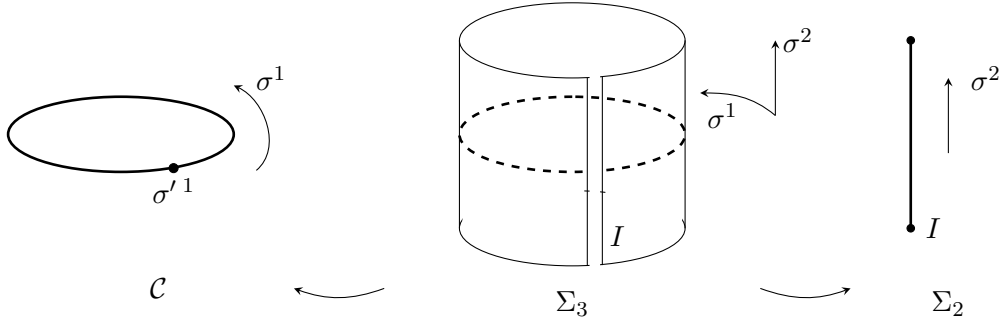
\begin{center}
%									% TARGET SPACE %
%
\begin{tikzpicture}[scale=1,>=stealth]
%
										%  M %
\draw[rotate=90] (0,0)
ellipse (0.5 and 1.5);
\draw[rotate=90] (2.5,0)
ellipse (0.5 and 1.5);
\draw[dashed,line width=1,rotate=90] (1.25,0)
ellipse (0.5 and 1.5);
\draw (0.2,-0.5)--(0.2,2);	
\draw (0.4,-0.49)--(0.4,2.02);	
\draw (1.5,0)--(1.5,2.5);
\draw (-1.5,0)--(-1.5,2.5);
										% erasing %
\draw[line width=20, white] (0.44,0.35)--(1.46,0.35);
\draw[line width=20, white] (0.16,0.35)--(-1.46,0.35);
\draw[line width=10, white] (0.23,2.0)--(0.37,2.0);
\draw[line width=68, white] (0.23,0.49)--(0.37,0.49);
										%  S %
\draw [line width=1] (4.5,0)--(4.5,2.5);
										%  C %
\draw[line width=1,rotate=90] (1.25,6)
ellipse (0.5 and 1.5);
										%  arrows %
\draw[->] (-4.2,0.77)  to [out=40, in=-25] (-4.5,1.9);
\draw[->] (-2.5,-0.8)  to [out=200, in=-20] (-3.7,-0.8);
\draw[->] (2.5,-0.8)  to [out=-20, in=200] (3.7,-0.8);

\draw[->] (2.7,1.5)  to [out=90, in=270] (2.7,2.5);
\draw[->] (2.7,1.5)  to [out=140, in=0] (1.7,1.8);
\draw[->] (5,1)  to [out=90, in=270] (5,2);
										% points %
\draw [fill=black] (4.5,0) circle (.5mm);
\draw [fill=black] (4.5,2.5) circle (.5mm);
\draw [fill=black] (-5.3,0.8) circle (.6mm);
\node at(-5.3,0.5) {$\sigma'\,^1$};
\node at (-4,2) {$\sigma^1$};
\node at (2,1.5) {$\sigma^1$};
\node at (3,2.5) {$\sigma^2$};
\node at (5.5,2) {$\sigma^2$};
\node at (-5.5,-0.8) {$\mc C$};
\node at (0,-1) {$\Sigma_3$};
\node at (0.6,-0.1) {$I$};
\node at (5,-1) {$\Sigma_2$};
\node at (4.8,0) {$I$};

\end{tikzpicture}
\end{center}
We do not display the temporal direction as it plays no role. The
mapping $\Sigma_3\to\Ccal$ is obtained by restriction of the domain of the
membrane path integral to fields which are independent of $\sigma^2$;
this gives a dimensional reduction of the membrane fields to closed
string fields which is reminescent of the Kaluza--Klein reduction of
M-theory to Type~IIA string theory. The mapping $\Sigma_3\to \Sigma_2$
is a restriction of field variables in the membrane path integral to
the cut $I$ of the spatial membrane cylinder; via reparametrization of
the membrane worldvolume, it also defines a map to the disk $\Sigma_2$
viewed as the complex upper half-plane with boundary the real line
$\real$, where the endpoints of the cut at $\pm\, \infty$ are mapped
to finite values. These two restrictions of the field domain in the membrane
path integral define the open/closed string duality that we were
after; in a certain sense it represents a sort of transmutation
between D-branes and fluxes. It is somewhat in line with the recent analysis
of~\cite{Davidovic:2012de} which demonstrates how non-geometric
doubled space coordinates arise as solutions to Neumann boundary
conditions in open string theory on flux backgrounds. Note that in
order to ensure independence of the specific location of the branch
cut $I$, it is important to assume that the $R$-flux is constant;
however, this restriction is no longer needed after we take the
$2+1$-dimensional Courant sigma-model as the fundamental model for
closed strings in the $R$-flux background.

Considering that the endpoints are at $\pm\,\infty$, it is natural to choose the
boundary conditions for the open string on the cut $I$ to coincide with those of~\cite{Cattaneo:1999fm}. In
this sense, the twisted boundary conditions (\ref{ZX3twisted}) on
$Q$-space can be made compatible with the Cattaneo--Felder boundary
conditions for the open twisted Poisson sigma-model. In the following
we will take the topological limit of (\ref{SR2expl}) where $g\ll R$;
this essentially decouples the open string modes from
the closed string modes. Then the propagator of the topological sigma
model is given by
\beq
\big\langle X^I(w)\, \eta_J(z)\big\rangle =
\mbox{$\frac{\ii\hbar}{2\pi}$}\, \delta^I{}_J\, \dd_z\phi^h(z,w) \ ,
\eeq
where $\hbar$ is a formal expansion parameter, the harmonic angle function
\beq
\phi^h(z,w):=\frac1{2\ii}\, \log \frac{(z-w)\,
  (z-\overline{w}\,)}{(\,\overline{z}-w)\,
  (\,\overline{z}-\overline{w}\,)}
\eeq
for $z,w\in\complex$ is the Green's function for the Laplacian on the
disk with Neumann boundary conditions, and $\dd_z:=\dd z\,
\frac\partial{\partial z}+\dd\overline{z}\, \frac\partial{\partial\overline{z}}$.
In this case, the
Feynman diagram expansion of suitable observables in the sigma-model
reproduces Kontsevich's graphical expansion for global deformation quantization of our twisted
Poisson structure~\cite{Konts:1997,Cattaneo:1999fm,Cattaneo:2001ys,Cattaneo:2001vu}, which we will take as our proposal for the
quantization of the $R$-flux
background.
In Section~\ref{Defquant} we shall compute the following schematic functional
integrals, whose precise meaning will be explained later on and whose
precise definitions can be found
in~\cite{Cattaneo:1999fm,Cattaneo:2001ys,Cattaneo:2001vu}. For $x\in
T^*M$, functions $f_i\in C^\infty(T^*M)$, and a collection of $n\geq1$
multivector fields ${\cal X}_r=\frac1{k_r!}\, {\cal X}_r^{I_1\dots
  I_r}(x)\, \partial_{I_1}\wedge\cdots\wedge \partial_{I_{k_r}} \in
C^\infty(T^*M, \bigwedge^{k_r}T(T^*M))$ of
degree $k_r$, define
\beq
U_n({\cal X}_1,\dots,{\cal X}_n)(f_1,\dots,f_m)(x) =\int \,
\e^{\frac\ii\hbar\, S_R^{(2)}}\, \frac\ii\hbar\, S_{{\cal X}_1}\,
\cdots\frac\ii\hbar\, S_{{\cal X}_n}\, \mathcal{O}_x(f_1,\dots, f_m) \
,
\eeq
where $m=2-2n+\sum_r\, k_r$, $S_{{\cal X}_r}=\oint_{\Sigma_2}\, \frac1{k_r!}\, {\cal X}_r^{I_1\dots
  I_r}(X)\,\eta_{I_1}\cdots\eta_{I_r}$, and $\mathcal{O}_x(f_1,\dots, f_m)$ are the boundary observables
\beq
\mathcal{O}_x(f_1,\dots, f_m) = \int_{X(\infty)=x}\,
\Big[f_1\big(X(q_1)\big)\cdots f_m\big(X(q_m)\big)\Big]^{(m-2)}
\eeq
with $1= q_1>q_2>\cdots> q_m=0$ and $\infty$ distinct points on the
boundary of the disk $\partial\Sigma_2$; the path integrals are
weighted with the full gauge-fixed action and the integrations taken
over all fields including ghosts. In particular, for functions $f,g\in
C^\infty(T^*M)$ one may define a star product by the functional
integral
\beq
(f\star g)(x) = \int_{X(\infty)=x}\, f\big(X(1)\big)\, g\big(X(0)\big)
\, \e^{\frac\ii\hbar\, S_R^{(2)}} \ ,
\eeq
whose properties will be thoroughly investigated in what follows.

\subsection{Twisted and higher Poisson structures\label{twistedhigher}}

We close this section with some general remarks about the twisted
Poisson structures we have derived, which will serve to help understand
some of the higher structures that will arise in our discussions about quantization.
Consider the algebra $\CV^\sharp = C^\infty(T^*M ,\bigwedge^\sharp
T(T^*M))$ of multivector fields on the cotangent bundle of the
target space $M$. Let $H=\frac16\, H_{IJK}(x)\, \dd x^I\wedge\dd
x^J\wedge \dd x^K$ be the closed three-form (\ref{Hmom}) on $T^*M$; it
extends by the Leibniz rule to give a ternary bracket $[-,-,-]_H$ on $\CV^\sharp$
of degree~$1$. Together with the Schouten--Nijenhuis bracket $[-,-]_{\rm S}$, it
defines an $L_\infty$-structure on $\CV^\sharp$ with zero differential,
generalizing the canonical differential graded Lie algebra structure
in the case of vanishing $R$-flux (see Appendix~\ref{HigherLA} for the
relevant definitions and background material). On the subspace
$C^\infty(T^*M)$ of smooth functions on $T^*M$, the $H$-twisted
Poisson structure (\ref{Bivec}) naturally defines a 2-term $L_\infty$-algebra $\big(V_1\xrightarrow{\
  \sfd \ }V_0\big)$ where $V_1=C^\infty(T^*M)$, $V_0$ is the space of
vector fields $\CX\in
C^\infty(T^*M,T(T^*M) )$ which preserve $\Theta$ in the sense that
$\Lcal_{\CX} \Theta=0$ where $\Lcal_{\CX}$ is the Lie
derivative along $\CX$, and 
$\sfd=\dd_\Theta=-[-,\Theta]_{\rm S}$ is the Lichnerowicz differential which sends a function $f\in C^\infty(T^*M)$ to
its Hamiltonian vector field $\CX_f=\Theta(\dd f,-)$~\cite{Royt3}.
The derived
bracket (\ref{derivedbracket}) on $V_1$
is just the quasi-Poisson bracket on $C^\infty(T^*M)$ determined by
$\Theta$ as
\beq
\{f,g\}_\Theta:= [\sfd f,g]_{\rm S}=\Theta(\dd f,\dd g) \ .
\eeq
The associated Jacobiator (\ref{derivedJac}) can be written as
\beq
\{f,g,h\}_\Theta= H(\CX_f,\CX_g,\CX_h) \ .
\label{Jactwisted}\eeq
Note that here the differential $\sfd$ is not nilpotent, and the
right-hand side of (\ref{Jactwisted}) can be expressed in terms of
$\sfd^2\neq0$; this is reminescent of a covariant derivative that does
not square to zero when the curvature is non-zero.

The corresponding
commutation relations in the associated semistrict Lie 2-algebra
$\CCV$ are (see Appendix~\ref{Linftyalg})
\bea
[\CX,\CY]_\CCV&=& [\CX,\CY]_{T(T^*M)} \ , \nonumber \\[4pt] \big[(\CX,f)\,,\,
(\CY,g)\big]_\CCV &=& \big([\CX,\CY]_{T(T^*M)} \,,\,\CX(g)-\CY(f) +
\{f,g\}_\Theta\big) \ ,
\eea
while the Jacobiator is
\beq
[\CX,\CY,\CZ]_\CCV = \big(\big[[\CX,\CY]_{T(T^*M)},\CZ\big]_{T(T^*M)} \,,\, H(\CX,\CY,\CZ) \big)
\eeq
for $\CX,\CY,\CZ\in C^\infty(T^*M,T(T^*M) )$ and $f,g\in C^\infty(T^*M)$. At linear
order, denoting the generators $(\partial_I ,0)$ and $(0,x^I)$
by $\mbp_I$ and $\mx^I$ for simplicity, we have
\beq
[\mbp_I,\mbp_J]_\CCV=0 \ , \qquad [\mbp_I,\mx^J]_\CCV=
\delta_I{}^J \qquad \mbox{and} \qquad [\mx^I,\mx^J]_\CCV= \Theta^{IJ}
\eeq
together with
\beq
[\mbp_I,\mbp_J,\mbp_K]_\CCV= H_{IJK} \ .
\eeq
In the following we will quantize this Lie
2-algebra.

As a side observation, it is intriguing to note that the twisted
Poisson brackets (\ref{Alg}) on the phase space $T^*M$ have an
alternative interpretation as a higher Poisson structure on the
configuration space $M$ (see
Appendix~\ref{Higherbrackets}); in this setting we regard the momenta
$p_i$ as the degree~$0$ generators $\partial_i$ of the multivector field algebra
$V^\sharp= C^\infty(M,\bigwedge^\sharp
TM)$. We take a degree~$3$
multivector field $R=R^{ijk}\, \partial_i\wedge\partial_j\wedge
\partial_k$, where $R^{ijk}$ is a constant $R$-flux on $M$. Then the
non-trivial derived brackets (\ref{higherderived}) of $R$ are
generated by
\bea
\sfd x^i \ := \ \{x^i\}_R &=& R^{ijk}\, \partial_j\wedge
\partial_k \ , \nonumber \\[4pt]
\{x^i,x^j\}_R &=& R^{ijk}\, \partial_k \ , \nonumber \\[4pt]
\{x^i,x^j,x^k\}_R &=& R^{ijk} \ ,
\eea
with all other brackets vanishing at linear order in $x^i$ and
$\partial_i$. These higher Poisson brackets define a 2-term
$L_\infty$-algebra structure on $V^\sharp$.

\newsection{Formal deformation quantization\label{Defquant}}

As discussed in Section~\ref{BCs}, a suitable perturbation expansion
of the membrane/string sigma-model of Section~\ref{TTHA} motivates
an approach to the quantum geometry of the $R$-flux background based on
deformation quantization.
In~\cite{Konts:1997}, Kontsevich constructs a deformation quantization
of an arbitrary Poisson structure, based on a graphical calculus which
is reproduced by the Feynman diagram expansion of the open Poisson
sigma-model on a disk~\cite{Cattaneo:1999fm}. In this section we shall follow
this prescription to derive a \emph{nonassociative} star 
product deformation of the usual pointwise product of functions on
$T^*M$ along
the direction of a generic twisted Poisson bivector $\Theta$, and describe its 
derivation properties. We then restrict to the case of constant
$R$-flux where we derive an explicit closed formula for the star
product and its associator, giving a quantization of the 2-brackets
(\ref{Alg}) and the 3-brackets (\ref{Jacobiator}) respectively. We
apply this formalism to derive Seiberg--Witten maps relating
nonassociative and associative deformations, and also add fluctuations
to the $R$-flux background. We
further explain how the 3-product proposed in~\cite{Blumenhagen:2011ph}
fits into our formalism.

\subsection{Star product\label{StarProd}}

Kontsevich's formalism relies on the constuction of the \emph{formality map}. 
The formality map is a sequence of $L_\infty$-morphisms $U_{n}$, $n\in\zed_{\geq0}$ that map 
tensor products of $n$ multivector fields to $m$-differential operators on
the manifold $T^*M$;
it defines an $L_\infty$-quasi-isomorphism between the differential
graded Lie algebra of multivector fields equiped with zero
differential and the
Schouten--Nijenhuis bracket (see Appendix~\ref{SNbrackets}), and the differential graded Lie algebra of
multidifferential operators equipped with the Hochschild differential
and the Gerstenhaber bracket (see
Appendix~\ref{Definitions}). Consider a collection of multivectors
${\cal X}_i$ of degree $k_i$ for
$i=1,\dots,n$. Then $U_n ({\cal X}_1, \dots ,{\cal X}_n)$ is a multidifferential operator whose degree $m$ is determined 
by the relation
\be
m=2-2n+\sum_{i=1}^n\, {k_i} \ . \label{Condition}
\ee
In particular, $U_0$ yields the usual pointwise product of functions while
$U_1$ is the Hochschild--Kostant--Rosenberg
map which takes a $k$-vector field to a $k$-differential operator defined by
\be
U_1\big({\cal X}^{I_1\dots I_k}\, \p_{I_1}\wg\cdots \wg\p_{I_k} \big)(f_1,
\dots,f_k)=\frac{1}{k!} \, \sum_{\sigma\in S_k}\, {\sgn(\sigma)}\,
{\cal X}^{I_{\sigma(1)}\dots I_{\sigma(k)}}\,
\p_{I_{\sigma(1)}}f_1\cdots\p_{I_{\sigma(k)}} f_k
\ee
for $f_i\in C^\infty(T^*M)$.
When the multivector fields ${\cal X}_i$ are all set equal to the bivector $\Theta$, the 
star product of functions $f,g\in C^\infty(T^*M)$ is given by the formal power series 
\be
 f\star g :=\sum_{n=0}^{\infty}\, {\frac{(\ii\hbar)^n}{n!} \,
   U_{n}(\Theta, \dots ,\Theta )} (f,g)\equiv \Phi(\Theta) (f,g) \ , \label{Kontsevich}
\ee
where $\hbar$ is a formal deformation parameter and $U_{n}(\Theta, \dots
,\Theta )$ is a bidifferential operator by \eqref{Condition}.

Kontsevich introduced a convenient diagrammatic representation on the 
upper hyperbolic half-plane $\mathbb{H}$ that provides all possible (admissible) 
differential operators to each order of the expansion \eqref{Kontsevich}, and 
thus determines the formality map $U_n$. Kontsevich diagrams encode the
rules for contracting indices and positioning partial
derivatives. Each diagram $\Gamma$ consists of:
\begin{enumerate}
\item Edges $e$ that are geodesics in $\mathbb{H}$ and represent 
partial derivatives;
\item A set $q_1, \dots , q_m \in \real$ of \emph{grounded} vertices 
that represent functions; and
\item A set $p_1, \dots ,p_n\in\mathbb H\,\backslash\,\real$ 
of \emph{aerial} vertices that represent the $k_i$-vector fields ${\cal X}_i$,
and thus $k_i$ edges may emanate from them. 
\end{enumerate} 
Here the real line $\real$ is the boundary of $\mathbb{H}$. An edge emanating from a given point $p_i$ is labelled as $e_i^{k_i}$. Edges 
that start from a vertex $v$ can land on any other vertex apart from $v$, while 
the condition $2n+m-2\geq 0$ must be satisfied. The multidifferential
operator
\beq
U_{n}({\cal X}_1, \dots ,{\cal X}_n) := \sum_{\Gamma\in G_n}\, w_\Gamma\,
D_\Gamma({\cal X}_1,\dots,{\cal X}_n)
\label{UnDGamma}\eeq
is calculated by summing over operators $D_\Gamma({\cal X}_1,\dots,{\cal X}_n) $ in the class $G_n$ of all $n$-th order admissible diagrams $\Gamma$, each contributing with 
weight $w_\Gamma$ given by the integral \cite{Kath:1998}
\be
w_\Gamma =\frac{1}{(2\pi)^{2n+m-2}}\, \int_{{\mathbb H}_n}\,
{\bigwedge_{i=1}^n \, \Big(\dd\phi^h_{e^1_i}\wg \cdots \wg \dd\phi^h_{e^{k_i}_i} \Big)}\ ,\label{Weight}
\ee
where $\quat_n$ is the space of pairwise distinct points $p_i\in\quat$
and the role of the harmonic angles $\phi^h_{e^{k_i}_i}$ is explained in Appendix~\ref{Graphs}.

In this setting, the diagrams for the bivector (quasi-Poisson bracket)
$\Theta (f,g)=\frac12\, \Theta^{IJ} \, \p_I f\, \p_J g$ and the trivector (Jacobiator)
$\Pi(f,g,h)=\frac 1 3  \, \Pi^{IJK}\, \p_I f \, \p_J g\, \p_K h$ contributions are
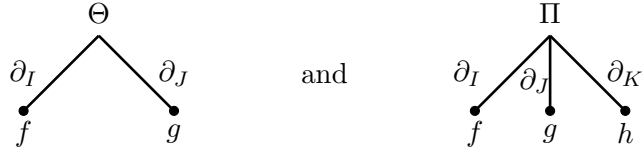
\be
%						% BI-VECTOR WEDGE %
%
\begin{tikzpicture}[scale=1,>=stealth, transform shape]
							% wedge %
\draw [line width=1] (3,1.5)--(2,0.5);
\draw [line width=1] (4,0.5)--(3,1.5);
\draw [fill=black] (2,0.5) circle (.6mm);
%\draw [fill=black] (3,1.5) circle (.6mm);
\draw [fill=black] (4,0.5) circle (.6mm);
\node at (2,1) {$\p_I$};
\node at (4,1) {$\p_J$};
\node at (2,0.2) {$f$};
\node at (4,0.2) {$g$};
\node at (3,1.8) {$\Theta$};
\node at (6,1) {and};
%						% TRI-VECTOR WEDGE %
%
\draw [line width=1] (8,0.5)--(9,1.5);
\draw [line width=1] (9,0.5)--(9,1.5);
\draw [line width=1] (10,0.5)--(9,1.5);
\draw [fill=black] (8,0.5) circle (.6mm);
\draw [fill=black] (9,0.5) circle (.6mm);
\draw [fill=black] (10,0.5) circle (.6mm);
\node at (7.9,1) {$\p_I$};
\node at (8.8,0.9) {$\p_J$};
\node at (10,1) {$\p_K$};
\node at (8,0.2) {$f$};
\node at (9,0.2) {$g$};
\node at (10,0.2) {$h$};
\node at (9,1.8) {$\Pi$};
%\node at (10.5,.5) {,};
\end{tikzpicture}
\nn \ee
which we will call the wedge and triple wedge respectively. The
geodesics here have been 
drawn as straight lines for the sake of clarity. Computing
\eqref{Kontsevich} then
provides the nonassociative star product deformation along the quasi-Poisson 
structure $\Theta$. 

Kontsevich's construction allows for one or more multivectors to be inserted 
in $U_{n}(\Theta, \dots ,\Theta )$. In our case, inserting the 
trivector $\Pi=[\Theta,\Theta]_{\rm S}$ that we acquired from the Schouten--Nijenhuis bracket is of particular interest since it encodes the nonassociativity of our star product. 
Then on functions $f,g,h\in C^\infty(T^*M)$ the series
\eqref{Kontsevich} is replaced with
\be
 [f,g,h]_\star :=\sum_{n=0}^\infty\, {\frac{(\ii\hbar)^n}{n!}\, U_{n+1}(\Pi,\Theta,\dots,\Theta)}(f,g,h) \equiv \Phi(\Pi) (f, g, h)\label{Bullet}
\ee
which, as we show in Section \ref{DerivationProp}, is the associator for the star product \eqref{Kontsevich}.
The condition \eqref{Condition} now implies that $U_{n+1}(\Pi,\Theta,\dots,\Theta)$ 
is a tridifferential operator. The map $U_{n+1}$ is calculated as in (\ref{UnDGamma}); 
this time though the integrations for the diagrams that give the
associated weights (\ref{Weight}) are much 
more involved since the edges of the triple wedge can land on any other wedge. 
Restricting to constant $R$-flux $R^{ijk}$ cures this problem, making the derivation of an 
explicit expression possible; this will be analysed in Section~\ref{ConstantR}. 

\subsection{Derivation properties and associator\label{DerivationProp}}

In order to define $L_\infty$-morphisms, the maps $ U_n$ 
must satisfy for $n\geq 1$ the \emph{formality conditions}~\cite{Konts:1997,Manchon:2000hy,Jurco:2001my,Szabo:2006wx}
\eqa
&&\dd_{\mu_2} U_n ({\cal X}_1,\dots ,{\cal X}_n) +\frac 1 2 \ \sum_{\substack
  {{\cal I}\sqcup {\cal J}=(1,\dots,n)\\ {\cal I},\cal
    J\neq\emptyset}}{\varepsilon_{\cal X}({\cal I},{\cal J})\,
  \big[U_{|{\cal I}|}({\cal X}_{{\cal I}})\,, \, U_{|{\cal J}|}({\cal X}_{{\cal J}} ) \big]_{\rm G}} \nn\\[4pt]
&& \qquad \qquad \qquad \ = \ \sum_{i<j}\, {(-1)^{\alpha_{ij}} \, U_{n-1}
  \big([{\cal X}_i,{\cal X}_j]_{\rm S} ,{\cal X}_1,\dots,\widehat {\cal X}_i,\dots , \widehat
  {\cal X}_j,\dots ,{\cal X}_n\big) } \ , \label{Formality}
\eqaend
where $\dd_{\mu_2} U_n:= -[U_n ,\mu_2]_{\rm G}$ with 
$\mu_n: C^\infty(T^*M)^{\otimes n}\to C^\infty(T^*M)$ the usual
commutative and associative
pointwise product of $n$ functions, $[-,-]_{\rm G}$ 
denotes the {Gerstenhaber bracket} defined in
Appendix~\ref{Definitions}, and for a multi-index ${\cal I}=(i_1,\dots,i_k)$
we denote ${\cal X}_{{\cal I}} :={\cal X}_{i_1}\wedge\dots \wedge {\cal X}_{i_k}$ and
$|{\cal I}|:=k$; the sign factor
$\varepsilon_{\cal X}({\cal I},{\cal J})$ is the ``Quillen sign'' associated with the
partition $({\cal I},{\cal J})$ of the integer $n$,
$(-1)^{\alpha_{ij}}$ is a prescribed sign rule arising from the
$L_\infty$-structure (see Appendix~\ref{Linftyalg}), and the hats denote omitted
multivectors. Formality follows from the Ward--Takahashi identities
for the Lie algebroid gauge symmetry of the Poisson sigma-model in the
BV formalism. In our case of interest, the conditions \eqref{Formality} reduce to
\be
\dd_\star \Phi(\Theta) = \ii\hbar\,\Phi(\dd_\Theta \Theta) \ , \label{Formality2}
\ee
where the {coboundary operators} are $\dd_\star = -[-
,\star]_{\rm G} $ and $\dd_\Theta=-[-,\Theta]_{\rm S}$ with 
$\dd_\Theta\Theta=\Pi$. Using \eqref{Formality2} and \eqref{Dstar} 
we derive a formula for the associator (\ref{Bullet}) given by
\be
 [f,g,h]_\star = \mbox{$\frac {2\ii}{\hbar}$} \, \big( (f\star g)\star h - f\star (g\star h)\big)  \ , \label{Associator}
\ee
which is non-zero since the product $\star$ is not associative.
This formula provides an exact formal expression which can be calculated up to 
any order in the deformation parameter $\hbar$ using Kontsevich diagrams.

The formality conditions give rise to derivation properties. Using \eqref{Bullet} 
we can define a new function $\underline{f}$ for every function $f$ by~\cite{Jurco:2001my,Szabo:2006wx}
\be
\underline{f} = f +\frac {(\ii\hbar)^2}{2}\, U_3 (f,\Theta,\Theta)
+\sum_{n=3}^\infty\, {\frac{(\ii\hbar)^n}{n!}\, U_{n+1}(f,\Theta,\dots,\Theta)} \ .
\ee
The formality condition is then $\dd_\star \, \underline{f}= \ii \hbar \, \Phi (\dd_\Theta f)$, 
which tells us that the Hamiltonian vector field $\dd_\Theta f$ is mapped to the inner 
derivation
\be
\dd_\star \, \underline{f} =\mbox{$\frac {\ii}{\hbar}$}\, [\,
\underline{f}\, ,-]_\star \ ,
\ee
where $[f,g]_\star:=f\star g- g\star f$ is the star commutator of
functions $f,g\in C^\infty(T^*M)$.
Similarly, a new vector field $\underline{{\cal X}}$ for any vector field ${\cal X}$ is defined by
\be
\underline{{\cal X}} = {\cal X} +\frac {(\ii\hbar)^2}{2}\, U_3 ({\cal X},\Theta,\Theta)
+\sum_{n=3}^\infty\, {\frac{(\ii\hbar)^n}{n!}\, U_{n+1}({\cal X},\Theta,\dots
  ,\Theta)} \ .
\label{newvectorfield}\ee
The formality condition is now $\dd_\star \, \underline{{\cal X}} = \ii \hbar \,
\Phi \, (\dd_\Theta {\cal X})$. 
$\dd_\Theta$-closed vector fields ${\cal X}$ preserve the twisted Poisson structure, i.e. $\dd_\Theta {\cal X}=0$. 
The formality condition then implies the derivation property
\be
\underline{{\cal X}} ( f \star g ) =  \underline{{\cal X}}( f) \star g + f \star \underline{{\cal X}}(g)
\ee
for $f,g\in C^\infty(T^*M)$.

Finally, we consider the formality condition
\be
\dd_\star \Phi(\Pi) = \ii\hbar\,\Phi(\dd_\Theta \Pi)   \label{Formality4} \ .
\ee
Using \eqref{Bullet}, we can express the left-hand side of this expression as
\bea
&& \dd_\star \Phi(\Pi) (f,g,h,k) \label{Derivation} \\[4pt] && \qquad\qquad \ = \ 
f \star [g,h,k]_\star - [ f \star g, h, k]_\star +   [f,  g\star h, k]_\star 
- [f, g, h\star k]_\star +[f,g,h]_\star \star k \ , \nonumber
\eea
while the Schouten--Nijenhuis bracket on the right-hand side is
\eqa
\dd_\Theta \Pi \ := \ [\Pi, \Theta]_{\rm S} &=& \mbox{$\frac{1}{24}$}\,
\big(\Theta^{LM} \, \p_M \Pi^{IJK} - \Theta^{IM} \, \p_M \Pi^{JKL} +
\Theta^{JM} \, \p_M \Pi^{KLI} - \Theta^{KM} \, \p_M \Pi^{LIJ} \nn \\
&& \qquad + \, \Pi^{IJM} \, \p_M \Theta^{KL} - \Pi^{JKM} \, \p_M
\Theta^{LI} + \Pi^{KLM} \, \p_M \Theta^{IJ} - \Pi^{LIM} \, \p_M \Theta^{JK} \nn\\
&& \qquad - \, \Pi^{IKM} \, \p_M \Theta^{JL} + \Pi^{JLM} \, \p_M
\Theta^{KI} \, \big) \, \p_I \wg \p_J \wg \p_K \wg \p_L \ . \label{Schouten}
\eqaend
Then \eqref{Formality4} relates these two expressions and gives the derivation property
for $f,g,h,k\in C^\infty(T^*M)$.

\subsection{Constant $R$-flux\label{ConstantR}}

We now turn our attention to the case of constant $R^{ijk}$
considered in Section~\ref{TTHA} and calculate the products 
we found in Section~\ref{StarProd} explicitly. Let us begin by computing \eqref{Kontsevich}. The 
zeroth order diagram is the usual pointwise multiplication. There is only one admissible first 
order diagram (the wedge) whose weight is found to be $\frac 12$ in Appendix~\ref{Graphs}, 
hence $U_1(\Theta)(f,g)=\frac 12\, \Theta^{IJ}\,\p_I f\,\p_Jg $. The admissible second order diagrams 
are
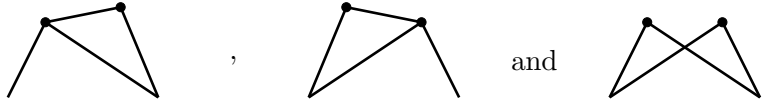
\be
%						% STAR PRODUCT - 2nd ORDER %
%
\begin{tikzpicture}[scale=1,>=stealth, transform shape]
							
\draw [line width=1] (0,0.5)--(0.5,1.5);
\draw [line width=1] (2,0.5)--(0.5,1.5);
\draw [line width=1](0.5,1.5)--(1.5,1.7);
\draw [line width=1] (2,0.5)--(1.5,1.7);
\draw [fill=black] (0.5,1.5) circle (.6mm);
\draw [fill=black] (1.5,1.7) circle (.6mm);
\node at (3,1) {,};
\draw [line width=1] (4,0.5)--(5.5,1.5);
\draw [line width=1] (6,0.5)--(5.5,1.5);
\draw [line width=1](4,0.5)--(4.5,1.7);
\draw [line width=1] (4.5,1.7)--(5.5,1.5);
\draw [fill=black] (4.5,1.7) circle (.6mm);
\draw [fill=black] (5.5,1.5) circle (.6mm);
\node at (7,1) {and};
\draw [line width=1] (8,0.5)--(8.5,1.5);
\draw [line width=1] (10,0.5)--(8.5,1.5);
\draw [line width=1] (8,0.5)--(9.5,1.5);
\draw [line width=1] (10,0.5)--(9.5,1.5);
\draw [fill=black] (8.5,1.5) circle (.6mm);
\draw [fill=black] (9.5,1.5) circle (.6mm);
%\node at (10.5,1) {.};
\end{tikzpicture}
\nn \ee
The first one represents $\Theta^{KL}\, \p_L\Theta^{IJ}\,\p_I f\,\p_K\p_J g=
R^{ijk}\,\p_i f\,\p_j\p_k g=0$ due to antisymmetry of $R^{ijk}$. The second 
diagram also vanishes for the same reason. Consequently, all higher order 
diagrams that contain these two subdiagrams are equal to zero. The third diagram is 
simply the product of two wedges, therefore its weight is $\frac 14$. Hence
$U_2(\Theta ,\Theta)(f,g) =\frac 14 \, \Theta^{IJ}\, \Theta^{KL}\,
\p_I\p_K f \, \p_J\p_L g$. 
Since wedges that land on wedges do not contribute to
$U_n(\Theta,\dots ,\Theta)$, 
there is only one admissible diagram to each order of the form
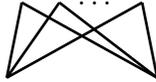
\be
%						% STAR PRODUCT - nth ORDER %
%
\begin{tikzpicture}[scale=1,>=stealth, transform shape]
							
\draw [line width=1] (0,0.5)--(0.2,1.5);
\draw [line width=1] (0,0.5)--(0.7,1.5);
\draw [line width=1] (0,0.5)--(1.8,1.5);
\draw [line width=1] (0.2,1.5)--(2,0.5);
\draw [line width=1] (0.7,1.5)--(2,0.5);
\draw [line width=1] (1.8,1.5)--(2,0.5);
\node at (1.2,1.5) {$\dots$};
%\node at (-6,1) { };
%\node at (2.5,0.7) {.};
\end{tikzpicture}
\nn \ee
Hence the star product for the constant $R$-flux background is given
by the Moyal type formula
\be
f\star g = \mu_2 \Big( \exp\big(\mbox{$\frac{\ii\hbar}{2}$} \, \Theta^{IJ}\,\p_I \otimes \p_J\big)  (f \otimes g) \Big) \ , \label{Star}
\ee
where as before $\mu_2$ is the pointwise multiplication map of functions.

The associator \eqref{Bullet} for constant $R^{ijk}$ can be computed either by 
calculating Kontsevich diagrams and summing the series or by using the star product 
\eqref{Star} to compute the left-hand side of \eqref{Associator}. Here we will 
follow the second approach, but before doing so it is instructive to calculate diagrams up 
to third order. A method for calculating Kontsevich diagrams involving
two functions for linear Poisson structures was developed
in~\cite{Kath:1998}; however we have found this setting unsuitable for calculations involving 
more than two grounded vertices and so we calculate diagrams in 
the usual manner. The lowest order admissible diagram is the triple wedge, whose weight 
is $\frac 16$ (see Appendix~\ref{Graphs}), thus $U_1(\Pi)=\frac 16 \,
\Pi^{IJK}\, \p_I f \,\p_J g \,\p_K h$. 
In $U_2 (\Pi,\Theta)$ a wedge is added, but since $R^{ijk}$ is constant, diagrams 
where the wedge lands on the trivector $\Pi$ are zero; thus all non-zero diagrams have 
weight $\frac {1}{12}$. Third order is more interesting as we now have two 
wedges that may land on each other. These diagrams are non-zero since the 
remaining edges all land on different functions. Calculating their weights
(see Appendix~\ref{Graphs}) we find that they combine to a trivector diagram according 
to the formula
\be
%						% TRIVECTOR FORMULA %
%
\begin{tikzpicture}[scale=1,>=stealth, transform shape]
							% trivector %
\draw [line width=1] (2.5,0.5)--(3.5,1.5);
\draw [line width=1] (3.5,0.5)--(3.5,1.5);
\draw [line width=1] (4.5,0.5)--(3.5,1.5);
							% + - = %
\node at (5.5,1) {$=$};
\node at (9.2,1) { $+$};
\node at (12,1) { $+$};
\node at (6,1) {{$\displaystyle{\frac 13} $}};
\node at (6.5,1) {\Bigg(};
\node at (14.7,1) {\Bigg)};
							% 1st graph %
\draw[line width=1] (7.8,1.5)--(6.8,0.5);
\draw [line width=1] (8.3,1)--(7.8,0.5);
\draw [line width=1] (8.8,0.5)--(7.8,1.5);
							% 2nd graph %
\draw [line width=1] (10.6,1.5)--(9.6,0.5);
\draw [line width=1] (10.6,1.5)--(10.1,1);
\draw [line width=1] (10.6,0.5)--(10.1,1);
\draw [line width=1] (11.6,0.5)--(10.6,1.5);
							% 3rd graph %
\draw [line width=1] (13.2,0.92)--(12.4,0.5);
\draw [line width=1] (13.34,0.99)--(13.9,1.3);
\draw [line width=1] (13.1,1.5)--(13.4,0.5);
\draw [line width=1] (13.1,1.5)--(13.9,1.3);
\draw [line width=1] (14.4,0.5)--(13.9,1.3);
\end{tikzpicture}
\nn \ee
which when  written out explicitly reproduces the formula (\ref{Trivector})
for the Schouten--Nijenhuis bracket $[\Theta, \Theta]_{\rm S}$ with constant 
$R^{ijk}$. 

To calculate the associator \eqref{Associator} explicitly to all
orders, we first observe that due to antisymmetry of $R^{ijk}$ the
star product (\ref{Star}) factorizes as
\be
f \star g =\mu_2\Big(\exp{\big(\mbox{$\frac{\ii\hbar}{2}$}\, R^{ijk}\, p_k \,
  \p_i \otimes \p_j\big)}\, \exp{\big[\mbox{$\frac{\ii \hbar}2$}\, \big(\p_i \otimes \tilde\p^i -
  \tilde\p^i \otimes \p_i \big)\big]}(f \otimes g ) \Big) =: f\star_p g \ , \label{Recovery}
\ee
where as before we write $\p_i=\frac{\p}{\p x^i}$ and
$\tilde\p^i=\frac{\p}{\p p_i}$. Here we denote the nonassociative
product $\star:=\star_p$ where $p$ is the dynamical momentum
variable. By replacing the dynamical variable $p$ with a constant
$\tilde p$ we obtain the associative Moyal star product
$\,\tilde\star\,:= \star_{\tilde p}$. Nonassociativity arises because
$\star$ acts non-trivially on the $p$-dependence of $\star:=\star_p$
in the associator.
Applying this on $(f\star g) \star h$ and $f\star (g\star h) $ using
antisymmetry of $R^{ijk}$ we find
\eqa
(f\star g) \star h \ := \ (f\star_p g) \star_p h &=& \Big[ \,\tilde\star\, \Big(\exp\big(\mbox{$\frac{\hbar^2}{4}$} \, R^{ijk}
\, \p_i \otimes \p_j \otimes \p_k\big) (f \otimes g \otimes h)\Big) \, 
\Big]_{\tilde p \to
  p} \
, \label{Assocbracketing} \\[4pt]
f\star (g\star h) \ := \ f\star_p (g \star_p h) &=& \Big[
\,\tilde\star\, \Big(\exp\big(- \mbox{$\frac{\hbar^2}{4}$} \, R^{ijk}
\, \p_i \otimes \p_j \otimes \p_k\big) (f \otimes g \otimes h)\Big) \, 
\Big]_{\tilde p \to
  p} \
, \nn
\eqaend
where no ordering is required on the right-hand sides due to
associativity of the Moyal product and the operation $[-]_{\tilde p\to p}$
reinstates the dynamical momentum dependence. Using (\ref{Associator}) we therefore find
\be
[f,g,h]_\star =\frac{4\ii}\hbar\, \Big[ \,\tilde\star\, \Big(
\sinh\big(\mbox{$\frac{\hbar^2}{4}$} \, R^{ijk} \, \p_i \otimes\p_j
\otimes \p_k \big)(f
\otimes g \otimes h)\Big)\, 
\Big]_{\tilde p \to
  p} \ . \label{Associator2}
\ee
This manner of regarding our nonassociative products is consistent
with the observation of Section~\ref{Rtwist} that only the momentum directions in the
membrane sigma-model are dynamical on~$T^*M$. From (\ref{Associator2})
it follows that our nonassociative star product is cyclic, i.e. the
associator is a total derivative and for Schwartz functions $f,g,h\in
C^\infty(T^*M)$ we have
\beq
\int_{T^*M}\, \dd^{2d}x \ [f,g,h]_\star(x) = 0 \ .
\label{starcyclic}\eeq
The cyclic property (\ref{starcyclic}) also holds for the nonassociative star products
derived from open string amplitudes in curved
backgrounds~\cite{Herbst:2001ai,Herbst:2003we} (see also~\cite{Ellwood:2006my}).

We conclude by writing out the derivation property 
\eqref{Derivation} explicitly for constant $R$-flux. The Schouten--Nijenhuis bracket 
\eqref{Schouten} now vanishes, and
therefore \eqref{Derivation} reduces to
\be
 [ f \star g , h, k]_\star -   [f,  g\star h,
 k]_\star + [f, g, h\star k]_\star = f \star [g,h,k]_\star +[f,g,h]_\star \star k \label{Derivation2}
\ee
for four functions $f$, $g$, $h$, and $k$,
while the remaining derivation properties we found in Section~\ref{DerivationProp} 
remain unaffected. We can interpret (\ref{Derivation2}) in the
following way. Just as the star commutator provides a quantization of
the twisted Poisson structure defined by the antisymmetric 2-bracket
$\{f,g\}_\Theta:=\Theta(\dd f,\dd g)$, in the sense that
$[f,g]_\star=2 \ii\hbar\, \{f,g\}_\Theta+\mc O(\hbar^2)$, the associator
(\ref{Associator2}) defines a quantization of the Nambu--Poisson
structure defined by the completely antisymmetric 3-bracket
\beq
\{f,g,h\}_\Theta:=\Pi(\dd f,\dd g, \dd h) \ ,
\label{NPRbracket}\eeq
in the sense that
\beq
[f,g,h]_\star=6\ii\hbar\, \{f,g,h\}_\Theta +\mc O\big(\hbar^2\big) \ .
\eeq
To this order, the star derivation property (\ref{Derivation2}) is just a consequence of
the usual Leibniz rule
\beq
\{f\,g,h,k\}_\Theta =f\, \{g,h,k\}_\Theta + \{f,h,k\}_\Theta \, g
\eeq
for the Nambu--Poisson bracket (\ref{NPRbracket}). Likewise, higher
derivation properties, when expanded in powers of $\hbar$, should encode the
fundamental identity
\bea
&& \big\{\{f_1,f_2,g\}_\Theta,h,k\big\}_\Theta +
\big\{g,\{f_1,f_2,h\}_\Theta,k\big\}_\Theta +
\big\{g,h,\{f_1,f_2,k\}_\Theta\big\}_\Theta \label{NPfundid} \\[4pt] &&
\qquad\qquad\qquad\qquad\qquad\qquad\qquad\qquad
\qquad\qquad\qquad\qquad \ = \ 
\big\{f_1,f_2,\{g,h,k\}_\Theta\big\}_\Theta \nn
\eea
for $f_i,g,h,k\in C^\infty(T^*M)$; we return to this issue
in Section~\ref{RQLie2group} where we will see that the equation (\ref{Derivation2}) can also be interpreted as the star 
product version of the pentagon identity (\ref{pentagonid}) for the Lie 2-group that we
encounter there.

\subsection{Seiberg--Witten maps\label{SWmaps}}

We will now apply the formalism of this section to analyze the effect
of adding fluctuations to the membrane boundary and the open string
endpoints. To start, we shall recall a few relevant facts of the open
string case with an ordinary Poisson structure, and then show how this
generalizes to the case of $H$-twisted Poisson structures and
ultimately to the membrane setting with $R$-flux.

By studying open strings in a closed string background, Seiberg and Witten~\cite{Seiberg:1999vs} found equivalent effective descriptions in terms of ordinary as well as noncommutative gauge theories. Realizing this, they proposed the existence of maps $\hat A(a)$ and $\hat \Lambda (\lambda, a)$ from an ordinary gauge potential $a_\mu$ and gauge parameter $\lambda$ to their noncommutative cousins $\hat A_\mu$ and $\hat \Lambda$, such that an ordinary infinitesimal gauge transformation $\delta_\lambda a_\mu = \partial_\mu \lambda$ induces its noncommutative analog $\delta_{\hat\Lambda} \hat A_\mu = \partial_\mu \hat \Lambda + \ii \hat \Lambda \star \hat A_\mu - \ii \hat A_\mu \star \hat \Lambda$. Further analysis has revealed~\cite{Jurco:2000fb,Jurco:2000fs}
that the Seiberg--Witten map can be interpreted as a special
generalized change of coordinates induced by an invertible linear
operator $\mathcal D$, which is a non-linear functional of the gauge
potential $a$ and maps ordinary spacetime coordinates $x^\mu$ to
covariant coordinates $\hat x^\mu = \mathcal D (x^\mu) = x^\mu +
\theta^{\mu\nu}\, \hat A_\nu(x)$. The covariantizing map transforms by
a star commutator with $\hat \Lambda$ under gauge transformations,
implying the appropriate noncommutative gauge transformation for $\hat
A_\mu$. For simplicity, we have written here equations for abelian
gauge fields and have focused on the case of constant Poisson
structure $\theta$. We will continue to focus on the abelian case, but
drop all other simplifying assumptions in the following.

A semi-classical version $\rho$ of the covariantizing map is given by
the flow generated by the vector field $a_\theta = \theta(a,-) =
\theta^{\mu\nu}\, a_\nu \, \partial_\mu$, which is obtained by
contracting the Poisson bivector field
$\theta=\frac12\,\theta^{\mu\nu}\, \partial_\mu\wedge \partial_\nu$ with the gauge potential
one-form $a = a_\mu(x)\, \dd x^\mu$. Under a gauge transformation,
$\rho$ transforms via the Poisson bracket with a semi-classical gauge
parameter $\hat \lambda$. The Poisson bivector $\theta$ is ``twisted''
by the two-form field strength $f = \dd a$ to a new Poisson bivector
$\theta' = \theta \, (1 + \hbar \, f \, \theta)^{-1} $ such that $\rho(\{f,g\}_{\theta'}) = \{\rho (f), \rho (g)\}_\theta$.
The quantum covariantizing map $\mathcal D$ is similarly obtained as
the ``flow'' of the differential operator $\underline{a_\theta}$, which
is the deformation quantization \eqref{newvectorfield} of $a_\theta$. Let~$\star$ and~$\star'$
likewise be the deformation quantizations along $\theta$ and $\theta'$
respectively according to \eqref{Kontsevich}. The covariantizing map
$\mathcal D$ relates these star products via $\mathcal D (f \star' g)
= \mathcal D f \star \mathcal D g$, i.e.\ it is an isomorphism of
associative algebras and noncommutative gauge transformations are
realised as
inner automorphisms. For the technical details of the construction, we
refer to \cite{Jurco:2001my,Jurco:2001kp}. In the following, we will
describe several covariantizing maps and the corresponding gauge
symmetries relevant for strings in the $R$-flux background.

The presence of a non-trivial closed three-form background $H$ leads
to a twisted Poisson structure: The bivector $\Theta$ fails to fulfill
the Jacobi identity and its Schouten--Nijenhuis bracket is
consequently non-zero:  $[\Theta,\Theta]_{\rm S}  = \hbar \,
\bigwedge^3 \Theta^\sharp (H)$, where we have introduced a factor
$\hbar$ to ensure formal convergence of all expressions in the ensuing
contruction. (Such a factor is understood to be implicitly included in
$\Theta$ in the rest of this article.)  From the point of view of
background fields and fluctuations, the structure we are dealing with
is a \emph{gerbe}: Given a suitable covering of the target space
manifold by contractible open patches (labelled by Greek indices $\alpha,\beta,
\ldots$), we can write $H$ in terms of local two-form fields $B_\alpha$
as $H = \dd B_\alpha$ on each patch. On the overlap of two patches,
the difference $B_\beta - B_\alpha =: F_{\alpha\beta}$ is closed,
hence exact and can be expressed in terms of one-form fields
$a_{\alpha\beta}$ as $F_{\alpha\beta} = \dd a_{\alpha\beta}$. On
triple overlaps we then encounter local gauge parameters
$\lambda_{\alpha\beta\gamma}$ that satisfy a suitable integrability
condition. This hierarchical description in terms of forms has a dual
description in terms of multivector fields that is suitable for
deformation quantization and leads to noncommutative gerbes in the
sense described in~\cite{Aschieri:2002fq}: The twisted Poisson
bivector $\Theta$ can be locally untwisted by the two-form fields
$B_\alpha$, leading to bonafide Poisson bivectors $\Theta_\alpha =
\Theta \, (1 - \hbar\, B_\alpha \, \Theta)^{-1}$. These local Poisson
tensors $\Theta_\alpha$ and the corresponding associative star
products $\star_\alpha$ are related by covariantizing maps computed
from~$a_{\alpha\beta}$.

As mentioned in Section~\ref{Rtwist}, the relevant geometric structure
in $R$-space is a gerbe in momentum space, with curvature (\ref{Hmom})
and 2-connection (\ref{Bfieldmomsp}). In the present paper, we are dealing with a topologically trivial
setting, so the forms and multivector fields are all globally
defined. Nevertheless, the constructions of twisted noncommutative
gauge theory and Seiberg--Witten maps are non-trivial and
interesting. On $T^*M$ the patch index $\alpha$ is replaced by a
constant momentum vector $\tilde p$ that parameterizes a degree of freedom in
the choice of Poisson structure~$\Theta_{\tilde p}$ and two-form
background field $B_{\tilde p}$. In matrix form, the pertinent
bivector and two-form fields are
\begin{equation}
\Theta = \begin{pmatrix} \hbar\, R^{ijk}\, p_k & \delta^i{}_j \\
  -\delta_i{}^j & 0 \end{pmatrix} \ , \qquad
\Theta_{\tilde p} = \begin{pmatrix} \hbar \, R^{ijk} \, \tilde p_k &
  \delta^i{}_j \\ -\delta_i{}^j & 0 \end{pmatrix} \qquad \mbox{and} \qquad
B_{\tilde p} = \begin{pmatrix} 0 & 0\\ 0 & R^{ijk}\, (p_k - \tilde
  p_k) \end{pmatrix} \ .
\end{equation}
They satisfy 
\begin{equation} \label{Schoutenb}
H=\dd B_{\tilde p} \ , \qquad [\Theta,\Theta]_{\rm S} = \hbar\,
\mbox{$\bigwedge^3$} \Theta^\sharp(H) \qquad \mbox{and} \qquad [\Theta_{\tilde
  p},\Theta_{\tilde p}]_{\rm S} = 0 \ ,
\end{equation}
together with
\begin{equation}\label{geometrics}
\Theta = \Theta_{\tilde p} \, (1 + \hbar \, B_{\tilde p} \,
\Theta_{\tilde p})^{-1}  \qquad \mbox{and} \qquad
\Theta_{\tilde p} = \Theta \, (1 - \hbar\, B_{\tilde p} \,
\Theta)^{-1} \ .
\end{equation}
The corresponding 1-connection is given by $a_{\tilde p,\tilde
  p\, '}=R^{ijk}\, p_i\, (\tilde p_k-\tilde p_k^{\,\prime})\, \dd p_j$.
Note that we cannot choose $\hbar \, B$ to be equal to $\Theta^{-1}$
as that would add terms of order $\hbar^{-1}$ to $H = \dd B$, which is
incompatible with \eqref{Schoutenb}, and it would lead to convergence
problems for the geometric series in \eqref{geometrics}. The
deformation quantizations along $\Theta$ and $\Theta_{\tilde p}$ yield
the nonassociative star product $\star$ and the associative star product $\star_{\tilde p}$ respectively. For the special case $\tilde p = 0$, $\Theta_0$ and $\star_0$ are respectively the canonical Poisson structure and associative star product on phase space.
For fixed three-form $H$, choices for $B$ can differ by any closed
(and hence exact) two-form $F=\dd A$. The corresponding choices of
Poisson structures and star products are related by covariantizing
maps constructed from gauge potentials
\beq
A = A_I(x)\, \dd x^I = a_i(x,p) \, \dd x^i + \tilde a^i(x,p)\, \dd p_i
\ . 
\eeq
Associated to these covariantizing maps are Seiberg--Witten maps as explained before. Gauge transformations $\delta_\lambda A = \dd\lambda$, where $\lambda = \lambda(x,p)$, induce a change of the covariantizing maps by a star commutator with $\hat\Lambda(\lambda,a)$.

So far we have discussed ordinary Seiberg--Witten maps for bonafide
Poisson bivectors. It would appear to be more natural to find also a
construction based directly on the twisted Poisson bivector
$\Theta$. Terms involving the non-zero Jacobiator (Schouten--Nijenhuis
bracket) usually spoil such a construction. In the present case it
turns out, however, that for the class of gauge potentials of the form
$A= (A_I) = (0, \tilde a^i(x,p))$ (i.e.\ with $a_i(x,p) = 0$) the
unwanted terms drop out, because $([\Theta,\Theta]_S)^{IJK} \, A_K$ is
proportional to $R^{ijk}\, a_k = 0$. The restriction thus imposed on
the class of admissible gauge potentials leads to a corresponding
restriction on the class of covariantizing maps. The admissible class
of maps is, however, still very large and actually quite interesting:
Evaluating $\Theta(A,-) = \Theta^{IJ} \, A_J\, \partial_I =
\delta^i{}_j\, \tilde a^j(x,p)\, \partial_i$ shows that any map
generated by a vector field of the form $\tilde a^i(x,p)
\, \partial_i$, which acts on configuration space and may even depend
on the momentum variables, is admissible. The associated class of
ordinary and noncommutative gauge transformations is more restricted:
The gauge parameters $\lambda$ and $\hat\Lambda$ may only depend on
the momenta $p$.

An interesting subclass of the covariantizing maps just described are
generated by gauge potentials of the form $A = R(a_2,-)$, where $a_2$ is
a two-form on configuration space and we have used the natural map
$\bigwedge^2 T^*M \rightarrow TM$ induced by the three-vector $R$; in
components $\tilde a^i(x) = R^{ijk} \, (a_2)_{jk}(x)$. The
semi-classical version of the resulting maps has been discussed in the
context of Nambu--Poisson structures on $p$-branes in~\cite{Jurco:2012yv}, where $a_2$ plays the role of a two-form gauge potential, and (noncommutative) gauge transformations are computed using Nambu--Poisson brackets and $x$-dependent one-form gauge parameters. (The restriction to gauge parameters that depend only on momenta is not needed here.) The present article provides a quantization of these Nambu--Poisson maps for membranes ($p=2$) with a constant Nambu--Poisson trivector $R$.

Another interesting special example concerns the relationship between
$\star_0$ and $\star_{\tilde p}$: The corresponding covariantizing map
$\mathcal D_{\tilde p}$ is constructed from $F_{\tilde p} = \dd
A_{\tilde p} = B_0 - B_{\tilde p}$, with a gauge potential defined by
\beq
A_{\tilde p} = \tilde a^j(p)\, \dd p_j = \mbox{$\frac{1}{2}$} \, R^{ijk}\,
p_i\, \tilde p_k\, \dd p_j
\label{Atildep}\eeq
up to gauge transformations
$\delta_{\tilde\lambda} \tilde a^j(p) = \tilde \partial^j \tilde
\lambda(p)$. It satisfies $f\star_{\tilde p} g = \mathcal D_{\tilde
  p}^{-1}\left(\mathcal D_{\tilde p} f \star_0 \mathcal D_{\tilde p} g\right)$.
Formally replacing the constant $\tilde p$ by the dynamical momentum
variable $p$ in this equation gives a Seiberg--Witten map from the
associative canonical star product $\star_0$ to the nonassociative
star product $\star := \star_{p}$ as
\begin{equation}
f \star g = \big[\mathcal D_{\tilde p}^{-1}\left(\mathcal D_{\tilde p}
  f \star_0 \mathcal D_{\tilde p} g\right)\big]_{\tilde p \rightarrow
  p} \ .
\end{equation}
In view of the foregoing discussion, this can also be written as
\begin{equation}
f\star g= \big[\mathcal D_{\tilde p} (f \star g)\big]_{\tilde p \rightarrow p} =
\big[\mathcal D_{\tilde p} f \star_0 \mathcal D_{\tilde p}
g\big]_{\tilde p \rightarrow p} \ ,
\end{equation}
since the underlying vector field $\Theta(A_{\tilde p},-)$ from
(\ref{Atildep}) vanishes when $\tilde p\to p$ and the covariantizing
map becomes trivial.
In the given gauge, the vector field $\Theta(A_{\tilde p},-)$ receives
no quantum corrections from deformation quantization and the
Seiberg--Witten map can thus be computed explicitly in closed form. This is one of the very rare cases where this is possible. 

From the point of view of noncommutative gauge theory as well as
noncommutative string geometry, gauge transformations preserve
star products. Expressed in terms of gauge fields, gauge
transformations correspond to different choices of one-form potentials
$A$ that preserve the curvature two-form $F = \dd A$. From the
membrane point of
view, however, the three-form $H$ is the fundamental global
quantity and gauge transformations correspond to different choices of
two-form potentials $B$ that preserve the gerbe curvature $H = \dd B$. The role of the
gauge parameter is taken by a one-form gauge potential $A$. As we have
discussed, such one-forms generate covariantizing maps $\mathcal
D$. These maps preserve associativity (as well as
nonassociativity). The collection of these maps describes the gauge
degrees of freedom of our system. Concretely, our construction for the
twisted Poisson structure $\Theta$ yielded covariantizing maps
$\mathcal D_\xi$  for all vector fields $\xi$ on configuration
space. This evidently generates a huge gauge degree of freedom
generated by quantized general coordinate transformations. This point
adds some credibility to the terminology ``nonassociative gravity'' that was
coined in~\cite{Blumenhagen:2010hj} to describe the quantum geometry of closed strings in non-geometric flux backgrounds.

\subsection{Closed string vertex operators and 3-product}

We close this section by comparing our associator with the ternary product for closed strings
propagating in a background with constant $R$-flux which was proposed 
in~\cite{Blumenhagen:2011ph}. Here the authors perform a linearized
conformal field theory analysis of the three-point function of tachyon
vertex operators in a flat background with constant $H$-flux. After applying
three T-dualities they arrive at a nonassociative algebra of closed
string vertex operators in the $R$-flux background, from which they propose a deformation of the pointwise product of functions via a 3-product of the form
\be
f\bullet g \bullet h = f\,g\,h + R^{ijk}\,\p_if\,\p_j g\, \p_k h
+\mc O\big(R^2\big) \ . \label{Triprod}
\ee
In the light of the above analysis, we are able to explain this result
analytically and 
relate it to our expressions. Expanding either of the two bracketings
\eqref{Assocbracketing} to linear order we have
\be
f\star g\star h = f\,g\,h + R'\,^{ijk} \,
\big[ \p_i f \,\tilde\star\, \p_j g
\,\tilde\star\, \p_k h\big]_{\tilde p\to p} +\mc O\big(R'\,^2\big) \ , \label{Associator3}
\ee
from which we conclude that \eqref{Triprod} agrees with
\eqref{Associator3} to first order in $R':= \pm\, \frac{\hbar^2}4 \, R$, but
without the Moyal star product between the derivatives of $f$, $g$ and $h$, and without the dependence
on the dynamical variable $p$. For functions that are independent of $p$, the two formulas agree at linear order. 
The main difference between the two formulas stems from our consideration of the cotangent bundle of $M$
as the effective target space geometry of closed strings in the $R$-flux
compactification.

This also explains the proposal
of~\cite{Blumenhagen:2011ph} that the binary product of functions is
the usual pointwise multiplication, as for closed strings only three
and higher point correlation functions experience the effect of the
flux background. Setting $p=\tilde p=0$ corresponds to the sector of zero winding number in
the T-dual $Q$-space frame. It truncates phase space to the
original configuration space $M$ and recovers the usual commutative pointwise
product $f\,\tilde\star\, g=f\, g= f\star g$, consistent with the
fact that only extended closed strings with non-trivial winding number
(dual momentum) are sensitive to the noncommutative deformation in the
$Q$-flux background (see Section~\ref{BCs}); nevertheless, this sector still retains a
non-trivial associator (\ref{Associator2}) of fields in the
nonassociative $R$-flux background as in~\cite{Blumenhagen:2010hj,Blumenhagen:2011ph}. Moreover, as
in~\cite{Blumenhagen:2011ph}, higher order associators are not simply
related to successive applications of (\ref{Associator2}). 
Together with the cyclic property (\ref{starcyclic}), we see therefore that in this sector our deformation
quantization approach agrees with the 3-product of~\cite{Blumenhagen:2011ph}.
Similarly to~\cite{Takhtajan:1993vr}, the authors
of~\cite{Blumenhagen:2011ph} conjecture an all orders ternary product
obtained by exponentiation of the trivector $R$ as a straightforward
generalization of the Moyal--Weyl formula. Our results confirm  this
conjecture insofar that the exponential of $R$ is indeed part of the
correct all order expressions (\ref{Assocbracketing}).

As we have discussed, the twisted Poisson structures we have found
naturally give rise to an $L_\infty$-structure on the algebra
$C^\infty(T^*M)$. In~\cite{Cornalba:2001sm} it is shown that
correlators of
\emph{open} string vertex operators in a non-constant $H$-flux
background endow the Kontsevich deformation of the algebra of
functions on $M$ with the structure of an $A_\infty$-algebra (see
Appendix~\ref{Linftyalg}), or more precisely an $A_\infty$-space, which are the
natural algebras that appear in generic open-closed string field
theories; in particular, the corresponding star commutator algebra is
an $L_\infty$-algebra. The
correlators of closed string vertex operators computed
in~\cite{Blumenhagen:2011ph} also exhibit dilogarithmic singularities
analogous to those found in~\cite{Cornalba:2001sm} (see also~\cite{Herbst:2001ai}), and it would be
interesting to see if they lead to an analogous
$A_\infty$-structure; indeed in~\cite{Aldi:2011cf} it is shown that
the reflection identity for the Rogers dilogarithm relates four-point
correlation functions to two-point correlators in a manner reminescent
of an associator, while the pentagonal identity is related to a
factorization property of five-point functions which is reminescent of the
higher coherence relation for the associator. The similarity between
open and closed string correlators is also noted
in~\cite{Blumenhagen:2011yv}. In Section~\ref{BCHquant} we will see such
structures emerging rather directly in the full quantized algebra of functions.

\newsection{Strict deformation quantization\label{BCHquant}}

In~\cite{Kath:1998}, Kathotia compares the two canonical deformation
quantizations of the linear Kirillov--Poisson structure on the vector
space $W=\frg^*$, where $\frg$ is a finite-dimensional Lie
algebra; these quantizations are provided by the Kontsevich formalism
and the associated Lie group convolution algebra. Let us briefly recall how
this latter quantization scheme proceeds~\cite{Rieffel}. We first
Fourier transform functions on $W$ to obtain elements in
$C^\infty(\frg)$. We then identify $\frg$ with its integrating Lie
group $G$ in a neighbourhood
of the identity element via the exponential mapping. On $G$, we can
use the convolution product between functions induced by the group
multiplication and the Baker--Campbell--Hausdorff formula. We then perform the inverse operations to pullback
the result and obtain a star product on $C^\infty(W)$. For nilpotent Lie algebras, the
exponential map between $\frg$ and $G$ is a global diffeomorphism. In
this case, the above construction is equivalent to both Kontsevich's
deformation quantization and quantization via the universal enveloping
algebra of $\frg$~\cite{Kath:1998}; see also~\cite{Shoikhet} for a
comparison directly at the level of
formality maps. Since our twisted Poisson structure (\ref{Alg}) is
linear for constant $R$-flux, it is natural to ask if there is an analogous approach which
would provide an alternative quantization framework to the
combinatorial approach we took in Section~\ref{Defquant}. In this
section we shall develop such an approach based on integrating a
suitable Lie 2-algebra to a Lie 2-group which will define a
convolution algebra object in a braided monoidal category (see
Appendices~\ref{Linftyalg} and~\ref{Lie2groups} for the precise
definitions), and demonstrate that it is equivalent to the
quantization of Section~\ref{Defquant} which was based on our proposed
membrane
sigma model. Here we focus for
definiteness on
the case of configuration space $M=\torus^d$ which is a
$d$-dimensional torus with constant $R$-flux. This approach will then further clarify how the $R$-space
nonassociativity is realized by a 3-cocycle associated to a
nonassociative representation of the translation group, as arises in
the presence of a magnetic monopole~\cite{Jackiw:1984rd}, and its
relation to the topological nonassociative tori studied
in~\cite{Bouwknegt:2004ap}.

\subsection{$R$-space and $Q$-space Lie 2-algebras\label{RQLie2alg}}

Let $V\cong\real^{2d}$ be a vector space of dimension $2d$ with a
fixed choice of basis elements which we denote by
\beq
(\hat x^I)=(\hat x^1,\dots,\hat x^{2d})= (\hat
x^1,\dots,\hat x^d,\hat p_1,\dots,\hat p_d) \ ,
\eeq
where throughout this section we use
hats to distinguish abstract vector space and categorical elements
from the concrete coordinate
functions we used in previous sections. We define a bracket $[-,-]_R:V\wedge V\to V$
by the relations
\beq 
[\hat x^i ,\hat x^j ]_R= \ii R^{ijk}\, \hat p_k \ , \qquad [\hat p_i ,\hat p_j]_R =0
\qquad \mbox{and} \qquad [\hat x^i ,\hat p_j ]_R = \ii \hbar\, \delta^i_j = -
[\hat p_j,\hat x^i]_R \ , \label{Alghat}
\eeq
which is just an abstract presentation of the twisted Poisson brackets
(\ref{Alg}). This bracket defines a \emph{pre-Lie
  algebra} structure on $V$, i.e. it is antisymmetric but does not
satisfy the Jacobi identity; it leads to the non-vanishing Jacobiator
\beq
[\hat x^i,\hat x^j,\hat x^k]_R :=\mbox{$\frac13$}\, \big( \big[[\hat
x^i,\hat x^j]_R,\hat x^k \big]_R + \big[[\hat x^k,\hat x^i]_R,\hat
x^j\big]_R + \big[[\hat x^j,\hat x^k]_R ,\hat x^i\big]_R \big) =
\hbar\, R^{ijk} \ , \label{Jacobiatorhat}
\eeq
and all other Jacobiators vanish. Hence the bracket naturally defines
a Lie 2-algebra $\CCV$ (see Appendix~\ref{Linftyalg}). For
this, we set $V_0=V$, $V_1=V$, and let $\sfd:V_1\to V_0$ be the
identity map ${\rm id}_V$. Let $[-,-]:V_0\wedge V_0\to V_0$ and
$[-,-]:V_0\otimes V_1\to V_1$ be the bracket (\ref{Alghat}) of $V$, and
let $[-,-,-]:V_0\wedge V_0\wedge V_0\to V_1$ be the Jacobiator (\ref{Jacobiatorhat})
of $V$. Then $\big(V_1\xrightarrow{\
  \sfd\ }V_0\big)$ is a 2-term $L_\infty$-algebra canonically
associated to the pre-Lie algebra $V$. 

We can identify the twisted Poisson structure (\ref{Alg}) on the
algebra of functions $C^\infty(T^*M)$ with the natural
twisted Poisson structure on the dual $V^*$ of the pre-Lie algebra $V$ as
follows. We first identify linear functions on $V^*$ with elements of
$V$, and define $\{\hat v_1,\hat v_2\}_R(x):=\langle x,[\hat
v_1,\hat v_2]_R \rangle$,
where $\hat v_1,\hat v_2\in V$, $x\in V^*$ and $\langle
-,-\rangle:V^*\otimes V\to \real$ denotes the dual pairing. By imposing the
Leibniz identity, this defines a quasi-Poisson bracket that extends
to polynomial functions on $V^*$, which in turn are dense in
$C^\infty(V^*)$.

As we discuss in Appendix~\ref{Lie2groups}, there is no general
construction of Lie 2-groups from Lie 2-algebras, but we can build a
suitable integration map with some intuition provided from our
considerations in Section~\ref{TTHA}. For this, we will write
down an equivalent Lie 2-algebra for which a corresponding Lie 2-group
can be ``guessed''. We start by replacing the pre-Lie algebra $V$ with
a quadratic Lie algebra $\frg$ whose generators $\hat x^i,\hat{\tilde
  p}_j$, $i,j=1,\dots,d$ have
the Lie brackets
\beq
[\hat x^i,\hat x^j]_Q=\ii R^{ijk}\, \hat{\tilde p}_j \qquad \mbox{and}
\qquad [\hat x^i,\hat{\tilde p}_j]_Q=0= [\hat{\tilde p}_i ,\hat{\tilde
  p}_j ]_Q \ ,
\label{HeisRfluxalg}\eeq
together with the nondegenerate inner product defined by
\beq
\langle\hat x^i,\hat{\tilde p}_j\rangle= \delta^i{}_j  \qquad
\mbox{and} \qquad \langle \hat x^i,\hat x^j\rangle =0=
\langle\hat{\tilde p}_i,\hat{\tilde p}_j \rangle
\label{tildeinnerprod}\eeq 
which is invariant under the adjoint action and is of split signature. There are two ways to
think about this Lie algebra. Firstly, it is the reduction of the
Courant algebroid of Section~\ref{Rtwist} over a point; we may regard
$\frg\cong\real^d\oplus(\real^d)^*$ as the cotangent bundle
$T^*\real^d$ with its canonical symplectic structure. Secondly, it is
an abstract version of the $Q$-space Poisson brackets
(\ref{QfluxPoisson}), and in particular it coincides with the $\hbar=0$ limit of the
brackets given by (\ref{Alghat}) and (\ref{Jacobiatorhat}); in this way we will mimick the dynamical quantization of
Section~\ref{ConstantR} by first integrating the $d$-dimensional Heisenberg algebra
(\ref{HeisRfluxalg}) involving the ``non-dynamical momenta''
$\hat{\tilde p}_i$, and then making the momenta ``dynamical''
$\hat{\tilde p}_i\to \hat p_i$ to recover the T-dual pre-Lie algebra
(\ref{Alghat}) appropriate to the $R$-space frame with the non-trivial
Jacobiator~(\ref{Jacobiatorhat}).

Associated to the quadratic Lie algebra $\frg$ is a Lie 2-algebra
$\tilde\CCV$ corresponding to the 2-term $L_\infty$-algebra
\beq
\tilde
V \ = \ \big(\tilde V_1 = \real \ \xrightarrow{\
  \tilde \sfd \ } \ \tilde V_0 = \frg \big)
\label{quadLie2alg}\eeq
which is \emph{skeletal},
i.e. $\tilde\sfd=0$, with brackets
$[-,-]:\tilde V_0\wedge \tilde V_0\to\tilde V_0$ given by the Lie
bracket (\ref{HeisRfluxalg}) of $\frg$ and $[-,-]:\tilde V_0\otimes \tilde V_1\to \tilde V_1$
given by $[\hat v,c]=0$ for $\hat v\in\frg$, $c\in\real$, and
Jacobiator $[-,-,-]:\tilde V_0\wedge \tilde V_0\wedge\tilde V_0\to\tilde V_1$
given by $[\hat v_1,\hat v_2,\hat v_3]=\langle[\hat v_1,\hat
v_2]_Q,\hat v_3\rangle$ for $\hat v_i\in\frg$; this is just the
reduction over a point of the Lie 2-algebra structure
(\ref{CourantLie2alg}) canonically associated to the exact Courant
algebroid $C\to M$ of Section~\ref{Rtwist}. The corresponding classifying
triple is $(\frg,\real,j)$ where $\real$ is the trivial representation
of $\frg$ and the 3-cocycle $j:\frg\wedge\frg\wedge\frg\to\real$ is
given by
\beq
j(\hat v_1,\hat v_2,\hat v_3) = \big\langle[\hat v_1,\hat v_2]_Q\,,\,
\hat v_3\big\rangle \ .
\label{tildeJacobiator}\eeq
The cocycle condition (or equivalently the pentagonal coherence
relation (\ref{pentcoherence})) follows from adjoint-invariance of the inner product
and since $\frg$ acts trivially on $\real$; note that its only
non-trivial values on generators are given by
\beq
j(\hat x^i,\hat x^j,\hat x^k)=R^{ijk}
\label{JachatxR}\eeq
as in (\ref{Jacobiatorhat}). The cohomology of the Heisenberg Lie
algebra (\ref{HeisRfluxalg}) is described in~\cite{Santha}; in particular for degree~$3$ one has
\beq
\dim H^3(\frg,\real) = D :=\mbox{$\frac16$}\, d\, (d-1)\,(d-2)-d
\eeq
and the space
of 3-cocycles
\beq
Z^3(\frg,\real)=\mbox{$\bigwedge^3$}\big(\hat x_1^*,\dots,\hat
x_d^*\big)
\label{3cocyclespace}\eeq
is the vector
space of homogeneous elements of degree~$3$ of the Grassmann algebra
over the dual basis to $\hat x^1,\dots,\hat x^d$. It follows that the
Jacobiator (\ref{tildeJacobiator}) gives rise to a generator $[j]$ of
$H^3(\frg,\zed) = \zed^D$, and all generators are obtained via a
choice of basis for the space of totally antisymmetric 3-vectors as in
(\ref{JachatxR}) (modulo linear redefinitions of the central elements
$\hat{\tilde p}_1,\dots,\hat{\tilde p}_d$).

\subsection{Integrating Lie 2-groups\label{RQLie2group}}

The classifying data $(\frg,\real,j)$ of the Lie 2-algebra $\tilde\CCV$, with $\real= \fru(1)$ regarded as the one-dimensional
abelian Lie algebra, can be straightforwardly exponentiated to a
triple $(G,U(1),\varphi)$ corresponding to a special Lie 2-group $\CCG= (\CCG_0,\CCG_1)$ (see
Appendix~\ref{Lie2groups}), modulo one subtlety. The universal 2-step
nilpotent Lie algebra
$\frg$ of rank~$d$ integrates to the non-compact simply connected $d$-dimensional Heisenberg
group $G$, the associated free 2-step nilpotent Lie group. In
order to exponentiate the generator $[j]\in H^3(\frg,\real)$ induced by
the Jacobiator (\ref{tildeJacobiator}) of $\tilde\CCV$ to a \emph{compact} element
$[\varphi]\in H^3(G,U(1))$, it is necessary to restrict the space of
3-cocycles (\ref{3cocyclespace}) to a lattice $\Lambda\cong\zed^d$
of maximal rank in the linear span of the generators $\hat
x^1,\dots,\hat x^d$. This lattice injects into a cocompact
lattice $\Gamma$ in $G$; the resulting quotient $G/\Gamma$ is a
Heisenberg nilmanifold or ``double twisted torus'', familiar in $d=3$
dimensions as the doubled space of the
geometric T-dual to the three-torus with $H$-flux~\cite{Hull:2009sg}. We assume that the
lattice is equipped with a nondegenerate inner product which is given in a suitable
basis by $\eta=(\eta_{ab}):
\Lambda\otimes_\zed \Lambda\to\real$, $a,b=1,\dots,d$, with inverse $\eta^{-1} =(\eta^{ab}):
\Lambda^*\otimes_\zed \Lambda^*\to\real$, and a nondegenerate dual pairing
$\Sigma=(\Sigma_a{}^{i}):\Lambda\otimes_\real (\real^d)^*\to\real$ which is a
vielbein for the inner product, i.e. 
$\Sigma_a{}^{i}\, \delta_{ij}\, \Sigma_b{}^{j}= \eta_{ab}$.

With these restrictions understood, the Lie 2-algebra $\tilde\CCV$
given by (\ref{quadLie2alg}) integrates to the Lie 2-group
\beq
\xymatrix@C=10mm{
\CCG_1=G\times U(1) \ \ar@< 2pt>[r]^{ \ \ \ \ \sfs} \ar@< -2pt>[r]_{ \
  \ \ \ \sft} & \
\CCG_0= G
}
\label{quadLie2group}\eeq
having $U(1)$ as the group of automorphisms of its unit object $1$ in
$G$, in which the source and target maps $\sfs,\sft$ are both projections onto the first
factor, vertical multiplication is given by $(g,\zeta) \circ
(g,\zeta'\,)=(g,\zeta\,\zeta'\,)$ for $g\in G$ and $\zeta,\zeta'\in U(1)$, and horizontal multiplication
$\otimes$ given by group multiplication. The associator
\beq
\CCP_{g,h,k} \,:\, (g\otimes h)\otimes k \ \longrightarrow
\ g\otimes(h\otimes k)
\label{CCPghk}\eeq
is the automorphism given by
\beq
\CCP_{g,h,k} = \big( g\, h\, k \,,\, \varphi(g,h,k) \big) \ ,
\eeq
where we have integrated the Lie algebra 3-cocycle
(\ref{tildeJacobiator}) to the smooth normalised Lie group 3-cocycle $\varphi:G\times
G\times G \to U(1)$
with
\beq
j(\hat v_1,\hat v_2,\hat v_3)= \left. \frac{\partial^3}{\partial
    t_1\, \partial t_2\, \partial t_3} \right|_{t_i=0}\varphi\big(\exp t_1\, \hat v_1\,,\,\exp t_2\, \hat
v_2\,,\,\exp t_3\, \hat v_3\big)
\eeq
for all $\hat v_i\in\Lambda$. All other structure maps of the Lie 2-group $\CCG$ are identity isomorphisms. Finally, to make the
transformation to ``dynamical'' momentum variables $\hat{\tilde
  p}_i\to\hat p_i$, and hence integrate our original Lie 2-algebra $\CCV$
with brackets
(\ref{Alghat}) and (\ref{Jacobiatorhat}), we endow $\CCG$ with a braiding
\beq
\CCB_{g,h} \,:\, g\otimes h \ \longrightarrow \ h\otimes g
\eeq
which is the automorphism given by
\beq
\CCB_{g,h} = \big( g\, h \,,\, \beta(g,h) \big) \ ,
\eeq
where we have integrated the inner product
(\ref{tildeinnerprod}) to the smooth normalised map $\beta:G\times
G\to U(1)$
with
\beq
\langle \hat v_1,\hat v_2\rangle = \left. \frac{\partial^2}{\partial
    t_1\, \partial t_2} \right|_{t_i=0}\beta\big(\exp t_1\, \hat
v_1,\exp t_2\, \hat v_2\big)
\eeq
for all $\hat v_i\in\Lambda$. The braided monoidal category $\CCG$ is
then the Lie 2-group that integrates the Lie 2-algebra $\CCV$.

We can make this construction somewhat more concrete and explicit in a way
that will be suitable to our ensuing constructions. For this, we
formally exponentiate the Lie 2-algebra generators to define
\be
\hat Z^a = \exp\big( 2\pi \ii (\Sigma^{-1})_i{}^{a}\, \hat x^i \big)
\qquad\trm{and}\qquad \hat P_\xi = \exp\big(  \ii  \xi^i\, \hat p_i \big) \label{GlobalZ}
\ee
for $a=1,\dots,d$ and $\xi=(\xi^i)\in\real^d$. We may compute exterior products
$\otimes:\CCG\times\CCG\to \CCG$ of the elements (\ref{GlobalZ})
in the Lie
2-group $\CCG$ by formally applying the Baker--Campbell--Hausdorff formula
using the brackets (\ref{Alghat}) and (\ref{Jacobiatorhat}); since the
bracket functor in this case is nilpotent, the Hausdorff series is
still applicable to the sole finite non-vanishing order that we require it
without any need of the Jacobi identity. The commutation relations are
then given by
\eqa
\hat Z^a \otimes \hat Z^b &=& \hat P_{\xi_R^{ab}} \otimes
\hat Z^b \otimes \hat Z^a \ , \label{CommutatorsZZ}\\[4pt]
\hat Z^a \otimes \hat P_\xi &=&\e^{2\pi \ii \hbar\, 
(\Sigma^{-1})_i{}^{a} \, \xi^i} \ \hat P_\xi \otimes \hat Z^a
\ , \label{CommutatorsZP}\\[4pt]
\hat P_\xi \otimes \hat P_{\xi'} &=& \hat P_{\xi'} \otimes \hat P_{\xi} \ , \label{CommutatorsPP}
\eqaend
where $\xi_R^{ab}\in\real^d$ is given by
\be
\big(\xi^{ab}_R\big)^i =-4\pi^2 \,(\Sigma^{-1})_j{}^{a}\,R^{ijk}\, (\Sigma^{-1})_k{}^{b} \ .
\ee
In (\ref{CommutatorsZP}) we recognize the non-trivial braiding
isomorphism $\CCB_{\hat Z^a,\hat P_\xi}$ on 2-group objects given by
the map $\beta: \real^d\times
\Lambda^* \to U(1)$ whose only non-trivial values are
\beq
\beta(\xi,m) =\e^{2\pi\ii \hbar\, \xi^i\, (\Sigma^{-1})_i{}^a\, m_a}
\eeq
for $\xi= (\xi^i)\in\real^d$ and $m=(m_a)\in\Lambda^*\cong\zed^d$,
while the remaining commutation relations in
(\ref{CommutatorsZZ})--(\ref{CommutatorsPP}) are those of the rank~$d$
Heisenberg group $G$. The non-trivial associators follow by applying the 
Baker--Campbell--Hausdorff formula once more to find
\be 
\big( \hat Z^a \otimes \hat Z^b \big) \otimes \hat Z^c =\e^{-2 \pi
  \ii \hbar\, R^{abc}} \ \hat Z^a \otimes \big( \hat Z^b \otimes \hat Z^c \big) \ , \label{ZZZ} 
\ee
where
\be
R^{abc}= 2\pi^2\, R^{ijk} \, (\Sigma^{-1})_i{}^{a}\, (\Sigma^{-1})_j{}^{b}\,(\Sigma^{-1})_k{}^{c}
\ee
are the dimensionless nonassociativity $R$-flux parameters. 
This expression is the Lie 2-group version of the ``cyclic double commutator''
that was calculated in \cite{Blumenhagen:2010hj}, which we
recognise as the action of the non-trivial associator isomorphism $\CCP_{\hat
  Z^a,\hat Z^b,\hat Z^c}$ on 2-group objects. The corresponding
3-cocycle can be regarded as a group homomorphism or tricharacter
$\varphi: \Lambda^*\times\Lambda^*\times \Lambda^*\to U(1)$ defined by
\beq
\varphi(m,n,q)=\e^{-2\pi\ii \hbar\, R^{abc}\, m_a\, n_b \, q_c} \ .
\label{varphimnq}\eeq
This map is normalised, i.e. $\varphi(m,n,q)=1$ if either of $m$, $n$, or $q$
is $0$; this implies that the two obvious maps from $\hat
Z^a\otimes(1\otimes\hat Z^b)=\CCP\big((\hat Z^a\otimes 1)\otimes\hat
Z^b\big)$ to $\hat Z^a\otimes\hat Z^b$ are consistent. It is also
skew-symmetric, i.e. $\varphi(m,n,q)=\varphi(n,m,q)^{-1}=\varphi(m,q,n)^{-1}= \varphi(q,n,m)^{-1}$, and it obeys the required pentagonal cocycle
identity
\beq
\varphi(m,n,q)\, \varphi(m,n+q,r)\, \varphi(n,q,r) = \varphi(m+n,
q,r)\, \varphi(m,n,q+r)
\label{varphipentagon}\eeq
for $m,n,q,r\in\Lambda^*$, which is equivalent to the pentagon identity \eqref{pentagonid} of the
category $\CCG$. The pentagon identity can also be derived explicitly by
iterating the above calculations to find the non-trivial higher nonassociativity relations
\bea
\hat Z^a \otimes \big( \hat Z^b \otimes (\hat Z^c \otimes \hat Z^d)\big) &= &\e^{2\pi\ii\hbar\,
  R^{bcd}}\ \hat Z^a \otimes \big{(}(\hat Z^b \otimes \hat Z^c)
\otimes \hat Z^d \big{)} \nn \\[4pt]
&=& \e^{2\pi\ii\hbar\, (R^{abc} + R^{abd})} \ \big(\hat Z^a  \otimes
\hat Z^b \big) \otimes \big( \hat Z^c \otimes \hat Z^d \big)
\label{Pentagon} \\[4pt]
&=& \e^{2\pi\ii\hbar\, (R^{acd} + R^{abd}+ R^{bcd})} \ \big{(} \hat Z^a \otimes (\hat Z^b \otimes \hat Z^c)\big{)}\otimes \hat Z^d \nn\\[4pt]
&=& \e^{2\pi\ii\hbar\, (R^{abc} +  R^{acd} + R^{abd} +  R^{bcd})} \ \big{(}( \hat Z^a \otimes \hat Z^b
) \otimes \hat Z^c\big{)} \otimes \hat Z^d \ . \nn
\eea

As discussed in Appendix~\ref{Lie2groups}, MacLane's coherence theorem
implies that these relations automatically imply all higher
associativity relations in the category $\CCG$. This is particularly
interesting from the perspective of the quantization of Nambu--Poisson
structures that we discussed in Section~\ref{ConstantR}: As the
fundamental identity (\ref{NPfundid}) should be encoded in the coherence relations
involving five objects, our categorical approach automatically encodes
its quantization. This should therefore help to alleviate at least
some of the
difficulties that arise in implementing the fundamental identity for
Nambu--Poisson brackets at the quantum level (see
e.g.~\cite{DeBellis:2010pf} for a discussion).

\subsection{Convolution algebra
  objects\label{Lie2groupalg}}

We will now apply this categorical formalism to the deformation
quantization of the algebra of functions $C^\infty(T^*M)$ on
$T^*M=\torus^d\times(\real^d)^*$, regarded as the algebra $C^\infty(V^*)$
as explained before. Here $\torus^d=\real^d/\Lambda$, and the $d\times
d$ invertible matrix $\Sigma=(\Sigma_a{}^{i})$ defines the periods 
of the directions of the $d$-torus $M= \torus^d$, i.e. $x^i\sim x^i+\Sigma_a{}^i$, $a=1,\dots,d$ 
for each $i=1,\dots,d$; in particular, the (inverse) metric of
$\torus^d$ is given by
$\Sigma_a{}^{i}\, \delta^{ab}\, \Sigma_b{}^{j}= g^{ij}$. We embed
$C^\infty(T^*M)$ as an algebra object $\alg$ of
the Lie 2-group $\CCG$ via a categorification of the Weyl
quantization map, see e.g.~\cite{Szabo:2001kg}; it is defined as the
linear isomorphism on $C^\infty(T^*M)$ 
given on the dense set of plane waves by
\eqa
\CCW\big(\e^{\ii k_I\, x^I}\big) = \hat W(m,\xi) := \exp\big(\ii k_I\, \hat x^I\big)
\ ,
\label{CCWdef}\eqaend
and extended by linearity; here
\beq
(k_I)= (k_1,\dots,k_{2d}) = (k_1,\dots,k_d,\xi^1,\dots,\xi^d)
\eeq
with
\beq
k_i=2\pi\,(\Sigma^{-1})_i{}^a\, m_a \ , \qquad m=(m_a)\in\Lambda^*
\eeq
the quantized Fourier momenta appropriate to smooth single-valued
functions on $\torus^d$.
We regard (\ref{CCWdef}) as an object in a suitable enrichment of the Lie 2-group $\CCG$
to a linear category over $\complex$, which we think of as an analog of a
convolution group algebra generated by the operators (\ref{GlobalZ}). 
This map can be applied to an arbitrary Schwartz function $f$ on
$\torus^d \times (\real^d)^*$ by expanding $f$ in its Fourier transformation
\be
 f(x,p)=\sum_{m\in\Lambda^*} \, \e^{2\pi \ii (\Sigma^{-1})_i{}^{a} \, m_a
   \, x^i} \ \int_{\real^d}\, {\frac{\dd^d \xi}{(2\pi)^d} \ f_m(\xi)\,
     \e^{\ii \xi^i  \, p_i}} \ ,
\ee
where the inverse Fourier transform is given by
\be
f_m (\xi) =\frac1{|\det\Sigma|}\, \int_{\torus^d}\, {\dd^d x \
  \e^{-2\pi \ii (\Sigma^{-1})_i{}^{a} \, m_a \, x^i}} \
\int_{(\real^d)^*}\, \dd^d p \ \e^{-\ii \xi^i \, p_i}\, f(x,p) \ .
\ee
We then set
\beq
\CCW(f):= \sum_{m\in\Lambda^*} \ \int_{\real^d}\,
\frac{\dd^d\xi}{(2\pi)^d} \ f_m(\xi) \ \hat W(m,\xi) \ .
\label{hatWf}\eeq

The convolution product $\circledast$ of two functions $f,g\in C^\infty(T^*M)$ is
defined via the horizontal product of two quantized functions as
\beq
\CCW(f\circledast g):= \CCW(f) \otimes \CCW(g)
\label{Whatprod}\eeq
in the 2-group $\CCG$ and the inverse map $\CCW^{-1}$ from
(\ref{CCWdef}). Another straightforward application of the
Baker--Campbell--Hausdorff formula as in
(\ref{CommutatorsZZ})--(\ref{CommutatorsPP}) yields the 2-group
multiplication law
\beq
\hat W(m,\xi)\otimes\hat W(n,\lambda)= \e^{\pi\ii \hbar\, (\Sigma^{-1})_i{}^a\,
  (m_a\,\lambda^i-n_a\, \xi^i)} \ \hat W\big(m+n\,,\,\xi+\lambda-R^{abc}\, m_a\,
n_b\, \Sigma_c\big) \ ,
\label{2groupmultlaw}\eeq
and we obtain
\eqa 
(f\circledast g)(x, p) &=&\sum_{m,n\in \Lambda^*} \ {\int_{\real^d}\,
  {\frac{\dd^d \xi}{(2\pi)^{d}} \ \int_{\real^d}\,
    \frac{\dd^d\lambda}{(2\pi)^d} \ f_n(\lambda)\, g_{m-n}(\xi-\lambda)\,
    \e^{2\pi \ii m_a\, (\Sigma^{-1})_i{}^a\, x^i + \ii \xi^i \, p_i}}}\nn\\
&&\qquad\qquad\qquad \times\, \e^{- \pi\ii \,
  (\Sigma^{-1})_i{}^a\, (\hbar\,(m_a \, \lambda^i - n_a \, \xi^i) - 2\pi\, 
  (\Sigma^{-1})_j{}^b\, m_a\, n_b\, R^{ijk}\, p_k) } \ . \label{BCHStar}
\eqaend
After introducing a factor of $\hbar$ as in (\ref{Schoutenb}), this formula is identical to the star product \eqref{Recovery} that we
found by formal deformation 
quantization along the twisted Poisson structure $\Theta$, and hence
the two quantizations are equivalent in this particular case. This
result is a Lie 2-algebra version of Kathotia's
theorem~\cite[Section~5]{Kath:1998} which asserts the equivalence
between Kontsevich's deformation quantization and the group
convolution algebra quantization of the dual of a nilpotent Lie
algebra. The crux of this theorem does not rely on the Jacobi
identity, and is easily applied to our pre-Lie algebra: By a trivial
relabelling of the generators, the commutation relations
(\ref{Alghat}) satisfy the conclusions
of~\cite[Theorem~5.2.1]{Kath:1998}. It is tempting to conjecture that
the Lie 2-group convolution algebra quantization that we have
developed in this section is equivalent to Kontsevich's deformation
quantization along the linear twisted Poisson bivector field on the dual
of any nilpotent pre-Lie algebra. It would be interesting to similarly
characterise the nonassociative quantizations of generic semistrict
nilpotent Lie 2-algebras, but these questions lie beyond the scope of
the present paper.

We conclude by establishing that the algebra of functions
$\alg= C^\infty(\torus^d\times(\real^d)^* )$ endowed with the nonassociative product $\circledast$
is really an algebra object of the Lie 2-group $\CCG$, i.e. it
satisfies the associativity relation (\ref{catassoc}) of the
category. Using the multiplication law (\ref{2groupmultlaw}) of the
2-group $\CCG$ we compute triple products of the operators
(\ref{CCWdef}) to get
\bea
&& \big(\hat W(m,\xi)\otimes\hat W(n,\lambda)\big) \otimes \hat
W(q,\eta) \nn
\\[4pt] && \qquad  \qquad \ = \ \e^{\pi\ii\hbar\, R^{abc}\, m_a\, n_a\, q_c}\,
\e^{\pi\ii\hbar\,(\Sigma^{-1})_i{}^a\,
  (m_a\,\lambda^i-n_a\,\xi^i+(m+n)_a\,\eta^i-q_a\, (\xi+\lambda)^i)} \label{tripleWhat}
\\ && \qquad \qquad \qquad \qquad \times \, \hat W\big(m+n+q\,,\,\xi+\lambda+\eta-R^{abc}\,
m_a\, n_b\,\Sigma_c-R^{abc}\, (m+n)_a\, q_b\, \Sigma_c\big) \ . \nn
\eea
A completely analogous calculation for the other ordering shows that
\bea
\hat W(m,\xi)\otimes\big(\hat W(n,\lambda) \otimes\hat
W(q,\eta) \big) &=& \varphi(m,n,q) \ \big(\hat W(m,\xi) \otimes\hat
W(n,\lambda)\big) \otimes \hat W(q,\eta) \nn\\[4pt] &=& \CCP\big[\big(\hat W(m,\xi)\otimes\hat W(n,\lambda) \big) \otimes\hat
W(q,\eta)\big] \ ,
\label{hatWassoc}\eea
where $\varphi$ is the 3-cocycle (\ref{varphimnq}) and we have used
(\ref{ZZZ}) to identify the application of the associator isomorphism
$\CCP$ to
2-group objects (\ref{CCWdef}); this formula is
extended to operators (\ref{hatWf}) in the usual way using
linearity. Using (\ref{tripleWhat}) and the quantization map
(\ref{CCWdef}), (\ref{Whatprod}) we now compute the triple convolution product of
functions $f,g,h\in C^\infty(T^*M)$ to get
\bea
\big(f\circledast (g\circledast h)\big)(x,p) &=&
\sum_{m,n,q\in\Lambda^*}\, \e^{-\pi\ii\hbar\, R^{abc}\, m_a\, n_b\,
  q_c}\, \e^{2\pi\ii (\Sigma^{-1})_i{}^a\, m_a\, x^i} \
\int_{\real^d}\, \frac{\dd^d\xi}{(2\pi)^d} \ \e^{\ii \xi^i\, p_i} \nn \\
&& \qquad \times \ \int_{\real^d}\, \frac{\dd^d\lambda}{(2\pi)^d} \
\int_{\real^d}\, \frac{\dd^d\eta}{(2\pi)^d} \
f_{m-n-q}(\xi-\lambda-\eta)\, g_n(\lambda)\, h_q(\eta) \nn\\ && \qquad\qquad
\times \, \e^{\pi\ii
  \hbar\, (\Sigma^{-1})_i{}^a\, ((m-q)_a\, \lambda^i-n_a\,
  (\xi-\eta)^i+m_a\, \eta^i-q_a\, \xi^i)} \nn \\ && \qquad\qquad\qquad \times\, \e^{\ii
  R^{abc}\, ((m-q)_a\, n_b+m_a\, q_b)\, \Sigma_c{}^i\, p_i} \ ,
\eea
which agrees with the corresponding formula of (\ref{Assocbracketing}). From (\ref{hatWassoc}) it follows that 
\beq
\big(f\circledast(g\circledast h)\big)(x,p)= \CCP\big((f\circledast
g)\circledast h\big)(x,p)
\eeq
as required, where here $\CCP((f\circledast g)\circledast h)$ is
short-hand notation for the composition of morphisms on the right-hand side of (\ref{catassoc})
applied to $(f\otimes g)\otimes h$.

\subsection{Monopole backgrounds and topological nonassociative
  tori\label{Lie2groupcomp}}

We conclude this section by comparing our noncommutative and
nonassociative deformation of the cotangent bundle
$T^*M=\torus^d\times (\real^d)^* $ with some other appearences of
nonassociativity in the literature. The relations
(\ref{2groupmultlaw}) (or (\ref{CommutatorsZZ})) are reminescent of
those obeyed by the gauge invariant operators which generate a
projective representation of the translation group in the background
field of a Dirac monopole~\cite{Jackiw:1984rd} (see
also~\cite{Nesterov:2004bn}), where the projective phase is a
2-cochain determined by the magnetic flux through a 2-simplex; in our
case this flux is proportional to $\xi_R(m,n)\in\real^d$ where
\beq
\xi_R(m,n)^i=-R^{abc}\, m_a\, n_b\, \Sigma_c{}^i \ ,
\label{ximni}\eeq
and
it arises as the gerbe 2-holonomy of the $B$-field (\ref{Bfieldmomsp})
through the triangle at $p$ formed by the lattice vectors $m,n\in\Lambda^*\subset(\real^d)^*$
in momentum space. The triple product relation (\ref{hatWassoc}) (or
(\ref{ZZZ})) is reminescent of the nonassociativity relation which arises from
the 3-cocycle proportional to
the flux through the 3-simplex enclosing the monopole; in our case the
3-cocycle (\ref{varphimnq}) is determined by the gerbe $H$-flux
(\ref{Hmom}) through the tetrahedron at $p$ formed by the lattice
vectors $m$, $n$ and $q$ in momentum space. See~\cite{Ellwood:2006my}
for an open string
realization of the monopole background in terms of D0-branes in $H$-space, or equivalently
D3-branes in $R$-space.

Let us now compare our construction with the nonassociative tori
discussed
in~\cite{Bouwknegt:2004ap,Bouwknegt:2007sk,Hannabuss:2010tp}. For
$n,q\in\Lambda^*$, we use the 3-cocycle (\ref{varphimnq}) to define unitary operators $\hat U_{n,q}$ on the
Hilbert space $\ell^2(\Lambda^*)$ of square-summable sequences $f_m$ on
the momentum lattice $\Lambda^*$ of $M=\torus^d$ by
\beq
\big(\hat U_{n,q}f\big)_m=\varphi(m,n,q)\, f_m \ .
\label{hatUnqfm}\eeq
These operators obey the composition law
\beq
\varphi(m,n,q)\, \hat U_{m,n}\, \hat U_{m+n,q} = \alpha_m\big(\hat
U_{n,q}\big)\, \hat U_{m,n+q} \ ,
\eeq
where $\alpha_m$ is the adjoint action by the regular
representation $f_n\mapsto f_{n+m}$ of lattice translations by $m\in\Lambda^*$. One then defines the twisted
convolution product
\beq
(f\circledast_\varphi g)_m=\sum_{n\in\Lambda^*}\, f_n\,\alpha_n(g_{m-n})\, \hat
U_{n,m-n}
\label{twistedconvprod}\eeq
on the algebra $C^\infty(\Lambda^*,\CK)$, where
$\CK=\CK\big(\ell^2(\Lambda^*)\big)$ is the algebra of compact operators
on $\ell^2(\Lambda^*)$. This defines a
nonassociative twisted crossed-product algebra
$\CK\big(\ell^2(\Lambda^*)\big)\rtimes_\varphi\Lambda^*$ which is
identified with the algebra of functions on the nonassociative
torus. When $\varphi=1$ ($R=0$), the operators $\hat U_{n,q}$ all act
as the identity operator on $\ell^2(\Lambda^*)$ and $\alpha_m$ can be
taken to be the identity; then $\circledast_{\varphi=1}$ is just the usual
convolution product on the algebra $C^\infty(\torus^d)\otimes\CK$ of
stabilized functions on the torus $\torus^d$, which is Morita
equivalent to the usual commutative algebra $C^\infty(\torus^d)$. In
the general case, by~\cite[Proposition~3.1]{Bouwknegt:2004ap} the twisted convolution product
$\circledast_\varphi$ satisfies (\ref{catassoc}) and hence makes
$\CK\big(\ell^2(\Lambda^*)\big)\rtimes_\varphi\Lambda^*$ an algebra
object of
the tensor category $\CCG$; in~\cite{Hannabuss:2010tp} it is shown
that this defines a \emph{strict} (i.e. non-formal) nonassociative
deformation quantization.

We can identify a covariant representation of $\big(\Lambda^*\,,\,
\CK(\ell^2(\Lambda^*)) \big)$ by using the commutation relations
(\ref{CommutatorsZP}) to identify the generators of translations in
the lattice $\Lambda^*$ as the operators $\hat W(m,0)$ for
$m\in\Lambda^*$. From (\ref{2groupmultlaw}) we may then identify
operators through the 2-group multiplication law
\beq
\hat W(m,0)\otimes\hat W(n,0) := \hat U_{m,n}\otimes \hat W(m+n,0) \ ,
\eeq
where
\beq
\hat U_{m,n}=\hat P_{\xi_R(m,n)}
\eeq
and we have used the Baker--Campbell--Hausdorff formula together with
antisymmetry of $R^{abc}$. By (\ref{hatWassoc}), it follows
from~\cite[Section~3]{Bouwknegt:2004ap} that these operators coincide
with the ones introduced in (\ref{hatUnqfm}). This correspondence is
completely analogous to that found
in~\cite[Section~5.2]{Ellwood:2006my} via an open string analysis of
D3-branes in $R$-space; in particular, our representation of the
operators $\hat U_{m,n}$ is determined by the surface holonomy
(\ref{ximni}) of the pertinent $B$-field as
in~\cite{Ellwood:2006my}. However, in our picture,
the meaning of the stabilization by the algebra of compact operators
$\CK$ is clear: It represents precisely the additional cotangent
degrees of freedom through the unitary momentum operators $\hat P_\xi\in\CK$
from (\ref{GlobalZ}).

We close by commenting on how our nonassociative algebras may be related to associative ones which can be represented as operator
algebras on separable Hilbert spaces, hence justifying some of the
constructions above. In the context of open strings in non-trivial
$H$-flux backgrounds, it was shown in~\cite{Ho:2000fv,Ho:2001qk} how
to map the nonassociative algebra of functions equiped with the
Kontsevich star product to an associative algebra by enlarging the
deformed configuration space to a deformed phase space; the resulting
algebra is interpreted as an algebra of pseudo-differential operators
as now both coordinates $x^I$ and derivatives $\partial_I$
appear. This mapping is the analog of the Bopp shift which maps the
Heisenberg commutation relations onto trivial commuting variables
when viewed as a subalgebra of extended canonical phase space
commutation relations. In~\cite{Chatzistavrakidis:2012yp} such Bopp
shifts are used to map noncommutative twisted tori onto commutative
tori with the same phase space nonassociativity. In our case, the
resulting algebra should be compared with the Lie 2-algebra
constructed in Section~\ref{twistedhigher} which has underlying
associative coordinate algebra. The construction
of~\cite{Ho:2000fv,Ho:2001qk} is simply a physical implementation of
MacLane's coherence theorem, which states that any monoidal category
is equivalent to a strict monoidal category in which the associativity
isomorphism (\ref{CCPghk}) is simply the obvious identification by
rebracketing $(g\otimes h)\otimes k\mapsto g\otimes(h\otimes k)$. In
the present case, it is shown in~\cite{Bouwknegt:2007sk} that the
equivalence functor is obtained by applying $\CCP^{-1}$ to
(\ref{twistedconvprod}) and it takes an algebra object $\alg$ to the
associative crossed product
algebra $\alg\rtimes\Lambda^*$; this augmented algebra is in a sense
the ``exponentiation'' of the extended algebras of~\cite{Ho:2000fv,Ho:2001qk}.

\subsection*{Acknowledgments}

We thank B.~Jur\v{c}o, A.~Konechny and C.~S\"amann for helpful
discussions. This work was supported in part by the Consolidated Grant ST/J000310/1
from the UK Science and Technology Facilities Council. The work of
D.M. is supported by 
the Greek National Scholarship Foundation.

\appendix

\newsection{Higher Lie algebra structures\label{HigherLA}}

In this appendix we collect the pertinent mathematical material on
higher structures which are used extensively in the main text.

\subsection{Lie 2-algebras\label{Linftyalg}}

\noindent
{\bf Homotopy Lie algebras. \ }
An \emph{$L_\infty$-algebra} or \emph{strong homotopy Lie algebra} is
a graded vector space $V$ together with a collection of totally
(graded) antisymmetric $n$-brackets $[-,\dots,-]:\bigwedge^nV \to V$, $n\geq1$ of degree $n-2$ satisfying the \emph{higher} or \emph{homotopy Jacobi identities}
\beq
\sum_{i=1}^n \ \sum_{\sigma\in{\rm Sh}(i,n-i)}\, (-1)^{\alpha(\sigma)}\, \big[[v_{\sigma(1)},\dots,v_{\sigma(i)}],v_{\sigma(i+1)},\dots,v_{\sigma(n)}\big]= 0
\label{higherJac}\eeq
for each $n\geq1$. Here $(-1)^{\alpha(\sigma)}$ is a prescribed sign rule for permuting homogeneous elements $v_1,\dots,v_n\in V$, while ${\rm Sh}(i,n-i)$ is the set of permutations $\sigma\in S_n$ which preserve the orders of the first $i$ elements and of the last $n-i$ elements, i.e. $\sigma(1)<\dots<\sigma(i)$ and $\sigma(i+1)<\dots<\sigma(n)$ for $i=1,\dots,n$. 

Denote the $1$-bracket by $\sfd:=[-]$. It has degree $-1$ and the generalized Jacobi identity (\ref{higherJac}) for $n=1$ reads
\beq
\sfd^2=0 \ ,
\eeq
which implies that $\sfd:V\to V$ is a differential making $V$ into
a chain complex. For $n=2$ one has
\beq
\sfd[v,w]=[\sfd v,w]+(-1)^{|v|}\, [v,\sfd w] \ ,
\eeq
which implies that $\sfd$ is a derivation with respect to the
antisymmetric $2$-bracket $[-,-]:V\wedge V\to V$. The bracket $[-,-]$
satisfies the usual Jacobi identity only up to a homotopy correction; from (\ref{higherJac}) with $n=3$ we obtain
\bea
&& (-1)^{|v|\,|u|}\,\big[[v,w],u\big]+(-1)^{|w|\,|u|}\,\big[[u,v],w\big] +(-1)^{|v|\,|w|}\, \big[[w,u],v\big] \label{LinftyJac} \\[4pt] && \qquad \ = \ (-1)^{|v|\,|u| +1}\, \big(\sfd[v,w,u]+[\sfd v,w,u]+(-1)^{|v|}\,[v,\sfd w,u]+(-1)^{|v|+|w|}\, [v,w,\sfd u]\big) \ , \nonumber
\eea
which implies that the \emph{Jacobiator} $[-,-,-]:V\wedge V\wedge
V\to V$ is a chain homotopy map. For $n>3$, the identities
(\ref{higherJac}) impose extra coherence relations on this homotopy
and all higher homotopies.

If $V$ has trivial grading, then an $L_\infty$-algebra is simply an
ordinary Lie algebra. More generally, an $L_\infty$-algebra with vanishing $n$-brackets for all $n\geq3$ is a differential graded Lie algebra.

A \emph{$2$-term $L_\infty$-algebra} is a strong homotopy Lie algebra
with underlying graded vector space $V=V_0\oplus V_1$ concentrated in
degrees $0$ and $1$; it has vanishing $n$-brackets for $n>3$ and the
only non-trivial identities in (\ref{higherJac}) occur for
$n=1,2,3,4$. It
may be regarded as a 2-term chain complex $V=\big(V_1\xrightarrow{\
  \sfd\ }V_0\big)$ whose bracket $[-,-]:V_i\otimes V_j\to V_{i+j}$,
$i+j=0,1$, is a chain map and whose Jacobiator $[-,-,-]:V_0\wedge
V_0\wedge V_0\to V_{1}$ is a chain homotopy from the chain map
\beq
V_0\wedge V_0\wedge V_0 \ \longrightarrow \ V_{1} \ , \qquad v\wedge w\wedge u \ \longmapsto \ \big[v,[w,u]\big]
\eeq
to the chain map
\beq
V_0\wedge V_0\wedge V_0 \ \longrightarrow \ V_{1} \ , \qquad v\wedge
w\wedge u \ \longmapsto \ \big[[v,w] ,u\big]+\big[w,[v,u]\big]
\eeq
satisfying the coherence condition
\bea
&&
\big[v,[w,u,s]\big]+\big[v,[w,u],s\big]+\big[v,u,[w,s]\big] +\big[[v,w,u],s]+\big[u,[v,w,s]\big]
\label{pentcoherence} \\[4pt] && \qquad \ = \ \big[v,w,[u,s]\big]+\big[[v,w],u,s\big] +
\big[w,[v,u,s]\big]+\big[w,[v,u],s\big]+ \big[w,u,[v,s]\big] \ . \nonumber
\eea
This higher Jacobi identity relates the two ways of using the Jacobiator to
rebracket the expression $[[[s,v],w],u]$.

A related notion is that of an \emph{$A_\infty$-algebra}, or
\emph{homotopy associative algebra}, which is a graded vector space
$A$ endowed with a family of $n$-multiplication operations
$\mu_n:A^{\otimes n}\to A$ of degree $n-2$, $n\geq1$ obeying the higher or
homotopy associativity relations
\beq
\sum_{j+k+l=n} \, (-1)^{\sigma}\,
\mu_n\circ\big(\Id_{A^{\otimes j}}\otimes \mu_k\otimes
\Id_{A^{\otimes l}}\big) = 0 \ .
\eeq
The first two relations
\bea
\sfd^2=0 \qquad \mbox{and} \qquad \sfd \mu_2(a,b)=\mu_2(\sfd
a,b)+(-1)^{|a|}\, \mu_2(a,\sfd b)
\eea
for $a,b\in A$ make $A$ into a chain complex with differential
$\sfd:=\mu_1$ which is a graded
derivation of the binary product $\mu_2$. The third relation states that the product $\mu_2$ is associative up to
the homotopy $\mu_3$, and so on. From an $A_\infty$-algebra structure
on $A$ one
constructs an $L_\infty$-algebra structure through the antisymmetric
$n$-brackets
\beq
[a_1,\dots,a_n]:= \sum_{\sigma\in S_n}\, {\rm sgn}(\sigma)\,
\mu_n(a_{\sigma(1)} ,\dots,a_{\sigma(n)})
\eeq
for $a_1,\dots,a_n\in A$. However, in general there is no converse
enveloping algebra type procedure to construct an $A_\infty$-structure
from an $L_\infty$-structure.

\medskip

\noindent
{\bf Lie 2-algebras. \ }
2-term $L_\infty$-algebras are the same things as Lie
2-algebras~\cite[Theorem~36]{Baez:2003fs}, which are categorified
versions of Lie algebras in which the Jacobi identity is replaced by a
Jacobiator isomorphism. For this, recall that a
\emph{$2$-vector space} is a linear category $\CCV=(\CCV_0,\CCV_1)$ consisting of a
vector space of objects $\CCV_0$ and a vector space of morphisms $\CCV_1$,
together with source and target maps $\sfs,\sft:\CCV_1\rightrightarrows \CCV_0$ sending a
morphism to its domain and range, and an inclusion map $\unit:\CCV_0\to
\CCV_1$, $v\mapsto\unit_v$, sending an object to its identity
morphism; the set of composable morphisms is
$\CCV_1\times_{\CCV_0}\CCV_1=\{(v_1,w_1)\in\CCV_1\times\CCV_1 \ | \
\sfs(w_1)=\sft(v_1)\}$. These maps are all linear and compatible in the usual sense with the composition $\circ:\CCV_1\times_{\CCV_0}\CCV_1\to
\CCV_1$ in the category.

A \emph{Lie
  $2$-algebra} is a $2$-vector space $\CCV$ together with an
antisymmetric bilinear bracket
functor $[-,-]_\CCV:\CCV\times \CCV\to \CCV$ and a natural
antisymmetric trilinear Jacobiator
isomorphism on objects satisfying a higher Jacobi identity. A Lie
2-algebra $\CCV$ is \emph{strict} if its Jacobiator is the identity isomorphism;
in that case both $\CCV_0$ and $\CCV_1$ are Lie algebras, and each
operation on the category is a homomorphism of Lie algebras. Otherwise
$\CCV$ is \emph{semistrict}; this is the case of relevance to this paper.

Given a 2-term
$L_\infty$-algebra $V=\big(V_1\xrightarrow{\
  \sfd\ }V_0\big)$, we construct a 2-vector space
$\CCV$ with vector spaces of objects and morphisms given by
$\CCV_0=V_0$ and $\CCV_1= V_0\oplus V_1$. A morphism $f=(v_0,v_1)$ in $\CCV_1$ with $v_0\in V_0$ and $v_1\in V_1$ has
source and target given by $\sfs(v_0,v_1)=v_0 $ and $\sft(v_0,v_1)=v_0
+\sfd v_1$,
while the object inclusion is $\unit_v=(v,0)$. The composition of
two morphisms $f=(v_0,v_1)$ and $f'=(v_0+\sfd v_1,v_1'\,)$ in $\CCV_1$ is $f\circ
f':=(v_0,v_1+v_1'\,)$. The bracket functor $[-,-]_\CCV:\CCV\times \CCV\to
\CCV$ is defined on objects $v,v'\in\CCV_0$ by $[v,v'\, ]_\CCV=[v,v'\,
]$, where $[-,-]$ denotes the bracket in the $L_\infty$-algebra $V$. The bracket of morphisms $f=(v_0,v_1)$ and $f'=(v_0',v_1'\,)$ in $\CCV_1$ is given by
\bea
[f,f'\,]_\CCV= \big([v_0,v_0'\,]\,,\,[v_1,v_0'\, ]+[v_0 +\sfd v_1,v_1'\,]\big) =
\big([v_0,v_0'\,]\,,\, [v_0,v_1'\,]+[v_1,v_0'+\sfd v_1'\, ]\big) \ .
\label{ffprimeCCV}\eea
Finally, the Jacobiator for $\CCV$ is defined on $v,w,u\in\CCV_0$ by
\beq
[v,w,u]_\CCV:= \big(\big[[v,w],u\big]\,,\, [v,w,u]\big) \ , 
\eeq
with source $\sfs([v,w,u]_\CCV)=[[v,w],u]$ and target
$\sft([v,w,u]_\CCV)=[v,[w,u]]+ [[v,u],w]$ by (\ref{LinftyJac}).

The skew-symmetric bracket
\beq
[v_1,v_1'\,]=[\sfd v_1,v_1'\,]=[v_1,\sfd v_1'\,] \ ,
\label{derivedbracket}\eeq
defined on elements $v_1,v_1'\in V_1$ which figures in the formula
(\ref{ffprimeCCV}), is called the \emph{derived bracket}. It satisfies
the Jacobiator identity
\beq
\big[v_1,[v_1',v_1^{\prime\prime}\,]\big] -
\big[[v_1,v_1'\,],v_1^{\prime\prime}\, \big] -
\big[v_1',[v_1,v_1^{\prime\prime}\,]\big] = [\sfd v_1,\sfd v_1',\sfd
v_1^{\prime\prime}\, ] \ .
\label{derivedJac}\eeq

\medskip

\noindent
{\bf Classification of Lie 2-algebras. \ }
There is a bijective correspondence between semistrict Lie 2-algebras and
certain classifying ``Postnikov data''~\cite{Baez:2003fs}, analogous to
the Faulkner construction of 3-Lie algebras. The data in question
are triples $(\frg,W,j)$ consisting of a Lie algebra
$\frg$, a representation of $\frg$ on a vector space $W$, and a
3-cocycle $j$ on $\frg$ with values in $W$; the isomorphism classes
are parametrized by elements $[j]\in H^3(\frg,W)$ of the degree~3 Lie algebra cohomology . 

For a Lie 2-algebra $\CCV$ obtained from a 2-term $L_\infty$-algebra
$V=\big(V_1\xrightarrow{\
  \sfd\ }V_0\big)$, the corresponding triple $(\frg,W,j)$ is
constructed by firstly setting
$\frg=\ker(\sfd)\subseteq V_1$; since $\sfd=0$ on $\frg$ the 2-bracket of the
$L_\infty$-structure satisfies the Jacobi identity exactly and makes
$\frg$ into a Lie algebra. Now let $W={\rm coker}(\sfd)\subseteq V_0$,
and use the 2-bracket to define an action $\frg\otimes W\to W$ by
$g\triangleright w=[g,w]$ for $g\in\frg$, $w\in W$; in this
correspondence $W$ is regarded as the abelian Lie algebra of
endomorphisms of the zero object of $\CCV$. Finally, the
Jacobiator of the
$L_\infty$-structure gives a map $[-,-,-]:\frg\wedge \frg\wedge
\frg\to W$ which is a Chevalley--Eilenberg 3-cocycle $j$ whose cohomology class $[j]\in
H^3(\frg,W)$ is the obstruction to $\CCV$ being functorially equivalent
to a strict Lie 2-algebra, or equivalently to a differential $\Z_2$-graded
Lie algebra.

\subsection {Gerstenhaber brackets\label{Definitions}}

Consider the Hochschild complex $H^n(\mc A, \mc A)=\textrm{Hom}_\complex(\mc A^{\otimes n},\mc A)$ 
of an algebra $\mc A$ with product
$\star\in H^2(\mc A, \mc A)$. The space of $n$-cochains
$C^n(\mc A, \mc A)=\textrm{Hom}_\complex(\, \bigwedge^n\mc A ,
\mc A)$ is constructed by antisymmetrization, and the Hochschild coboundary
operator $\dd_\star : C^n(\mc A, \mc A) \rightarrow C^{n+1}(\mc A, \mc A)$ is defined by 
\eqa
\dd_\star\mc C (f_1,\dots,f_{n+1}) &=& f_1 \star \mc C ( f_2, \dots
,f_{n+1}) + \sum_{i=1}^n\, (-1)^i\, \mc C(f_1,\dots,f_i\star f_{i+1},\dots,f_{n+1}) \nn\\
&&+\, (-1)^{n+1}\, \mc C ( f_1, \dots ,f_n) \star f_{n+1}
\label{Dstar}\eqaend
for $\mc C\in C^n(\mc A, \mc A)$ and $f_i\in\alg$.
From the product on $\alg$ we construct a cup product $\star :
H^{n_1}(\mc A, \mc A)\otimes H^{n_2}(\mc A, \mc A)\rightarrow H^{n_1 +n_2}(\mc A, \mc A)$ by
\be
(\mc C_1 \star \mc C_2)(f_1,\dots ,f_{n_1+n_2}) := \mc C_1 (f_1,\dots
,f_{n_1}) \star \mc C_2(f_{n_1+1},\dots ,f_{n_1 +n_2})
\ee
for $\mc C_1\in C^{n_1}(\mc A, \mc A)$ and $\mc C_2 \in C^{n_2}(\mc A, \mc A)$.

The \emph{Gerstenhaber bracket} of $\mc C_1\in C^{n_1}(\mc A, \mc A)$ and $\mc C_2 \in C^{n_2}(\mc A, \mc A)$ is defined by
\be 
[\mc C_1,\mc C_2]_{\rm G}=\mc C_1 \circ \mc C_2-(-1)^{(n_1+1)\,
  (n_2+1)}\,  \mc C_2 \circ \mc C_1 \label{Gerstenhaber}
\ee
in $C^{n_1+n_2-1}(\mc A, \mc A)$, where the composition product is defined as
\eqa
&& (\mc C_1 \circ \mc C_2)(f_1,\dots ,f_{n_1+n_2-1}) \nn\\[4pt] && \qquad
\ = \ \mc C_1\big(\mc
C_2(f_1, \dots ,f_{n_2}),f_{n_2+1}, \dots ,f_{n_1+n_2-1}\big)  \\ 
&& \qquad \qquad \qquad +\, \sum_{i=1}^{n_1-2}\, (-1)^{i\, n_2}\, \mc
C_1\big(f_1,\dots,f_i,\mc
C_2(f_{i+1},\dots,f_{i+n_2}),f_{i+n_2+1},\dots, f_{n_1+n_2-1}\big) \nn\\
&& \qquad \qquad \qquad \qquad \qquad +\, (-1)^{(n_1+1)\, (n_2+1)} \, \mc C_1\big(f_1,\dots
,f_{n_1-1}, \mc C_2(f_{n_1}, \dots ,f_{n_1+n_2-1}) \big) \nn
\eqaend
for $f_i\in\alg$. The coboundary operator is then given by
\be
\dd_\star\mc C =-[\mc C, \star ]_{\rm G} \ .
\ee
The associativity of the product $\star\in C^2(\mc A, \mc A)$ may be expressed by using
\be
[\star, \star]_{\rm G} (f,g,h)=2\, \big((f\star g)\star h -f\star (g\star
h)\big) \ .
\ee
Associativity is thus equivalent to $\dd_\star\star=[\star,\star]_{\rm
  G} =0$ or $\dd_\star^2=0$; in that case, the \emph{Gerstenhaber
  algebra} $\big(C^\sharp(\alg,\alg) ,\dd_\star,[-,-]_{\rm G} \big)$ is a differential
graded Lie algebra.

\subsection{Schouten--Nijenhuis brackets\label{SNbrackets}}

Let ${\cal V}^\sharp= C^\infty({\cal M},\bigwedge^\sharp T{\cal M})$ be the graded-commutative
algebra of multivector fields on a smooth manifold ${\cal M}$; notice that
${\cal V}^\sharp$ contains the associative algebra ${\cal V}^0=C^\infty({\cal M})$ of smooth complex
functions on ${\cal M}$. The usual Lie bracket of vector fields $[-,-]_{T{\cal M}}$
extends to the canonical
\emph{Schouten--Nijenhuis bracket} $[-,-]_{\rm S}$ on ${\cal V}^\sharp$. It gives ${\cal V}^\sharp$ the structure of
a differential graded
{Gerstenhaber algebra} with vanishing differential, i.e. $[-,-]_{\rm
  S}$ is a graded Lie bracket of degree
$-1$ satisfying the graded Leibniz rule with respect to the
associative (graded-commutative) exterior
product. Given homogeneous multivector fields ${{\cal X}}={\cal X}^{I_1\dots
  I_{|{\cal X}|}}\, \partial_{I_1}\wedge \dots\wedge \partial_{I_{|{\cal X}|}}$ and
${\cal Y}={\cal Y}^{I_1\dots I_{|{\cal Y}|}}\, \partial_{I_1}\wedge \dots\wedge
\partial_{I_{|{\cal Y}|}}$, it is defined by
\bea
[{\cal X},{\cal Y}]_{\rm S}= (-1)^{|{\cal X}|-1}\, {\cal X}\diamond {\cal Y}-(-1)^{|{\cal X}|\, (|{\cal Y}|-1)}\, {\cal Y}\diamond {\cal X}
\eea
in ${\cal V}^{|{\cal X}|+|{\cal Y}|-1}$, where
\bea
{\cal X}\diamond {\cal Y}:= \sum_{l=1}^{|{\cal X}|}\, (-1)^{l-1}\, {\cal X}^{I_1\dots I_{|{\cal X}|}}\,
\partial_l{\cal Y}^{J_1\dots J_{|{\cal Y}|}} \, \partial_{I_1}\wedge \cdots \wedge
  \widehat{\partial_{I_l}} \wedge\cdots \wedge
  \partial_{I_{|{\cal X}|}}\wedge \partial_{J_1}\wedge
  \cdots \wedge \partial_{J_{|{\cal Y}|}}
\eea
and the hat indicates an omitted derivative.

The condition for a bivector $\Theta=\frac12\, \Theta^{IJ}\,
\partial_I\wedge\partial_J$ to define a Poisson structure on
$C^\infty({\cal M})$ can be expressed through
\beq
[\Theta,\Theta]_{\rm S} = \mbox{$\frac1{3!}$}\,\big(\Theta^{IL} \,
\partial_L \Theta^{JK}+\Theta^{JL}\, \partial_L\Theta^{KI} +\Theta^{KL}\,
\partial_L\Theta^{IJ}\big)\,
\partial_I\wedge\partial_J\wedge\partial_K \ .
\eeq
The corresponding antisymmetric bracket $\{f,g\}_\Theta:=\Theta(\dd
f,\dd g)$ for $f,g\in C^\infty({\cal M})$ 
satisfies the Jacobi identity on $C^\infty({\cal M})$ if and only if
$[\Theta,\Theta]_{\rm S}=0$, and thus defines a Poisson bracket. In
terms of the Lichnerowicz coboundary operator $\dd_\Theta :{\cal V}^n\to
{\cal V}^{n+1}$ defined by
\beq
\dd_\Theta=-[-,\Theta]_{\rm S} \ ,
\eeq
the Poisson condition can be
expressed as $\dd_\Theta\Theta=0$ or $\dd_\Theta^2=0$. The Poisson
bracket extends to the cotangent bundle $T^*{\cal M}$ where it encodes the
Schouten--Nijenhuis bracket of multivector fields.

\subsection{Higher derived brackets\label{Higherbrackets}}

Let $\Pi\in {\cal V}^\sharp =C^\infty({\cal M},\bigwedge^\sharp T{\cal M})$ be a multivector field satisfying
$[\Pi,\Pi]_{\rm S}=0$. Following~\cite{Voronov}, we define the \emph{$n$-th
  derived bracket} of $\Pi$ as
\beq
\{{\cal X}_1,\dots,{\cal X}_n\}_\Pi:= [\cdots [\,[\Pi,{\cal X}_1]_{\rm S},{\cal X}_2]_{\rm S},\dots, {\cal X}_n ]_{\rm S}
\label{higherderived}\eeq
for ${\cal X}_i\in {\cal V}^\sharp$ and $n\geq 1$. Then the sequence of brackets
$\{-,\dots,-\}_\Pi$ defines a \emph{higher Poisson structure} on
${\cal V}^\sharp$. Each derived bracket strictly
obeys a generalized Leibniz rule with respect to the exterior product on
${\cal V}^\sharp$, i.e. $\{-,\dots,-\}_\Pi$ is a derivation in each
argument. By~\cite[Corollary~1]{Voronov}, this sequence of higher
Poisson brackets
gives ${\cal V}^\sharp$ the structure of an $L_\infty$-algebra; the full countable
tower of homotopy Jacobi identities is equivalent to the requirement
$[\Pi,\Pi]_{\rm S}=0$. 

In this correspondence we use a parity $\Z_2$-grading defined as the multivector
degree modulo~$2$, and then apply the parity reversion functor. Hence we
introduce the total $\Z_2$-grading ${\cal V}^\sharp={\cal V}_0\oplus {\cal V}_1$ where
${\cal V}_0=C^\infty({\cal M},\bigwedge^{\rm odd}T{\cal M})$ and
${\cal V}_1=C^\infty({\cal M},\bigwedge^{\rm even}T{\cal M})$. Owing to the generalized
Leibniz rule, in examples it suffices to display the bracket at linear order in the
generators of ${\cal V}^\sharp$, with $|1|=1=|x^I|$ and $|\partial_I|=0$.

\subsection{Courant algebroids\label{NWQuant}}

\noindent
{\bf Lie algebroids. \ }
A {\em Lie algebroid} over a smooth manifold ${\cal M}$ is a vector bundle
$E\to {\cal M}$ endowed with a Lie bracket $[-,-]_E$ on
smooth sections of $E$ and a bundle morphism $\rho:E\rightarrow
T{\cal M}$, called the {\em anchor map}, which is compatible with the Lie
bracket on sections, i.e. the tangent map to $\rho$ is a Lie algebra homomorphism,
\begin{equation}
\dd\rho_{[\psi_1,{\psi_2}]_E} =[\dd \rho_{\psi_1},\dd
\rho_{\psi_2}]_{T{\cal M}}~, \qquad
   {\psi_1},{\psi_2}\in C^ \infty({\cal M},E)~,
\end{equation}
and a Leibniz rule is satisfied when multiplying sections of $E$ by smooth functions on ${\cal M}$,
\begin{equation}\label{eq:LALeibniz}
[{\psi_1},f\,
{\psi_2}]_E=f\,[{\psi_1},{\psi_2}]_E+\rho_{\psi_1}(f)\,
{\psi_2}~,\qquad {\psi_1},{\psi_2}\in C^\infty({\cal M},E)~, \qquad f\in C^\infty({\cal M})~.
\end{equation}
Equivalently, a Lie algebroid is a vector bundle $E\to {\cal M}$ endowed with
a differential $\dd_E$ of degree $+1$ on the free graded-commutative algebra
$\bigwedge_{C^\infty({\cal M})}^\sharp C^\infty({\cal M},E)^*$ over
$C^\infty({\cal M})$. For $\omega\in\bigwedge^{n-1}_{C^\infty({\cal M})}
C^\infty({\cal M},E)^*$ and $\psi_i\in C^\infty({\cal M},E)$, the differential
$\dd_E$ is given here by
\bea
\dd_E\, \omega(\psi_1,\dots,\psi_n) &=& \sum_{\sigma\in S_n}\, \Big(
\rho_{\psi_{\sigma(1)}}\,
\big(\omega(\psi_{\sigma(2)},\dots,\psi_{\sigma(n)})
\big) \nonumber \\ && \qquad \qquad +\, \omega\big([\psi_{\sigma(1)},
\psi_{\sigma(2)}]_E ,\psi_{\sigma(3)},\dots,\psi_{\sigma(n)}\big) \Big) \
  . 
\eea
This defines a differential graded algebra
\beq
\CE(E)=\big(\, \mbox{$\bigwedge_{C^\infty({\cal M})}^\sharp$}
C^\infty({\cal M},E)^*\,,\, \dd_E\big)
\label{CEalg}\eeq
which dually has the structure of a Gerstenhaber algebra with
the Lie bracket on $C^\infty({\cal M},E)$ extended as a biderivation with
$[\psi,f]_E=\psi(\dd_Ef) $ for $\psi\in C^\infty({\cal M},E)$ and $f\in
C^\infty({\cal M})$; this bracket
generalizes the Schouten--Nijenhuis bracket of multivector fields. The
pair (\ref{CEalg}) is called the
\emph{Chevalley--Eilenberg algebra} of the Lie algebroid. It is the
complex which computes Lie algebroid cohomology.

A Lie algebroid over a point is just a Lie algebra (with trivial
anchor map), and (\ref{CEalg})
is the usual Chevalley--Eilenberg algebra which computes Lie algebra
cohomology. More generally, Lie algebra bundles provide natural
examples of Lie algebroids.

The tangent Lie algebroid over a manifold ${\cal M}$ is $E=T{\cal M}$ with the
identity anchor map $\rho=\Id_{T{\cal M}}$ and the usual Lie bracket on
vector fields. In this case $\CE(T{\cal M})=\big(\Omega^\sharp({\cal M})\,,\,
\dd\big)$ is the usual de~Rham complex.

Any bivector field $\Theta$ on ${\cal M}$ induces a map 
$\Theta^\sharp:T^*{\cal M} \rightarrow T{\cal M}$ via contraction together with a
bracket on $C^\infty({\cal M},T^*{\cal M})=\Omega^1({\cal M})$ called the Koszul bracket
\begin{equation}
 [\alpha,\beta]_\Theta:=\Lcal_{\Theta^\sharp (\alpha)}\beta-\Lcal_{\Theta^\sharp
   (\beta)}\alpha-\dd\Theta(\alpha,\beta)
\end{equation}
for $\alpha,\beta\in\Omega^1({\cal M})$, where $\Lcal$ denotes the Lie
derivative. Then $E=T^*{\cal M}$, $\rho=\Theta^\sharp$, and $[-,-]_E=[-,-]_\Theta$ defines a Lie
algebroid on ${\cal M}$ if and only if the Schouten--Nijenhuis bracket of $\Theta$ vanishes,
i.e. $\Theta$ defines a Poisson structure on ${\cal M}$. In this case
$\dd_{T^*{\cal M}}=\dd_\Theta= [\Theta,-]_{\rm S}$ is the Lichnerowicz differential and
the Chevalley--Eilenberg algebra (\ref{CEalg}) computes the Poisson
cohomology of~${\cal M}$.

\medskip

\noindent
{\bf Courant algebroids. \ }
The higher structures which arise in this paper, such as twisted
Poisson structures, require a higher extension of the notion of Lie
algebroid. For this, consider a vector bundle $E\rightarrow {\cal M}$ over a smooth manifold ${\cal M}$
equiped with a metric $\langle-,-\rangle$ and an antisymmetric bracket
$[-,-]_E :  C^\infty({\cal M},E)\wedge C^\infty({\cal M},E)\to  C^\infty({\cal M},E)$, together with an anchor map $\rho:E\rightarrow T{\cal M}$. We define the {\em Jacobiator} $J: C^\infty({\cal M},E)\wedge C^\infty({\cal M},E)\wedge C^\infty({\cal M},E)\rightarrow C^\infty({\cal M},E)$ by
\begin{equation}
 J(\psi_1,\psi_2,\psi_3)=\big[[\psi_1,\psi_2]_E\,,\,\psi_3\big]_E+\big[[\psi_2,\psi_3]_E \,,\,\psi_1\big]_E+\big[[\psi_3,\psi_1]_E \,,\,\psi_2\big]_E~,
\end{equation}
a ternary map $[-,-,-]_E: C^\infty({\cal M},E)\wedge C^\infty({\cal M},E)\wedge C^\infty({\cal M},E)\rightarrow  C^\infty({\cal M})$ by
\begin{equation}
 [\psi_1,\psi_2,\psi_3]_E =\mbox{$\frac1{3!}$}\,\big(
 \big\langle[\psi_1,\psi_2]_E\,,\,\psi_3\big\rangle+\big\langle[\psi_2,\psi_3]_E\,,\, \psi_1\big\rangle+\big\langle[\psi_3,\psi_1]_E\,,\, \psi_2\big\rangle \big) ~,
\end{equation}
and the pullback $\sfd: C^\infty({\cal M})\to  C^\infty({\cal M},E)$ of the exterior derivative $\dd$ via the
adjoint map $\rho^*$ by
\begin{equation}
 \langle \sfd f,\psi \rangle=\rho_\psi(f) \ ,
\end{equation}
where $f\in  C^\infty({\cal M})$ and $\psi,\psi_i\in C^\infty({\cal M},E)$; this map
defines a flat connection, $\sfd^2=0$.

Such a vector bundle is called a {\em Courant algebroid}~\cite{Liu:1997aa} if the following conditions are satisfied:
\begin{itemize}
 \item[(i)] The Jacobi identity holds up to an exact expression: \ $J(\psi_1,\psi_2,\psi_3)=\sfd [\psi_1,\psi_2,\psi_3]_E$;
 \item[(ii)] The anchor map $\rho$ is compatible with the bracket: \
   $\rho_{[\psi_1,\psi_2]_E} =[\rho_{\psi_1} ,\rho_{\psi_2} ]_{T{\cal M}} $;
 \item[(iii)] There is a Leibniz rule: \ $[\psi_1,f\, \psi_2]_E=f\,
   [\psi_1,\psi_2]_E+\rho_{\psi_1}( f)\, \psi_2-\mbox{$\frac12$}\, \langle
   \psi_1,\psi_2\rangle \, \sfd f$;
 \item[(iv)] $\langle \sfd f,\sfd g\rangle=0$;
 \item[(v)] $\rho_\psi\big(\langle \psi_1,\psi_2\rangle \big) =\big\langle[\psi
   ,\psi_1]_E+\mbox{$\frac12$}\, \sfd\langle \psi
   ,\psi_1\rangle\,,\,\psi_2 \big\rangle+ \big\langle \psi_1 \,,\,[\psi
   ,\psi_2]_E+\mbox{$\frac12$}\, \sfd\langle \psi ,\psi_2\rangle \big\rangle$;
\end{itemize}
where $\psi,\psi_i \in C^\infty({\cal M},E)$ and
$f,g\in C^\infty({\cal M})$.

The graded differential Lie algebra (\ref{CEalg}) is now generalized to a
Lie 2-algebra: The structure maps $\sfd,[-,-]_E, [-,-,-]_E$ of the
Courant algebroid $E\to {\cal M}$ on the
complex
\beq
C^\infty({\cal M}) \ \xrightarrow{ \ \sfd \ } \ C^\infty({\cal M},E) \ ,
\label{CourantLie2alg}\eeq
extended as $[\psi,f]_E:= \frac12\,\langle\sfd f,\psi\rangle$ for
$\psi\in C^\infty({\cal M},E)$ and $f\in C^\infty({\cal M})$, define a 2-term
$L_\infty$-algebra~\cite{RoytWein}.

\subsection{Lie 2-groups\label{Lie2groups}}

A group is a monoid in which every element has an inverse; 2-groups
are categorifications of groups. For this, recall that 
a \emph{tensor} or \emph{monoidal category} is a category
$\CCC=(\CCC_0,\CCC_1)$ equipped with an exterior product $\otimes:\CCC\times
\CCC\to\CCC$ together with an identity object $\unit\in\CCC_0$ and
three natural functorial isomorphisms: The unity isomorphisms
$\unit_X:= \unit\otimes {X}\cong {X} \cong X \otimes\unit$ in $\CCC_1$ for all objects
$X\in\CCC_0$, and the associator isomorphisms
\beq
{\mathscr P}={\mathscr P}_{X,Y,Z}\,:\, (X\otimes Y)\otimes Z \ \xrightarrow{ \ \approx
  \ } \
X\otimes (Y\otimes Z)
\label{associso}\eeq
for all objects $X,Y,Z\in\CCC_0$. They satisfy the pentagon identities
\beq
(\unit_X\otimes {\mathscr P}_{Y,Z,W})\circ {\mathscr P}_{X,Y\otimes
  Z,W}\circ{\mathscr P}_{X,Y,Z\otimes\unit_W}= 
{\mathscr P}_{X,Y,Z\otimes W}\circ {\mathscr P}_{X\otimes Y,Z,W}
\label{pentagonid}\eeq
which state that the five ways of bracketing four objects commutes,
and also the triangle identities which state that the associator
isomorphism with $Y=\unit$ is compatible with the unity
isomorphims. For morphisms $\CCF:X\to Y$ and $\CCF':X'\to Y'$, their
exterior product is the morphism $\CCF\otimes\CCF':X\otimes X'\to
Y\otimes Y'$ in $\CCC_1$. By MacLane's coherence theorem, these identities ensure that all higher associators are
consistent.

We call $\CCC$ \emph{braided} when there are natural functorial isomorphisms
\beq
\CCB=\CCB_{X,Y} \,:\, X\otimes Y \ \xrightarrow{ \ \approx \ } \ Y\otimes X
\eeq
for any pair of objects $X,Y\in \CCC_0$, called commutativity relations. The braiding $\CCB_{X,Y}$ satisfies two conditions,
one expressing $\CCB_{X\otimes Y,Z}$ in terms of associativity relations
$\Id_X\otimes \CCB_{Y,Z}$ and $\CCB_{Z,X}\otimes\Id_Y$, and a similar one
for $\CCB_{X,Y\otimes Z}$.

An object $\alg\in\CCC_0$ in a tensor category $\CCC$ is an \emph{algebra} or
\emph{monoid object} if there is a
``multiplication'' morphism
$\circledast:\alg\otimes\alg\to \alg$, written $a\otimes b\mapsto a\circledast b$, which is associative in the category,
i.e. it satisfies the associativity condition
\beq
\circledast\circ(\circledast\otimes\Id_\alg)=\circledast\circ
(\Id_\alg\otimes\circledast)\circ{\mathscr P}_{\alg,\alg,\alg}
\label{catassoc}\eeq
as maps $(\alg\otimes\alg) \otimes\alg\to \alg$. By MacLane's
coherence theorem, we can
deal with nonassociative algebras in this way by expressing usual algebraic
operations as compositions of maps and doing the same in the monoidal
category with the relevant associator $\CCP$ inserted between any
three objects as needed in order to make sense of expressions. If in addition $\CCC$
is braided, then $\alg$ is \emph{commutative} if its product morphism
obeys
\beq
\circledast\circ\CCB_{\alg,\alg}=\circledast
\eeq
as maps $\alg\otimes\alg\to\alg$.
A \emph{group object}
is a monoid
object $\alg$ together with a ``unit'' morphism $1_\alg:\unit\to\alg$
satisfying the unit condition
\beq
\circledast\circ (1_\alg\otimes\Id_\alg)=\Id_\alg= \circledast\circ (\Id_\alg\otimes 1_\alg) \
,
\eeq
such that every element of $\alg$ has an inverse with respect to the
product morphism $\circledast$ and the identity $1_\alg$.

A \emph{$2$-group} is a monoidal category in which every object and
morphism has an inverse. A \emph{Lie $2$-group} is a pair
$\CCG=(\CCG_0,\CCG_1)$ of objects in the category of smooth manifolds
and smooth maps, with source and target maps
$\sfs,\sft:\CCG_1\rightrightarrows \CCG_0$, and a vertical
multiplication $\circ:\CCG_1\times \CCG_1\to \CCG_1$ of morphisms. In addition there is a
horizontal multiplication functor $\otimes:\CCG\times \CCG\to \CCG$ on
objects and morphisms, an
identity object $1$, and a contravariant inversion functor
$(-)^{-1}: \CCG\to\CCG$ together with natural isomorphisms provided by the
associator ${\mathscr P}_{g,h,k}:(g\otimes h)\otimes k\to g\otimes(h\otimes k)$, the
left and right units $1\otimes g\cong g \cong g\otimes1$, and the units
and counits $g\otimes g^{-1}\cong 1 \cong g^{-1}\otimes g$ obeying pentagon,
triangle and zig-zag identities; see~\cite[Section~7]{Baez} for details. If the
structure morphisms are all identity isomorphisms, the Lie 2-group
$\CCG$ is
called \emph{strict}; otherwise $\CCG$ is \emph{semistrict}.

A Lie 2-group $\CCG=(\CCG_0,\CCG_1)$ is \emph{special} if its source
and target morphisms
$\sfs,\sft:\CCG_1\rightrightarrows \CCG_0$ are equal, and the units and counits are all identity
isomorphisms. There is a bijective correspondence between special Lie
2-groups and triples $(G,H,\varphi)$ consisting of a Lie group $G$, an
action of $G$ as automorphisms of an abelian group $H$, and a
normalized smooth $3$-cocycle $\varphi:G\times G\times G\to H$; the
isomorphism classes are parameterized by elements $[\varphi]\in H^3(G,H)$
in the degree $3$ group cohomology with smooth cocycles.
Given a triple $(G,H,\varphi)$, the corresponding semistrict Lie 2-group
$\CCG=(\CCG_0,\CCG_1)$ has the Lie group $\CCG_0= G$ as the space of
objects, the semi-direct product Lie group $\CCG_1= G\ltimes H$ as the
space of morphisms, and the associator ${\mathscr P}$ is given by the
action of 
$\varphi$; the source and target maps
$\sfs,\sft:\CCG_1\rightrightarrows\CCG_0$ are both projection onto the
first factor of $G\times H$, while the
cocycle condition on $\varphi$ is equivalent to the pentagon
identities (\ref{pentagonid}). In this correspondence the abelian
group $H$ is the group of automorphisms of the identity object $1$ in
the monoidal category $\CCG$.

The exponential map takes an ordinary Lie algebra to
its integrating simply connected Lie group, while the tangent space at the identity of
an ordinary Lie group is the corresponding infinitesimal Lie
algebra. In marked contrast, there are no general constructions relating Lie 2-algebras
and Lie 2-groups. Integration/differentiation between \emph{strict} Lie
2-algebras and \emph{strict} Lie 2-groups is described
in~\cite{Baez,Baez:2003fs}; a general procedure for integrating
$L_\infty$-algebras is described in~\cite{Getzler,Henriques}. In the semistrict cases of interest to us
in this paper, given a triple $(G,H,\varphi)$ representing a special Lie 2-group $\CCG$ (with $H$ an abelian Lie
group), by differentiation
we obtain a triple $(\frg,W,j)$ representing a 2-term
$L_\infty$-algebra $V$ (with $W$
regarded as an abelian Lie algebra); in this case we call the Lie 2-group $\CCG$ an
\emph{integration} of the Lie 2-algebra $\CCV$ corresponding to $V$.

\setcounter{equation}{0}

\newsection{Weights of Kontsevich diagrams\label{Graphs}}

In this appendix we explain in some detail how to calculate the
weights (\ref{Weight}) of the diagrams that 
enter into Kontsevich's formula (\ref{Kontsevich}) and present some
representative examples of the computations.

The edges of a generic diagram $\Gamma$ between two vertices
$p,q\in\quat$ lie on semicircular geodesics $\ell(p,q)$
in the hyperbolic upper half-plane $\mathbb{H}$. 
The \emph{harmonic angle} $\phi^h=\phi^h(p,q)$ is defined to be the
angle between an edge $\ell(p,q)$ and the directed geodesic
$\ell(p,\infty)$ at $p$; it
may be integrated to provide the weight $w_\Gamma$ with which each multidifferential 
operator contibutes to the star product~(\ref{Kontsevich}). This is
depicted in the following diagram:

\medskip

%						% KONTSEVICH FORMULATION %
%
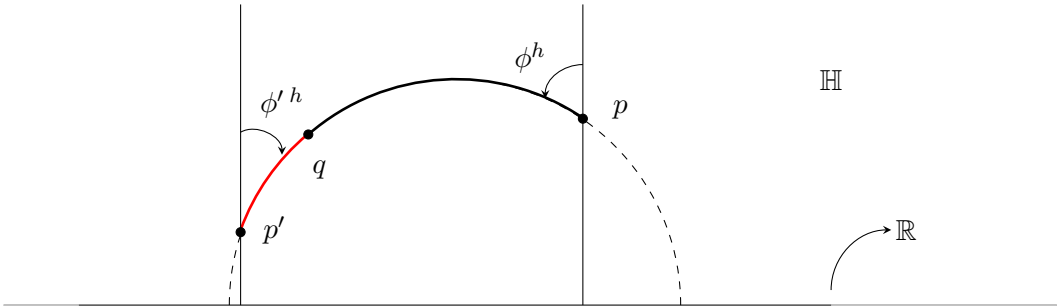
\usetikzlibrary{arrows,automata,backgrounds}
\begin{tikzpicture}[scale=1,>=stealth]\label{harmonic angle}
%
							% real line %
\draw (1,0)--(15,0);
\draw [white] (1,0)--(2,0);
\draw [white] (12,0)--(15,0);
							% graph %
\draw [line width=1] (5.05,2.27) arc (131:56:3cm);
\draw [line width=1,red] (4.15,1) arc (160:130:3cm);
\draw [dashed] (4,0) arc (180:160:3cm);
\draw [dashed] (10,0) arc (0:80:3cm);
							% angles %
\draw (8.7,0)--(8.7,4);
\node at (8,3.3) {$\phi^h$};
\draw[<-] (8.2,2.75)  to [out=90, in=180] (8.7,3.2);
\draw (4.15,0)--(4.15,4);
\node at (4.7,2.7) {$\phi'\,^h$};
\draw[->] (4.15,2.3)  to [out=30, in=90] (4.7,2);
							% points %
\node at (9.2,2.6) {$p$};
\draw [fill=black] (8.7,2.48) circle (.6mm);
\node at (5.2,1.8) {$q$};
\draw [fill=black] (5.05,2.27) circle (.6mm);
\node at (4.6,1) {$p'$};
\draw [fill=black] (4.15,0.97) circle (.6mm);
\node at (13,1) {$\mathbb{R}$};
\draw[->] (12,0.2)  to [out=90, in=180] (12.8,1);
\node at (12,3) {$\mathbb{H}$};
\end{tikzpicture}

\medskip

Angles in $\mathbb{H}$ are defined in the usual manner; thus 
$\phi^h, \ph\phi'\,^h\in [0,\pi]$ as points $p$ and $p'$ run along the
semicircle from the real axis
$\mathbb{R}$ (the boundary of $\mathbb{H}$) to $q$ in $\mathbb{H}$. It is important to note that the harmonic angle is measured 
\emph{counterclockwise}. This means that $\phi^h \in [0,2\pi]$ as we cross 
$q$ to integrate over $\mathbb{H}$ along the semicircle.

\medskip

\noindent
{\bf Bivector diagrams. \ }
As an example, let us calculate the weight of the wedge which corresponds to the 
twisted Poisson bracket $\Theta^{IJ}\,\p_I f\, \p_J g$; here we denote $\phi^h_{e^1_1}$ 
by $\theta_1$ and $\phi^h_{e^2_1}$ by~$\psi_1$:

\medskip

%						% TOTALLY GROUNDED WEDGE %
%
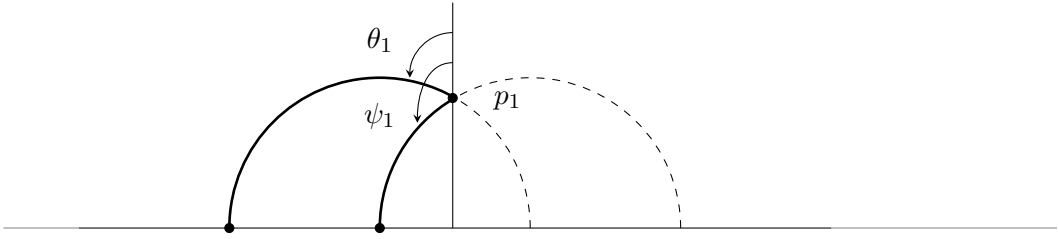
\usetikzlibrary{arrows,automata,backgrounds}
\begin{tikzpicture}[scale=1,>=stealth]\label{tot ground wedge}
%
							% real line %
\draw (1,0)--(15,0);
\draw [white] (1,0)--(2,0);
\draw [white] (12,0)--(15,0);
							% wedge %
\draw [line width=1] (4,0) arc (180:60:2cm);
\draw [dashed] (8,0) arc (0:180:2cm);
\draw [line width=1] (6,0) arc (180:120:2cm);
\draw [dashed] (10,0) arc (0:180:2cm);
							% angles %
\draw (6.97,0)--(6.97,3);
\node at (6,2.5) {$\theta_1$};
\draw[<-] (6.4,2)  to [out=90, in=180] (6.97,2.6);
\node at (6,1.5) {$\psi_1$};
\draw[<-] (6.5,1.4)  to [out=90, in=180] (6.97,2.2);
							% points %
\node at (7.7,1.7) {$p_1$};
\draw [fill=black] (6.97,1.73) circle (.6mm);
\draw [fill=black] (4,0) circle (.6mm);
\draw [fill=black] (6,0) circle (.6mm);
\end{tikzpicture}

\medskip

Integrating the two-form $\dd\theta_1\wedge \dd\psi_1$ over $\quat$, keeping in mind 
that $\psi_1 > \theta_1$, is straightforward and gives the weight
\be 
\frac{1}{(2\pi)^2}\, \int_0^{2\pi}\, {\dd\psi_1 \
    \int_0^{\psi_1}\, {\dd\theta_1}}= \frac{1}{(2\pi)^2}\, \int_0^{2\pi}\, {\dd\psi_1 \ \psi_1}=\frac 1 2 \ .
\ee
It is important to note here that changing the order of integration produces a minus 
sign since $\dd\psi_1 \wedge \dd\theta_1  = - \dd\theta_1\wedge \dd\psi_1$. This means 
that the topologically equivalent \emph{tractable} wedge
has weight equal to $-\frac 1 2$: A tractable diagram is one that has the derivatives assigned to its edges
 reversed, i.e. the tractable wedge corresponds to $\Theta^{IJ}\, \p_J
 f\, \p_I g$.

\medskip

\noindent
{\bf Triple product diagrams. \ }
Let us now calculate the weights of the following three diagrams which appear at order $\hbar^2$ when we
 star multiply three functions:

\medskip

%						% W-COMPUTABLE GROUNDED TRIVECTOR %
%
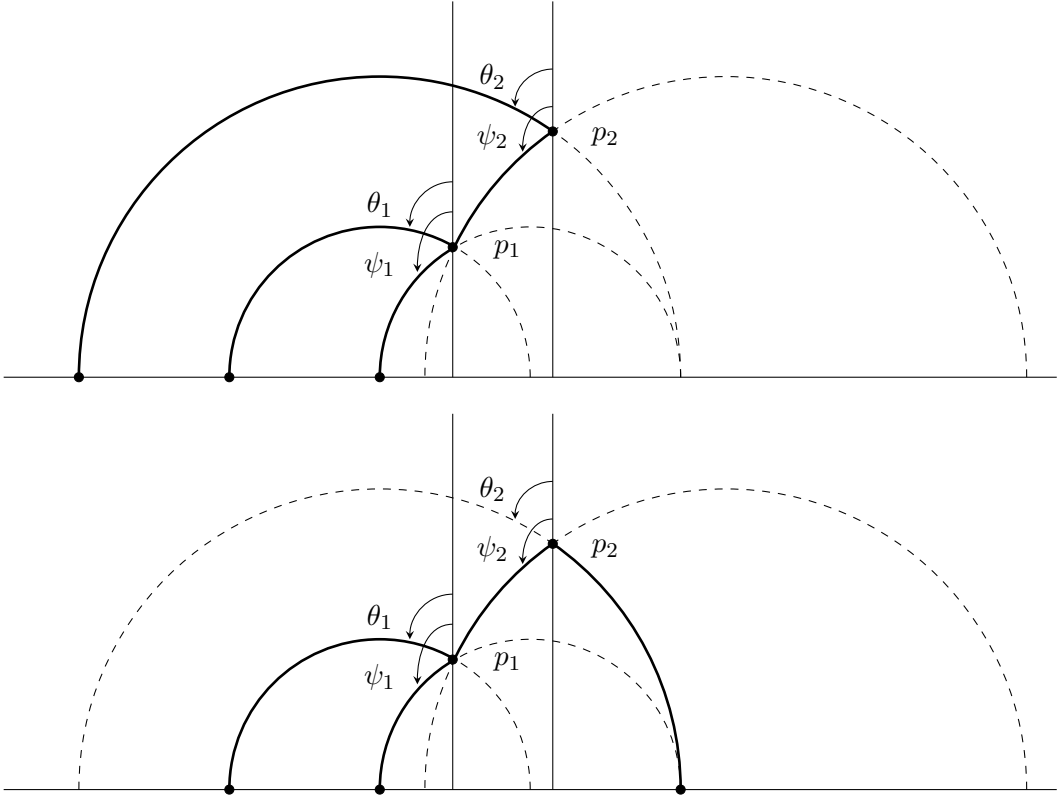
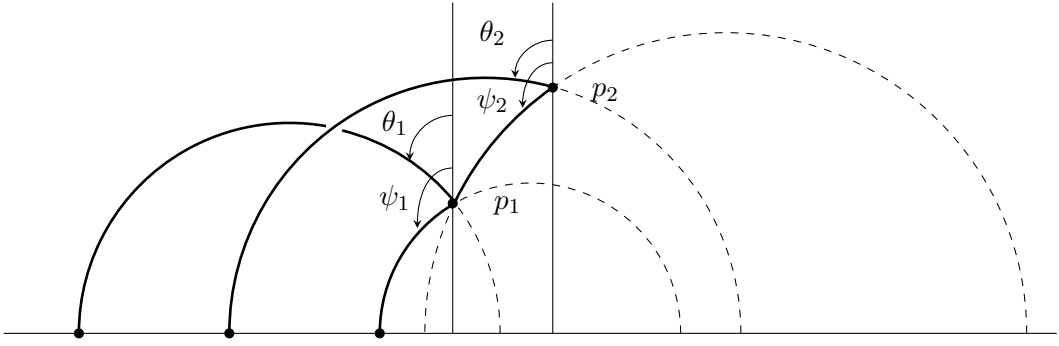
\begin{tikzpicture}[scale=1,>=stealth]
%
							% real line %
\draw (1,0)--(15,0);
%\draw [white] (1,0)--(2,0);
%\draw [white] (12,0)--(15,0);
							% grounded wedge %
\draw [line width=1] (4,0) arc (180:60:2cm);
\draw [dashed] (8,0) arc (0:180:2cm);
\draw [line width=1] (6,0) arc (180:120:2cm);
\draw [dashed] (10,0) arc (0:180:2cm);
							% with angles%
\draw (6.97,0)--(6.97,5);
\node at (6,2.3) {$\theta_1$};
\draw[<-] (6.4,2)  to [out=90, in=180] (6.97,2.6);
\node at (6,1.5) {$\psi_1$};
\draw[<-] (6.5,1.4)  to [out=90, in=180] (6.97,2.2);
							% semi-aerial wedge %
\draw [line width=1] (2,0) arc (180:55:4cm);
\draw [dashed] (10,0) arc (0:180:4cm);
\draw [line width=1] (7,1.75) arc (154:125:4cm);
\draw [dashed] (14.6,0) arc (0:180:4cm);
							% with angles%
\draw (8.3,0)--(8.3,5);
\node at (7.5,4) {$\theta_2$};
\draw[<-] (7.8,3.6)  to [out=90, in=180] (8.3,4.1);
\node at (7.5,3.2) {$\psi_2$};
\draw[<-] (7.9,3)  to [out=90, in=180] (8.3,3.6);
							% points %
\node at (7.7,1.7) {$p_1$};
\draw [fill=black] (6.97,1.73) circle (.6mm);
\node at (9,3.2) {$p_2$};
\draw [fill=black] (8.3,3.27) circle (.6mm);
\draw [fill=black] (2,0) circle (.6mm);
\draw [fill=black] (4,0) circle (.6mm);
\draw [fill=black] (6,0) circle (.6mm);
\end{tikzpicture}

\medskip

%						% RIGHT-TRACTABLE GROUNDED TRIVECTOR - {1} %
%
\begin{tikzpicture}[scale=1,>=stealth]
%
							% real line %
\draw (1,0)--(15,0);
%\draw [white] (1,0)--(2,0);
%\draw [white] (12,0)--(15,0);
							% grounded wedge %
\draw [line width=1] (4,0) arc (180:60:2cm);
\draw [dashed] (8,0) arc (0:180:2cm);
\draw [line width=1] (6,0) arc (180:120:2cm);
\draw [dashed] (10,0) arc (0:180:2cm);
							% with angles%
\draw (6.97,0)--(6.97,5);
\node at (6,2.3) {$\theta_1$};
\draw[<-] (6.4,2)  to [out=90, in=180] (6.97,2.6);
\node at (6,1.5) {$\psi_1$};
\draw[<-] (6.5,1.4)  to [out=90, in=180] (6.97,2.2);
							% semi-aerial wedge %
\draw [dashed] (2,0) arc (180:0:4cm);
\draw [line width=1] (10,0) arc (0:55:4cm);
\draw [line width=1] (7,1.75) arc (154:125:4cm);
\draw [dashed] (14.6,0) arc (0:180:4cm);
							% with angles%
\draw (8.3,0)--(8.3,5);
\node at (7.5,4) {$\theta_2$};
\draw[<-] (7.8,3.6)  to [out=90, in=180] (8.3,4.1);
\node at (7.5,3.2) {$\psi_2$};
\draw[<-] (7.9,3)  to [out=90, in=180] (8.3,3.6);
							% points %
\node at (7.7,1.7) {$p_1$};
\draw [fill=black] (6.97,1.73) circle (.6mm);
\node at (9,3.2) {$p_2$};
\draw [fill=black] (8.3,3.27) circle (.6mm);
\draw [fill=black] (4,0) circle (.6mm);
\draw [fill=black] (6,0) circle (.6mm);
\draw [fill=black] (10,0) circle (.6mm);
\end{tikzpicture}

\medskip

%						% LEFT-TRACTABLE GROUNDED TRIVECTOR - {2} %
%
\begin{tikzpicture}[scale=1,>=stealth]
%
							% real line %
\draw (1,0)--(15,0);
%\draw [white] (1,0)--(2,0);
%\draw [white] (12,0)--(15,0);
							% grounded wedge %
\draw [line width=1] (2,0) arc (180:80:2.8cm);
\draw [line width=1] (5.5,2.7) arc (76:39:2.8cm);
\draw [dashed] (7.6,0) arc (0:60:2.8cm);
\draw [line width=1] (6,0) arc (180:120:2cm);
\draw [dashed] (10,0) arc (0:180:2cm);
							% with angles%
\draw (6.97,0)--(6.97,4.4);
\node at (6.2,2.8) {$\theta_1$};
\draw[<-] (6.4,2.3)  to [out=90, in=180] (6.97,2.9);
\node at (6.2,1.8) {$\psi_1$};
\draw[<-] (6.5,1.4)  to [out=90, in=180] (6.97,2.2);
							% semi-aerial wedge %
\draw [line width=1] (4,0) arc (180:75:3.4cm);
\draw [dashed] (10.8,0) arc (0:180:3.4cm);
\draw [line width=1] (7,1.75) arc (154:125:4cm);
\draw [dashed] (14.6,0) arc (0:180:4cm);
							% with angles%
\draw (8.3,0)--(8.3,4.4);
\node at (7.5,4) {$\theta_2$};
\draw[<-] (7.8,3.4)  to [out=90, in=180] (8.3,3.9);
\node at (7.5,3.1) {$\psi_2$};
\draw[<-] (7.9,3)  to [out=90, in=180] (8.3,3.6);
							% points %
\node at (7.7,1.7) {$p_1$};
\draw [fill=black] (6.97,1.73) circle (.6mm);
\node at (9,3.2) {$p_2$};
\draw [fill=black] (8.3,3.27) circle (.6mm);
\draw [fill=black] (2,0) circle (.6mm);
\draw [fill=black] (4,0) circle (.6mm);
\draw [fill=black] (6,0) circle (.6mm);
\end{tikzpicture}

\medskip

For the first two diagrams we have $\psi_2 >\theta_2$ and $\psi_2
>\psi_1 >\theta_1$, and thus we get the weights
\be 
w_1=\frac{1}{(2\pi)^4}\, \int_0^{2\pi}\, {\dd\psi_1 \
  \int_0^{\psi_1}\, {\dd\theta_1 \ \int_{\psi_1}^{2\pi}{\dd\psi_2 \
      \int_0^{\psi_2}\, {\dd\theta_2}}}} =\frac{1}{8} 
\ee
for the first diagram and
\be 
w_2=\frac{1}{(2\pi)^4}\, \int_0^{2\pi}\, {\dd\psi_1 \
  \int_0^{\psi_1}\, {\dd\theta_1 \ \int_{\psi_1}^{2\pi}\, {\dd\psi_2 \ 
      \int_0^{\psi_2}\,- {\dd\theta_2}}}} =-\frac{1}{8} 
\ee
for the second diagram (which is tractable). The third diagram has $\psi_2 >\theta_2 >\theta_1$ 
and $\psi_2 >\psi_1 >\theta_1$ which gives the weight
\be 
w_3=\frac{1}{(2\pi)^4}\,\int_0^{2\pi} \, {\dd\psi_1 \ \int_0^{\psi_1}\,
  {\dd\theta_1  \ \int_0^{2\pi}\, {\dd\psi_2 \
      \int_{\theta_1}^{\psi_2}\, {\dd\theta_2}}}} =\frac{1}{12} \ .
\ee

\medskip

\noindent
{\bf Trivector diagrams. \ }
Finally, we calculate the weight of the following trivector diagram that
enters the associator $\Phi(\Pi)$:

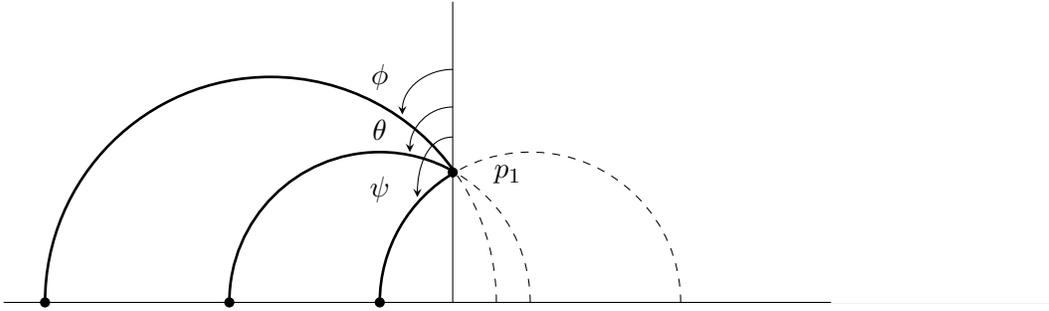
\medskip

%						% TOTALLY GROUNDED TRIVECTOR %
%
\begin{tikzpicture}[scale=1,>=stealth]
%
							% real line %
\draw (1,0)--(15,0);
%\draw [white] (1,0)--(2,0);
\draw [white] (12,0)--(15,0);
							% grounded graph %
\draw [line width=1] (4,0) arc (180:60:2cm);
\draw [dashed] (8,0) arc (0:180:2cm);
\draw [line width=1] (6,0) arc (180:120:2cm);
\draw [dashed] (10,0) arc (0:180:2cm);
\draw [line width=1] (1.55,0) arc (180:35:3cm);
\draw [dashed] (7.55,0) arc (0:180:3cm);
							% angles%
\draw (6.97,0)--(6.97,4);
\node at (6,2.3) {$\theta$};
\draw[<-] (6.4,2)  to [out=90, in=180] (6.97,2.6);
\node at (6,1.5) {$\psi$};
\draw[<-] (6.5,1.4)  to [out=90, in=180] (6.97,2.2);
\node at (6,3) {$\phi$};
\draw[<-] (6.3,2.5)  to [out=90, in=180] (6.97,3.1);
							% points %
\node at (7.7,1.7) {$p_1$};
\draw [fill=black] (6.97,1.73) circle (.6mm);
\draw [fill=black] (1.55,0) circle (.6mm);
\draw [fill=black] (4,0) circle (.6mm);
\draw [fill=black] (6,0) circle (.6mm);
\end{tikzpicture}

\medskip

Here $\psi >\theta >\phi$ and the formula \eqref{Weight} for the diagram weight gives~\cite{Konts:1997}
\eqa
w &=& \frac1{(2\pi)^3}\, \int_{\quat_3}\, \dd\phi\wedge
\dd\theta\wedge \dd\psi \ H(\psi-\theta)\, H(\theta-\phi)\,
H(\psi-\phi) \nn \\[4pt]  &=& \frac{1}{(2\pi)^3}\, \int_0^{2\pi}\, {\dd\psi \ \int_0^\psi\,
  {\dd\theta \ \int_0^\theta\, {\dd\phi}}} \ = \ \frac 1 6 \ ,
\eqaend
where $H$ denotes the Heaviside step function.

\end{document}